\documentclass[aps,prd,reprint,twocolumn,superscriptaddress,letterpaper,longbibliography]{revtex4-1}

\usepackage{amsmath}
\usepackage{amsfonts}
\usepackage{amssymb}
\usepackage{graphicx}
\usepackage{epsfig}
\usepackage{color}
\usepackage{multirow}
\usepackage{todonotes}
\usepackage{enumitem}
\usepackage{hyperref}
\usepackage[normalem]{ulem}

\def\bea{\begin{eqnarray}}
\def\eea{\end{eqnarray}}
\def\bef{\begin{flalign}}
\def\eef{\end{flalign}}
\def\nn{\nonumber}
\def\d{\mathrm{d}}

\def\la{\langle}
\def\ra{\rangle}
\def\({\left(}
\def\){\right)}
\def\[{\left[}
\def\]{\right]}
\def\<{\left\langle}
\def\>{\right\rangle}

\newcommand{\icol}[1]{% inline column vector
\left[\begin{smallmatrix}#1\end{smallmatrix}\right]%
}

\newcommand{\Tr}[1]{\text{Tr}\left[ #1 \right]}

\newcommand{\dd}{\text{d}}
\newcommand{\ket}[1]{\left\vert #1 \right\rangle}

\newcommand{\braket}[2]{\left.\left\langle #1 \right\vert #2 \right\rangle}

\newcommand{\braOket}[3]{\left\langle #1 \left\vert #2 \right\vert #3 \right\rangle}

\newcommand{\schrodinger}{Schr\"{o}dinger}

\begin{document}

%%%%%%%%%%%%%%%%%%%%%%%%%%%%%%%%%%%%%%%%%%%%%%%%%%

\title{Initial value formulation of a quantum damped harmonic oscillator}

\author{Nishant Agarwal}
\email{nishant\_agarwal@uml.edu}
\affiliation{Department of Physics and Applied Physics, University of Massachusetts, Lowell, MA 01854, USA}

\author{Yi-Zen Chu}
\email{yizen.chu@gmail.com}
\affiliation{Department of Physics, National Central University, Chungli 32001, Taiwan, Republic of China}
\affiliation{Center for High Energy and High Field Physics (CHiP), National Central University, Chungli 32001, Taiwan, Republic of China}

\date{\today}

\begin{abstract}
The in-in formalism and its influence functional generalization are widely used to describe the out-of-equilibrium dynamics of unitary and open quantum systems, respectively. In this paper, we build on these techniques to develop an effective theory of a quantum damped harmonic oscillator and use it to study initial state-dependence, decoherence, and thermalization. We first consider a Gaussian initial state and quadratic influence functional and obtain general equations for the Green's functions of the oscillator. We solve the equations in the specific case of time-local dissipation and use the resulting Green's functions to obtain the purity and unequal-time two-point correlations of the oscillator. We find that the dynamics must include a non-vanishing noise term to yield physical results for the purity and that the oscillator decoheres in time such that the late-time density operator is thermal. We show that the frequency spectrum or unequal-time correlations can, however, distinguish between the damped oscillator and an isolated oscillator in thermal equilibrium, and obtain a generalized fluctuation-dissipation relation for the damped oscillator. We briefly consider time-nonlocal dissipation as well, to show that the fluctuation-dissipation relation is satisfied for a specific choice of dissipation kernels. Lastly, we develop a double in-out path integral approach to go beyond Gaussian initial states and show that our equal-time results for time-local dissipation are in fact non-perturbative in the initial state.
\end{abstract}

\maketitle

%%%%%%%%%%%%%%%%%%%%%%%%%%%%%%%%%%%%%%%%%%%%%%%%%%

%-------------------------------
\section{Introduction}
%-------------------------------

Green's functions are extremely useful for calculating correlation functions of quantum systems, both in- and out-of-equilibrium. In an equilibrium quantum field theory (QFT), for example, we typically assume that the field is in the ground state in the infinite past and future and that any interaction is turned on and off adiabatically. Specifying both initial and final conditions picks out the Feynman Green's function as the primary Green's function, in terms of which we can obtain any time-ordered correlation function of the field. In an out-of-equilibrium QFT, on the other hand, we are typically interested in finite-time correlations, with the field initialized in any state at a finite initial time. Specifying initial conditions now picks out the retarded Green's function as the primary Green's function, in terms of which we can obtain any field correlations.

The Green's functions of a given quantum system are most readily obtained through the path integral approach. The standard in-out path integral, for example, allows us to obtain the generating functional and hence correlation functions in an equilibrium QFT. Its out-of-equilibrium generalization or the in-in path integral \cite{Schwinger:1960qe,Bakshi:1962dv,Bakshi:1963bn,Keldysh:1964ud,Jordan:1986ug} instead allows us to obtain the generating functional and hence correlation functions in an out-of-equilibrium QFT. While the original in-in path integral only describes unitary dynamics, i.e., the dynamics of a {\it closed} quantum system that is initialized in any state and evolves with a time-independent or time-dependent Hamiltonian, it can be further generalized to describe the non-Hamiltonian dynamics of an {\it open} quantum system by introducing an influence functional \cite{Feynman:1963,Feynman:2012}. The influence functional is the path integral analog of the quantum master equation \cite{Breuer:2002pc,Calzetta:2008} and simplifies the calculation of certain quantities in open quantum systems, such as Green's functions and unequal-time correlations \cite{Boyanovsky:2015xoa}.

The influence functional method has been used in a variety of problems since it was first developed. Amongst exactly solvable ones are the quantum damped harmonic oscillator (DHO) and linear Brownian motion, for which a standard reference is the textbook \cite{Weiss:2012}. An incomplete list of other problems for which it has been used is the study of quantum Brownian motion in different environments \cite{Hu:1991di,Hu:1993vs}, quantum transport in interacting nanojunctions \cite{Magazzu:2022}, decoherence in interacting QFTs \cite{Koksma:2009wa,Koksma:2011dy} and inflation \cite{Lombardo:2005iz}, entanglement in primordial correlations \cite{Boyanovsky:2016exa,Boyanovsky:2018soy}, coarse-graining in interacting QFTs \cite{Lombardo:1995fg,Agon:2014uxa}, and open holographic QFTs \cite{Jana:2020vyx}. Given the versatility of the method, in this paper, we revisit one of the simplest out-of-equilibrium quantum systems described by an influence functional -- a quantum DHO -- with a goal of developing an effective field theory-inspired approach to the problem. We are thus interested in constraining parameters that appear in the influence functional on physical grounds, without knowledge of the full microscopic model.

We first initialize the oscillator in a Gaussian state and evolve it with a general but quadratic influence functional while remaining agnostic to the source of dissipation. We then specialize to time-local dissipation, find exact solutions for the Green's functions in this case, and show that terms in the influence functional are constrained, due in large part to fluctuation-dissipation relations of {\it environment} degrees of freedom. We find, in particular, that the influence functional must contain a non-vanishing {\it noise} term, which leads, when set to zero, to a non-physical late-time purity of the oscillator. Using purity as an indicator of decoherence, we next show that the DHO decoheres in time and settles into a thermal state at a temperature defined by the dissipation parameters. We show that the frequency spectrum or unequal-time correlations can, however, distinguish between the DHO and an isolated oscillator in thermal equilibrium. We use unequal-time correlations in the late-time limit to obtain a generalized fluctuation-dissipation relation and show that it reduces to the usual relation only in the high-temperature regime. We further consider time-nonlocal dissipation with a specific choice of dissipation kernels, where the fluctuation-dissipation relation is satisfied at any temperature. Finally, we develop a double in-out path integral approach that allows us to obtain equal-time correlations for any initial state and show that our equal-time results for time-local dissipation are in fact non-perturbative in the initial state.

The paper is organized as follows. In Sec.\ \ref{sec:Zinin}, we review the in-in formalism in the context of a harmonic oscillator, deriving the corresponding generating functional in an Appendix for completeness, and discuss the initial state, which we choose to be Gaussian, and the influence functional, which we choose to be quadratic. In Sec.\ \ref{sec:greensfunctions}, we write the generating functional in terms of Green's functions, use them to obtain $n$-point correlations, again relegating details to an Appendix, find the initial conditions they must satisfy, and obtain their equations of motion. In Sec.\ \ref{sec:tldissipation}, we restrict to time-local dissipation and obtain exact solutions for the Green's functions in this case, highlighting the contribution from the noise term. We use these solutions to understand how the oscillator's purity evolves in time and whether it satisfies the fluctuation-dissipation relation at late times. In Sec.\ \ref{sec:tnldissipation}, we consider the fluctuation-dissipation relation for time-nonlocal dissipation with a specific choice of dissipation kernels. In Sec.\ \ref{sec:beyondgaussianis}, we develop a double in-out path integral approach that allows us to go beyond Gaussian initial states and use it to show that our results for equal-time correlations and purity in the case of time-local dissipation hold for any initial state. We end with a discussion in Sec.\ \ref{sec:discussion}.

%-------------------------------
\section{In-in generating functional}
\label{sec:Zinin}
%-------------------------------

In this section, we obtain the generating functional of a quantum harmonic oscillator that is initialized in a Gaussian initial state and undergoes non-unitary/dissipative evolution described by a quadratic dissipative action; also see Ref. \cite{BenTov:2021jsf} for a recent review, Ref. \cite{Um:2002ab} for an earlier review, and Ref. \cite{Markus:2022zbu} for related recent work. We denote the initial state of the oscillator at time $t_0$ by the density operator $\hat{\rho}(t_0)$ and first ignore dissipation, so that the oscillator evolves unitarily with the Hamiltonian $\hat{H} = \frac{1}{2m} \hat{P}^2 + \frac{1}{2} m \omega^2 \hat{X}^2$, $m$ being the mass of the oscillator, $\omega$ its frequency \footnote{We restrict to a time-independent $\omega$ in this paper. The problem with a more general $\omega(t)$, if not exactly solvable, may be solvable with the JWKB approximation if $\omega(t)$ is a slowly-varying function.}, $\hat{X}$ the position operator, and $\hat{P}$ the momentum operator. Say we evolve $\hat{\rho}(t_0)$ to a final time $t_f$, which we take to be later than any times of interest, in the presence of a source $J^+(t)$ on the forward branch (from $t_0$ to $t_f$) and $J^-(t)$ on the backward branch (from $t_f$ to $t_0$). The in-in generating functional is then defined as $Z[J^+,J^-] = {\rm Tr} \[ \hat{\rho}(t_f) \]_{J^+,J^-}$ and finite-time correlation functions of the Heisenberg picture operator $\hat{X}(t)$ can be obtained by taking functional derivatives of $Z[J^+,J^-]$ with respect to the two sources $J^{\pm}(t)$, which are typically set to zero at the end of the calculation.

It is convenient to write $Z[J^+,J^-]$ in the path integral representation as that allows us to express it in terms of Green's functions. Let us thus define eigenkets and eigenvalues of the Schr\"{o}dinger picture operator $\hat{X}_S$, which we denote with $|x\ra$ and $x$, so that $\hat{X}_S |x\ra = x |x\ra$. As shown in Appendix\ \ref{app:Zinin}, one finds that
\bea
    & & Z[J^+,J^-] \, = \, \int {\cal D}x^+ {\cal D}x^- \rho[x^+, x^-, t_0] \quad \nn \\
    & & \quad \times \, {\rm exp} \bigg[ i \left\{ S[x^+] + \int_t J^+ x^+ - S[x^-] - \int_t J^- x^- \right\} \bigg] \nn \\
    & & \quad \times \, \delta \[ x^+(t_f) - x^-(t_f) \] ,
\label{eq:Zinin}
\eea
where $\rho[x^+, x^-, t_0] \equiv \braOket{x^+}{\hat{\rho}(t_0)}{x^-}$ is a matrix element of $\hat{\rho}(t_0)$ in position basis with the functions $x^{\pm}(t)$ evaluated at $t_0$, $S[x^{\pm}]$ is the action of the oscillator, the shorthand $\int_t$ stands for $\int_{t_0}^{t_f} \d t$, the $\delta$-function at the end imposes the boundary condition at the turn-around point, and we have set $\hbar$ to unity. The action $S[x^{\pm}]$ is given by $\frac{1}{2} \int_t \big[ \( \dot{x}^{\pm} \)^2 - \omega^2 \( x^{\pm} \)^2 \big]$, where we have set the mass $m$ to unity for simplicity and without loss of generality and the dot denotes a derivative with time. Note that $Z[J,J] = 1$ since in this case the forward and backward evolution exactly cancels out or, in other words, $Z[J,J]$ is simply ${\rm Tr} \[ \hat{\rho}(t_f) \]$ for evolution in the presence of a source, which is normalized to unity.

Let us now introduce dissipation in the dynamics. Dissipation is described by an influence functional that leads to time-nonlocal terms on each of the forward and backward branches of evolution and additionally ties together terms on the two branches. We thus rewrite the generating functional of Eq.\ (\ref{eq:Zinin}) as
\bea
    & & Z_{\rm diss.}[J^+,J^-] \, = \, \int {\cal D}x^+ {\cal D}x^- \rho[x^+, x^-, t_0] \nn \\
    & & \quad \times \, {\rm exp} \bigg[ i \bigg\{ S[x^+] + \int_t J^+ x^+ - S[x^-] - \int_t J^- x^- \nn \\
    & & \qquad + \ S_{\rm IF}[x^+, x^-] \bigg\} \bigg] \delta \[ x^+(t_f) - x^-(t_f) \] ,
\label{eq:ZininIF}
\eea 
where we have added the influence functional $S_{\rm IF}[x^+, x^-]$ to the action. We must still have that $Z_{\rm diss.}[J,J] = 1$ since ${\rm Tr} \[ \hat{\rho}(t_f) \]$ for evolution in the presence of a source is normalized to unity for both unitary and non-unitary evolution.

In order to explicitly calculate the generating functional, we need an ansatz for the initial state and influence functional, which we discuss in the two subsections below.

%-------------------------------
\subsection{Initial state}
\label{subsec:is}
%-------------------------------

We choose a Gaussian initial state for the oscillator, parameterized as $\rho[x^+, x^-, t_0] = N {\rm exp} \left\{ iS_0[x^+, x^-] \right\}$, $N$ being a normalization constant chosen so that ${\rm Tr} \[ \hat{\rho}(t_0) \] = 1$ and \cite{Berges:2004yj,Agarwal:2012mq}
{\allowdisplaybreaks
\bea
    & & S_0[x^+, x^-] \, = \, p_0 (x^+ - x^-) + \frac{1}{2} \big[ A (x^+ - x_0)^2 \nn \\
    & & \qquad - \, A^* (x^- - x_0)^2 + 2iB (x^+ - x_0) (x^- - x_0) \big] \, , \qquad
\label{eq:initialS}
\eea}%
with the functions $x^{\pm}(t)$ evaluated at $t_0$, as mentioned earlier. Note that $B$ here must be real since $\hat{\rho}(t_0)$ is Hermitian. $A$, on the other hand, can be complex and we denote its real and imaginary parts with $A_R$ and $A_I$. The normalization $N$ is then found to be $N = \sqrt{(A_I + B)/\pi}$ with the condition that $A_I + B > 0$. $x_0$ and $p_0$ are the initial one-point functions $\big\la \hat{X}(t_0) \big\ra$ and $\big\la \hat{P}(t_0) \big\ra$, where we have used angular brackets to denote the expectation value in $\hat{\rho}(t_0)$. Lastly, $A$ and $B$ are related to the initial two-point correlators, that we denote as $c_{xx}$, $c_{xp}$, and $c_{pp}$ for convenience,
\bea
	\big\la \hat{X}^2(t_0) \big\ra_c & \equiv & c_{xx} \, = \, \frac{1}{2(A_I + B)} \, ,
\label{eq:phiphicorr} \\
	\frac{1}{2} \big\la \big\{ \hat{X}(t_0), \hat{P}(t_0) \big\} \big\ra_c & \equiv & c_{xp} \, = \, \frac{A_R}{2(A_I + B)} \, ,
\label{eq:phimomcorr} \\
	\big\la \hat{P}^2(t_0) \big\ra_c & \equiv & c_{pp} \, = \, \frac{A_R^2 + A_I^2 - B^2}{2(A_I + B)} \, . \quad
\label{eq:mommomcorr}
\eea
The subscript `$c$' here denotes connected correlators, for example, $\big\la \hat{X}^2 \big\ra_c = \big\la \hat{X}^2 \big\ra - \big\la \hat{X} \big\ra^2$, and $\{\cdot,\cdot\}$ is the anti-commutator. We can also calculate the purity of our initial state, which we denote ${\rm Pu}(t_0) = {\rm Tr}\[ \hat{\rho}^2(t_0) \]$, and relate it to the initial correlators as ${\rm Pu}(t_0) = \sqrt{\frac{A_I + B}{A_I - B}} = 0.5/\sqrt{c_{xx} c_{pp} - c_{xp}^2}$. Since purity must be between zero and one, we also have the condition that $B \le 0$, with the initial state being pure for $B = 0$ and mixed for $B < 0$.

%-------------------------------
\subsection{Influence functional}
\label{subsec:if}
%-------------------------------

We choose the influence functional to be quadratic and parameterize it as
\bea
	& & S_{\rm IF}[x^+, x^-] \, = \, \int_{t,t'} \big[ \gamma_1(t,t') x^+(t) x^+(t') \nn \\
	& & \qquad - \, \gamma_1^*(t,t') x^-(t) x^-(t') + \gamma_2^*(t,t') x^+(t) x^-(t') \nn \\
	& & \qquad - \, \gamma_2(t,t') x^-(t) x^+(t') \big] \, ,
\label{eq:dissaction}
\eea
where $\gamma_1(t,t')$ and $\gamma_2(t,t')$, which we refer to as dissipation kernels, are complex functions. The complex conjugates and signs in Eq.\ (\ref{eq:dissaction}) ensure that the generating functional is real or, equivalently, the density operator is Hermitian at all times. The dissipation kernels must further satisfy $\gamma_1(t,t') = \gamma_1(t',t)$ and $\gamma_2(t,t') = -\gamma_2^*(t',t)$ since the integration measure is symmetric under the interchange of $t$ and $t'$. We find in the next section that only two real functions contribute to dissipation and show later in the paper that they too are not independent of one another. We also note that in a microscopic derivation of the influence functional, we would typically take the oscillator to be coupled to some environment degrees of freedom, with the full system plus environment evolving unitarily. Performing a partial trace over the environment, we would find that the dissipation kernels here are related to correlation functions of the environment and additionally satisfy $\gamma_1(t,t') = \gamma_2(t,t') \theta(t-t') - \gamma_2^*(t,t') \theta(t'-t)$. We will not impose this constraint here, except when mapping to a specific microscopic model.

%%%%%%%%%%%%%%%%%%%%%%%%%%%%%%%%%%%%%%%%%%%%%%%%%%

%-------------------------------
\section{Green's functions}
\label{sec:greensfunctions}
%-------------------------------

In this section, we write the generating functional in terms of Green's functions and obtain the equations of motion and initial conditions that they satisfy. To introduce Green's functions, it is convenient to first express the integrand in Eq.\ (\ref{eq:ZininIF}) as an exponent. Since we have already written the initial state and influence functional as exponents in Secs.\ \ref{subsec:is} and \ref{subsec:if}, we only need to further express the $\delta$-function in Eq.\ (\ref{eq:Zinin}) as an exponent. Following \cite{Weinberg:2005vy}, the $\delta$-function can be written as
\bea
	& & \delta \big[ x^+(t_f) - x^-(t_f) \big] \nn \\
	& & \quad \, = \, \lim_{\epsilon \rightarrow 0} \frac{1}{\sqrt{\pi\epsilon}} \, {\rm exp} \[ -\frac{1}{\epsilon} \left\{ x^+(t_f) - x^-(t_f) \right\}^2 \] \nn \\
	& & \quad = \ \lim_{\epsilon \rightarrow 0} \frac{1}{\sqrt{\pi\epsilon}} \, {\rm exp} \bigg[ - \int_{t, t'} C(t,t') \left\{ x^+(t) - x^-(t) \right\} \nn \\
	& & \quad \qquad \times \, \left\{ x^+(t') - x^-(t') \right\} \bigg] \, ,
\eea
where $C(t,t') = \frac{2}{\epsilon} \delta(t-t') \delta(t'-t_f)$ and the extra factor of two is needed to cancel the factor of half that arises from evaluating the $\delta$-function at a limit of the integral. We will also integrate by parts the kinetic term in the action $S[x^{\pm}, J^{\pm}]$, to move both time derivatives to a single $x^{\pm}$. This generates boundary terms at $t_0$ and $t_f$, of which those at $t_f$ cancel out between the plus and minus branches given that $x^+(t_f) = x^-(t_f)$ and $\dot{x}^+(t_f) = \dot{x}^-(t_f)$, where the second condition is shown to follow from the first in Appendix\ \ref{app:Zinin}. We also note in Appendix\ \ref{app:Zinin} that, in the presence of dissipation, $\dot{x}^+(t_f) = \dot{x}^-(t_f)$ only holds for dissipation kernels $\gamma_1(t,t')$ and $\gamma_2(t,t')$ that are not proportional to $\d^2 \delta(t-t')/\d t'^2$. The boundary terms at $t_0$ remain, and we keep track of them below.

Putting everything together, we can now write the generating functional in Eq.\ (\ref{eq:ZininIF}) as
\bea
    & & Z_{\rm diss.}[J^+,J^-] \, = \, N \int {\cal D}x^+ {\cal D}x^- \ {\rm exp} \bigg[ -\frac{i}{2} \int_{t,t'} \nn \\
    & & \qquad \times \, \boldsymbol{x}^T(t) \boldsymbol{\cal O}(t,t') \boldsymbol{x}(t') + i \int_t \boldsymbol{J}_s^T(t) \boldsymbol{x}(t) \bigg] \, . \quad
\label{eq:Zininp}
\eea
The vectors $\boldsymbol{x}(t)$ and $\boldsymbol{J}_s(t)$ here are given by
\bea
	\boldsymbol{x}(t) \, = \, \[
	\begin{array}{c}
		x^+(t) \\
		x^-(t)
	\end{array} \] 
    \quad
    \text{and}
    \quad
	\boldsymbol{J}_s(t) \, = \, \[
	\begin{array}{c}
		J_s^+(t) \\
		-J_s^-(t)
	\end{array} \] ,
\label{eq:defxJs}
\eea
with $J_s^+(t) = J^+(t) + 2 \delta(t-t_0) [ p_0 - (A + iB)x_0 ]$ and $J_s^-(t) = J^-(t) + 2 \delta(t-t_0) [ p_0 - (A^* - iB)x_0 ]$ denoting {\it shifted} sources, and the superscript $T$ indicates a transpose. $\boldsymbol{\cal O}(t,t')$ is a $2 \times 2$ matrix of differential operators whose $11$ and $12$ components are given by
\begin{widetext}
\bea
    {\cal O}^{11}(t,t') & = & \delta(t-t') \[ \( \frac{\d^2}{\d t'^2} + \omega^2 \) + 2\delta(t'-t_0) \( \frac{\d}{\d t'} - A \) - \frac{4i}{\epsilon} \delta(t'-t_f) \] - 2\gamma_1(t,t') \, ,
\label{eq:O11} \\
    {\cal O}^{12}(t,t') & = & \delta(t-t') \[ -2iB \delta(t'-t_0) + \frac{4i}{\epsilon} \delta(t'-t_f) \] - 2\gamma_2^*(t,t') \, ,
\label{eq:O12}
\eea
\end{widetext}
the extra factors of $2$ in front of the $\delta(t'-t_0)$ terms again arising from evaluating the $\delta$-function at a limit of the integral. The other two components of $\boldsymbol{\cal O}(t,t')$ are related to these by ${\cal O}^{21}(t,t') = -{\cal O}^{12*}(t,t')$ and ${\cal O}^{22}(t,t') = -{\cal O}^{11*}(t,t')$.

Let us now define a $2 \times 2$ matrix $\boldsymbol{G}(t,t')$ that satisfies the following Green's function equation
\bea
	\int_{t''} \boldsymbol{\cal O}(t,t'') \boldsymbol{G}(t'',t') & = & -i \boldsymbol{I} \delta(t-t') \, ,
\label{eq:greenfn}
\eea
where
\bea
	\boldsymbol{G}(t,t') & = & \[
	\begin{array}{cc}
		G^{++}(t,t') & G^{+-}(t,t') \\
		G^{-+}(t,t') & G^{--}(t,t')
	\end{array} \]
\label{eq:Gk2by2}
\eea
and $\boldsymbol{I}$ is the $2 \times 2$ identity matrix. This leads to four equations of motion that couple the functions $G^{\pm\pm}(t,t')$. Since the $\epsilon$ that appears in $\boldsymbol{\cal O}(t,t')$ is arbitrarily small, all terms proportional to $1/\epsilon$ in Eq.\ \eqref{eq:greenfn} must cancel out, giving us the constraints
\bea
    G^{++}(t_f,t') & = & G^{-+}(t_f,t') \, ,
\label{eq:Gpptfconstraint} \\
    G^{--}(t_f,t') & = & G^{+-}(t_f,t') \, ,
\eea
for all $t_0 < t' < t_f$. Note that these constraints follow directly from the requirement that $x^+(t_f) = x^-(t_f)$. We similarly expect the derivatives at $t_f$ to also satisfy the same constraints, so that
\bea
    \partial_t G^{++}(t,t') \big|_{t=t_f} & = & \partial_t G^{-+}(t,t') \big|_{t=t_f} \, , \\
    \partial_t G^{--}(t,t') \big|_{t=t_f} & = & \partial_t G^{+-}(t,t') \big|_{t=t_f} \, ,
\label{eq:dGmmtfconstraint}
\eea
for all $t_0 < t' < t_f$, in analogy with the condition that $\dot{x}^+(t_f) = \dot{x}^-(t_f)$. One way to see this is to first write the solution for the {\it classical} field that satisfies the equation $\int_{t''} \boldsymbol{\cal O}(t,t'') \boldsymbol{x}_c(t'') = \boldsymbol{J}(t)$ in the presence of any source $\boldsymbol{J}(t) = \icol{J^+(t) \\ -J^-(t)}$: $\boldsymbol{x}_c(t) = {\rm homogeneous\ solution} + i \int_{t'} \boldsymbol{G}(t,t') \boldsymbol{J}(t')$. Then realizing that the boundary conditions are carried by the classical field, i.e., $x_c^+(t_f) = x_c^-(t_f)$ and $\dot{x}_c^+(t_f) = \dot{x}_c^-(t_f)$, all four constraints in Eqs.\ (\ref{eq:Gpptfconstraint})--(\ref{eq:dGmmtfconstraint}) follow. We note again, however, that this argument only holds for dissipation kernels that are not proportional to $\d^2 \delta(t-t')/\d t'^2$. All four constraints in Eqs.\ (\ref{eq:Gpptfconstraint})--(\ref{eq:dGmmtfconstraint}) are needed to write the generating functional in standard form. To do so, we first shift $\boldsymbol{x}(t)$ in Eq.\ (\ref{eq:Zininp}) to $\boldsymbol{x}_s(t) = \boldsymbol{x}(t) - i \int_{t'} \boldsymbol{G}(t,t') \boldsymbol{J}_s(t')$, then integrate by parts to move the derivatives from $x_s^{\pm}(t)$ to $G^{\pm\pm}(t,t')$, and lastly make use of the four constraints above to cancel the boundary terms. The generating functional can then be written as
\bea
	Z_{\rm diss.}[J^+,J^-] & = & Z_0 \, {\rm exp} \[ - \frac{1}{2} \int_{t,t'} \boldsymbol{J}_s^T(t) \boldsymbol{G}(t,t') \boldsymbol{J}_s(t') \] , \nn \\
\label{eq:ZininJGJ}
\eea
where the normalization $Z_0$ is chosen such that $Z_{\rm diss.}[J,J] = 1$ and it turns out to be unity in the case that the initial state is a coherent state. Further, since we can freely interchange the $t$ and $t'$ integrals in the exponent, the functions $G^{\pm\pm}(t,t')$ must also satisfy the conditions
\bea
    G^{++}(t,t') & = & G^{++}(t',t) \, , \\
    G^{+-}(t,t') & = & G^{-+}(t',t) \, , \\
    G^{--}(t,t') & = & G^{--}(t',t) \, ,
\eea
in addition to the constraints written earlier.

Writing the generating functional in the form of Eq.\ (\ref{eq:ZininJGJ}) makes it easy to obtain correlation functions as usual, and we discuss this further in the first subsection below. In the next two subsections, we obtain the initial conditions and equations of motion for $G^{\pm\pm}(t,t')$, and sketch how to solve the resulting equations.

%-------------------------------
\subsection{$n$-point correlations}
\label{subsec:npoint}
%-------------------------------

As shown in Appendix\ \ref{app:npoint} for the case of unitary dynamics, one- and two-point correlation functions of $\hat{X}(t)$ are easily obtained by taking functional derivatives of $Z[J^+, J^-]$ with respect to $J^{\pm}$. We can write similar expressions for the case when the oscillator is coupled to an environment, with the full system plus environment evolving unitarily, and then trace out the environment. Correlation functions in the presence of dissipation can, therefore, be obtained by taking functional derivatives of $Z_{\rm diss.}[J^+, J^-]$ instead. Generalizing the calculation in Appendix\ \ref{app:npoint} to the dissipative case and additionally beyond one- and two-point correlations gives
{\allowdisplaybreaks
\bea
	& & \big\la \bar{T} \big\{ \hat{X}(t_{n+1}) \ldots \hat{X}(t_{n+m}) \big\} T \big\{ \hat{X}(t_1) \ldots \hat{X}(t_n) \big\} \big\ra \nn \\
	& & \qquad = \, (-i)^n i^m \frac{\delta}{\delta J^+(t_1)} \ldots \frac{\delta}{\delta J^+(t_n)} \frac{\delta}{\delta J^-(t_{n+1})} \nn \\
	& & \qquad \qquad \ \times \, \ldots \frac{\delta}{\delta J^-(t_{n+m})} \, Z_{\rm diss.}[J^+,J^-] \Big|_{J^{\pm}=J} \, , \qquad
\label{eq:defnmpt}
\eea}%
in the presence of a source $J(t)$, where the first $n$ operators are time-ordered, denoted by $T$, and the next $m$ are anti-time-ordered, denoted by $\bar{T}$.

We now want to relate $n$-point correlations to the Green's functions by making use of Eq.\ (\ref{eq:ZininJGJ}). Let us first consider the one-point function $\big\la \hat{X}(t) \big\ra$. Setting $n = 1$ and $m = 0$ in Eq.\ (\ref{eq:defnmpt}) and using Eq.\ (\ref{eq:ZininJGJ}), we find that
\bea
	\la \hat{X}(t) \ra & = & i \int_{t'} \big[ G^{++}(t,t') J_s^+(t') \nn \\
	& & \qquad - \, G^{+-}(t,t') J_s^-(t') \big] \Big|_{J^{\pm} = J} \, .
\label{eq:def1pt}
\eea
We expect $\la \hat{X}(t) \ra$ to match the solution to the classical equation of motion. Let us next consider the two-point correlations $\big\la T \hat{X}(t) \hat{X}(t') \big\ra$ and $\big\la \hat{X}(t') \hat{X}(t) \big\ra$. These yield the functions $G^{++}(t,t')$ and $G^{+-}(t,t')$ plus a product of one-point expectation values and, therefore,
\bea
    \big\la T \hat{X}(t) \hat{X}(t') \big\ra_c & = & G^{++}(t,t') \, , \\
    \big\la \hat{X}(t') \hat{X}(t) \big\ra_c & = & G^{+-}(t,t') \, .
\eea
We can similarly show that $\big\la \bar{T} \hat{X}(t) \hat{X}(t') \big\ra_c = G^{--}(t,t')$ and $\big\la \hat{X}(t) \hat{X}(t') \big\ra_c = G^{-+}(t,t')$, where $\bar{T}$ denotes anti-time-ordering. Since these are also equal to the Hermitian conjugates of the above expressions, we further have that $G^{--}(t,t') = G^{++*}(t,t')$ and $G^{-+}(t,t') = G^{+-*}(t,t')$.

We can now write the functions $G^{\pm\pm}(t,t')$ in a more convenient form. Let us denote $G^{+-}(t,t')$ as $G^<(t,t')$ and $G^{-+}(t,t')$ as $G^>(t,t') = G^{<*}(t,t')$, and expand out $\big\la T \hat{X}(t) \hat{X}(t') \big\ra_c = \big\la \hat{X}(t) \hat{X}(t') \big\ra_c \theta(t-t') + \big\la \hat{X}(t') \hat{X}(t) \big\ra_c \theta(t'-t)$. Then we can write
\bea
    G^{++}(t,t') & = & G^>(t,t') \theta(t-t') + G^<(t,t') \theta(t'-t) \, , \quad \ \ 
\label{eq:Gppform} \\
    G^{+-}(t,t') & = & G^<(t,t') \, ,
\label{eq:Gpmform}
\eea
and $G^{--}(t,t')$ as the complex conjugate of Eq.\ (\ref{eq:Gppform}). Note that the functions $G^{\pm\pm}(t,t')$ written as above satisfy all constraints and conditions written earlier in this section. By using Eqs.\ (\ref{eq:Gppform}) and (\ref{eq:Gpmform}) and similar expressions for $G^{--}(t,t')$ and $G^{-+}(t,t')$, we can verify that they additionally satisfy
\bea
    G^{++}(t,t') - G^{+-}(t,t') - G^{-+}(t,t') + G^{--}(t,t') & = & 0 \, , 
\label{eq:Gsum0} \nn \\
\eea
which is needed to ensure that $Z_{\rm diss.}[J,J] = 1$ and will also turn out to be useful later in this section.

Let us lastly consider the $n$-point correlation $\big\la T \hat{X}(t_1) \ldots \hat{X}(t_n) \big\ra$ obtained by setting $m = 0$ in Eq.\ (\ref{eq:defnmpt}). Using again Eq.\ (\ref{eq:ZininJGJ}) for $Z_{\rm diss.}[J^+, J^-]$ and expanding out the $J^+(t_i)$ derivatives on the right hand side gives us a product of all $G^{++}(t_i,t_j)$ plus a sum of disconnected correlators for even $n$ and only a sum of disconnected correlators for odd $n$. Taking the disconnected correlators to the left hand side then turns it into a time-ordered product of $\hat{X}(t) - \big\la \hat{X}(t) \big\ra$ at times $t_1, \ldots, t_n$. The time-ordered product of $\hat{X}(t) - \big\la \hat{X}(t) \big\ra$, therefore, obeys Wick's theorem. We can also obtain this result in a simpler way, and it is thus worth discussing, by writing Eq.\ (\ref{eq:ZininJGJ}) in a slightly different form. On expanding out the shifted sources on the right hand side of Eq.\ (\ref{eq:ZininJGJ}) and rearranging the resulting expression, we can write it as
\bea
	& & {\rm exp} \[ -i \int_t \left\{ J^+(t) - J^-(t) \right\} \la \hat{X}(t) \ra \] Z_{\rm diss.}[J^+,J^-] \nn \\
	& & \qquad = \, {\rm exp} \[ - \frac{1}{2} \int_{t,t'} \boldsymbol{J}_c^T(t) \boldsymbol{G}(t,t') \boldsymbol{J}_c(t') \] ,
\label{eq:ZininJGJWicks}
\eea
where $\la \hat{X}(t) \ra$, given in Eq.\ (\ref{eq:def1pt}), is calculated in the presence of a source $J(t)$ and $\boldsymbol{J}_c(t) = \icol{J^+(t) \, - \, J(t) \\ -J^-(t) \, + \, J(t)}$. Taking derivatives of Eq.\ (\ref{eq:ZininJGJWicks}) with respect to $J^+(t_i)$ and setting $J^{\pm}(t) = J(t)$ now produces the time-ordered product of $\hat{X}(t) - \big\la \hat{X}(t) \big\ra$ at times $t_1, \ldots, t_n$ on the left hand side and a product of $G^{++}(t_i,t_j)$ for even $n$ and zero for odd $n$ on the right hand side, directly giving us Wick's theorem.

%-------------------------------
\subsection{Initial conditions}
\label{subsec:ic}
%-------------------------------

In the previous subsection, we showed that connected two-point correlations of $\hat{X}(t)$ can be identified with the functions $G^{\pm\pm}(t,t')$. We next want to argue that the Heisenberg picture operator $\hat{P}(t)$ is given by $\hat{P}(t) = \dot{{\hat{X}}}(t)$, so that we can additionally relate the connected two-point correlations that involve $\hat{P}(t)$ with time derivatives of $G^{\pm\pm}(t,t')$. If we substitute the form of $G^{++}(t,t')$ given in Eq.\ (\ref{eq:Gppform}) into its equation of motion given by Eq.\ (\ref{eq:greenfn}) and equate the $\delta$-function on both sides, we find that it yields the Wronskian condition $(\partial_t - \partial_{t'}) G^<(t,t') \big|_{t=t'} = i$ for those kernels $\gamma_1(t,t')$ that are not proportional to $\d^2 \delta(t-t')/\d t'^2$. We similarly need $\gamma_2(t,t')$ to not be proportional to $\d^2 \delta(t-t')/\d t'^2$, so that the $G^{+-}(t,t')$ equation does not spoil this condition. Since $G^<(t,t') = \big\la \hat{X}(t') \hat{X}(t) \big\ra_c$, the Wronskian condition further implies that $\big\la \big[ \hat{X}(t), \dot{\hat{X}}(t) \big] \big\ra = i$ which, together with the commutation relation $\big[ \hat{X}(t), \hat{P}(t) \big] = i$, {\it suggests} that indeed $\hat{P}(t) = \dot{{\hat{X}}}(t)$. We will restrict to those influence functionals that preserve $\hat{P}(t) = \dot{{\hat{X}}}(t)$ in this paper, but note that this is not guaranteed for more general influence functionals, specifically those that arise from derivative system-environment interactions.

With $\hat{P}(t) = \dot{{\hat{X}}}(t)$, we can directly transcribe the initial conditions given in Eqs.\ (\ref{eq:phiphicorr}), (\ref{eq:phimomcorr}), and (\ref{eq:mommomcorr}) to conditions on $G^{\pm\pm}(t,t')$, in particular on $G^<(t,t')$,
{\allowdisplaybreaks
\bea
    G^<(t_0,t_0) & = & c_{xx} \, ,
\label{eq:phiphicorr2} \\
    \frac{1}{2} (\partial_t + \partial_{t'}) G^<(t,t') \big|_{t=t'=t_0} & = & c_{xp} \, ,
\label{eq:phimomcorr2} \\
    \partial_t \partial_{t'} G^<(t,t') \big|_{t=t'=t_0} & = & c_{pp} \, .
\label{eq:mommomcorr2}
\eea}%
From the Wronskian condition, or the commutator of $\hat{X}(t)$ and $\hat{P}(t)$ at the initial time, we additionally have that
\bea
    (\partial_t - \partial_{t'}) G^<(t,t') \big|_{t=t'=t_0} & = & i \, .
\label{eq:wronskiant0}
\eea
As shown later, the three initial conditions in Eqs.\ (\ref{eq:phiphicorr2}), (\ref{eq:phimomcorr2}), and (\ref{eq:mommomcorr2}) are sufficient to solve the equations of motion for the symmetrized two-point correlation that we introduce in the next subsection.

It is also worth noting that the initial conditions written here are consistent with the $\delta(t'-t_0)$ terms in the equations of motion in Eq.\ (\ref{eq:greenfn}) and we show this next for completeness. Consider specifically the equations of motion of $G^{++}(t,t')$ and $G^{+-}(t,t')$ for $t_0 \le t, \, t' < t_f$, which are obtained by plugging the $11$ and $12$ components of $\boldsymbol{\cal O}(t,t')$, given in Eqs.\ (\ref{eq:O11}) and (\ref{eq:O12}), into Eq.\ (\ref{eq:greenfn}). On dropping the terms containing a $\delta$-function at $t_f$ since they cancel out and further dropping the dissipative terms assuming they are of a form that does not affect the initial conditions, the $G^{++}(t,t')$ and $G^{+-}(t,t')$ equations become
\begin{widetext}
\bea
	\[ \frac{\d^2}{\d t^2} + \omega^2 + 2\delta(t-t_0) \( \frac{\d}{\d t} - A \) \] G^{++}(t,t') - 2iB \delta(t-t_0) G^{-+}(t,t') & = & -i \delta(t-t') \, ,
\label{eq:Gppxapp} \\
	\[ \frac{\d^2}{\d t^2} + \omega^2 + 2\delta(t-t_0) \( \frac{\d}{\d t} - A \) \] G^{+-}(t,t') - 2iB \delta(t-t_0) G^{--}(t,t') & = & 0 \, .
\label{eq:Gpmxapp}
\eea
\end{widetext}
Let us first set $t' = t_0 + \epsilon$, for some small parameter $\epsilon$, in Eq.\ (\ref{eq:Gppxapp}) and integrate the equation over $t$ from $t_0$ to $t_0 + 2\epsilon$, taking care of factors of $1/2$ that arise from evaluating the $\delta$-function at a limit of the integral. Using the forms of $G^{\pm\pm}(t,t')$ given in Eqs.\ (\ref{eq:Gppform}), (\ref{eq:Gpmform}), and the following text in the resulting equation and further taking the limit $\epsilon \rightarrow 0$ gives
\bea
	\partial_t G^>(t,t_0) \big|_{t=t_0} - (A + iB) G^<(t_0,t_0) & = & -i \, .
\label{eq:intGppxapp}
\eea
Let us next integrate Eq.\ (\ref{eq:Gpmxapp}) over $t$ from $t_0$ to $t_0 + \epsilon$. Using again the forms of $G^{\pm\pm}(t,t')$ given in Eqs.\ (\ref{eq:Gppform}), (\ref{eq:Gpmform}), and the following text in the resulting equation and further taking the limit $\epsilon \rightarrow 0$ gives
\bea
	\partial_t G^<(t,t') \big|_{t=t_0} - A G^<(t_0,t') - iB G^{--}(t_0,t') & = & 0 \, , \nn \\
\label{eq:intGpmxapp}
\eea
for any $t'$. For the specific choice of $t' = t_0$ and using $G^<(t,t') = G^{>*}(t,t')$, adding Eqs.\ (\ref{eq:intGppxapp}) and (\ref{eq:intGpmxapp}) gives
\bea
	(\partial_t + \partial_{t'}) G^<(t,t') \big|_{t=t'=t_0} - 2(A + iB) G^<(t_0,t_0) & = & -i \, . \nn \\
\eea
Now equating first the imaginary parts and then the real parts on both sides of this equation, and using the definitions of $c_{xx}$ and $c_{xp}$ from Eqs.\ (\ref{eq:phiphicorr}) and (\ref{eq:phimomcorr}), yields the initial conditions given in Eqs.\ (\ref{eq:phiphicorr2}) and (\ref{eq:phimomcorr2}). Similarly, setting $t' = t_0$ and subtracting Eq.\ (\ref{eq:intGppxapp}) from Eq.\ (\ref{eq:intGpmxapp}) gives the commutator or Wronskian condition at the initial time given in Eq.\ (\ref{eq:wronskiant0}). Lastly, let us differentiate Eq.\ (\ref{eq:intGpmxapp}) with respect to $t'$ and set $t' = t_0 + \epsilon$. Using the previous initial conditions to simplify the resulting equation, taking the limit $\epsilon \rightarrow 0$, and using the definition of $c_{pp}$ from Eq.\ (\ref{eq:mommomcorr}) then gives the initial condition in Eq.\ (\ref{eq:mommomcorr2}).

%-------------------------------
\subsection{Equations of motion}
%-------------------------------

We next consider the equations of motion for $G^{\pm\pm}(t,t')$ that are obtained as noted in the previous subsection. We drop the terms containing a $\delta$-function at $t_0$ since these simply impose initial conditions on $G^{\pm\pm}(t,t')$ and those containing a $\delta$-function at $t_f$ since they cancel out as noted earlier. Then $G^{++}(t,t')$ and $G^{+-}(t,t')$ satisfy the following equations for $t_0 < t, \, t' < t_f$, 
\begin{widetext}
\bea
	\( \frac{\d^2}{\d t^2} + \omega^2 \) G^{++}(t,t') - 2\int_{t''} \big[ \gamma_1(t,t'') G^{++}(t'',t') + \gamma_2^*(t,t'') G^{-+}(t'',t') \big] & = & -i \delta(t-t') \, ,
\label{eq:Gppx} \\
	\( \frac{\d^2}{\d t^2} + \omega^2 \) G^{+-}(t,t') - 2\int_{t''} \big[ \gamma_1(t,t'') G^{+-}(t'',t') + \gamma_2^*(t,t'') G^{--}(t'',t') \big] & = & 0 \, ,
\label{eq:Gpmx}
\eea
\end{widetext}
while $G^{--}(t,t')$ and $G^{-+}(t,t')$ satisfy the complex conjugates of the above equations.

To solve these equations, it is simplest to first decouple them by rotating to a new basis,
\bea
	\boldsymbol{\xi}(t) \, = \, \boldsymbol{M} \boldsymbol{x}(t)
    \quad
    \text{with}
    \quad
	\boldsymbol{M} \, = \, 	
	\[ \begin{array}{cc}
		1/2 & 1/2 \\
		1 & -1
	\end{array} \] ,
\label{eq:xibasis}
\eea
where $\boldsymbol{\xi}(t)$ is a vector containing $\xi^{\pm}(t)$. The differential operators and Green's functions in the two bases are then related by $\boldsymbol{\cal O}^{\xi}(t,t') = \( \boldsymbol{M}^{-1} \)^{T} \boldsymbol{\cal O}(t,t') \boldsymbol{M}^{-1}$ and $\boldsymbol{G}^{\xi}(t,t') = \boldsymbol{M} \boldsymbol{G}(t,t') \boldsymbol{M}^T$, as found using Eqs.\ (\ref{eq:Zininp}) and (\ref{eq:greenfn}), respectively, where $\boldsymbol{G}^{\xi}(t,t')$ is a $2 \times 2$ matrix containing the functions $G^{\xi,\pm\pm}(t,t')$. It is instructive to write these functions explicitly,
\bea
    G^{\xi,++}(t,t') & = & \frac{1}{2} \[ G^>(t,t') + G^<(t,t') \] \, , \\
    G^{\xi,+-}(t,t') & = & \[ G^>(t,t') - G^<(t,t') \] \theta(t-t') \, , \\
    G^{\xi,-+}(t,t') & = & G^{\xi,+-}(t',t) \, , \\
    G^{\xi,--}(t,t') & = & 0 \, ,
\eea
where the last identity follows from Eq.\ (\ref{eq:Gsum0}). $G^{\xi,++}(t,t')$ is thus the symmetrized correlation and $G^{\xi,+-}(t,t')$ is the retarded Green's function of the theory.

We can now obtain the equations of motion for the functions $G^{\xi,\pm\pm}(t,t')$ in a similar way as we did for the functions $G^{\pm\pm}(t,t')$. Since $G^{\xi,--}(t,t')$ vanishes identically, its equation of motion yields the constraint
\bea
\label{GammsAreEqual}
    \int_{t''} \big[ \gamma_{1I}(t,t'') - \gamma_{2I}(t,t'') \big] G^{\xi,+-}(t'',t') & = & 0 \, ,
\eea
where the subscripts $I$ indicate imaginary parts as before. We take this to imply that $\gamma_{1I}(t,t') = \gamma_{2I}(t,t')$, although one can envision a more general class of functions that satisfy this constraint. The equations for $G^{\xi,++}(t,t')$ and $G^{\xi,+-}(t,t')$ then simplify to
\begin{widetext}
\bea
	\( \frac{\d^2}{\d t^2} + \omega^2 \) G^{\xi,++}(t,t') - 2\int_{t''} \big[ \gamma_{1R}(t,t'') + \gamma_{2R}(t,t'') \big] G^{\xi,++}(t'',t') & = & 2i\int_{t''} \gamma_{1I}(t,t'') G^{\xi,+-}(t',t'') \, ,
\label{eq:Gppxi} \\
	\( \frac{\d^2}{\d t^2} + \omega^2 \) G^{\xi,+-}(t,t') - 2\int_{t''} \big[ \gamma_{1R}(t,t'') + \gamma_{2R}(t,t'') \big] G^{\xi,+-}(t'',t') & = & -i \delta(t-t') \, ,
\label{eq:Gpmxi}
\eea
\end{widetext}
which can be solved for specific dissipation kernels. Note that only two real functions, $\gamma_{1R}(t,t') + \gamma_{2R}(t,t')$ and $\gamma_{1I}(t,t')$, contribute to dissipation. The solution to the second equation above gives us the retarded Green's function, $G^{\xi,+-}(t,t')$, of the theory. The first equation, on the other hand, yields the symmetrized correlation, $G^{\xi,++}(t,t')$, and can be solved by first writing it as the sum of a homogeneous piece, that we denote $G^{\xi,++[h]}(t,t')$, and a particular solution or {\it noise} piece, that we denote $G^{\xi,++[n]}(t,t')$. While the homogeneous piece is the solution to Eq.\ (\ref{eq:Gppxi}) with zero on the right hand side and satisfies the initial conditions written in the previous subsection, the noise piece is obtained by convolving the source term on the right hand side of Eq.\ (\ref{eq:Gppxi}) with the retarded Green's function and vanishes at the initial time (i.e., at $t' = t_0$). We will choose a specific form for the dissipation kernels in the next section that will allow us to explicitly solve for $G^{\xi,+-}(t,t')$ and $G^{\xi,++}(t,t')$ in that case.

%%%%%%%%%%%%%%%%%%%%%%%%%%%%%%%%%%%%%%%%%%%%%%%%%%

%-------------------------------
\section{Time-local dissipation}
\label{sec:tldissipation}
%-------------------------------

In this section, we specialize to the simple and well-studied case of time-local dissipation to understand whether the oscillator thermalizes. Time-local dissipation is obtained within the Caldeira-Leggett model, for example, when specializing to an Ohmic spectral density with infinite cutoff and the high-temperature regime \cite{Breuer:2002pc,Caldeira:1982iu}. First, to reduce the integrals on the left hand side of Eqs.\ (\ref{eq:Gppxi}) and (\ref{eq:Gpmxi}) to a simple damping term, we choose dissipation kernels such that $\gamma_{1R}(t,t') + \gamma_{2R}(t,t') = \gamma \d\delta(t-t')/\d t'$, where $\gamma$ is a (real) positive constant. Second, to reduce the integral on the right hand side of Eq.\ (\ref{eq:Gppxi}) to a local source term, we choose $\gamma_{1I}(t,t') = (\alpha/2) \delta(t-t')$, where $\alpha$ is a (real) constant. Eqs.\ (\ref{eq:Gppxi}) and (\ref{eq:Gpmxi}) then simplify to
\bea
	\( \frac{\d^2}{\d t^2} + \omega^2 + 2\gamma \frac{\d}{\d t} \) G^{\xi,++}(t,t') & = & i\alpha G^{\xi,+-}(t',t) \, , \quad \ \ 
\label{eq:Gpptl} \\
	\( \frac{\d^2}{\d t^2} + \omega^2 + 2\gamma \frac{\d}{\d t} \) G^{\xi,+-}(t,t') & = & -i \delta(t-t') \, .
\label{eq:Gpmtl}
\eea
At the moment, we have not imposed any conditions on the constants $\gamma$ and $\alpha$, except that $\gamma \ge 0$. We note that in the Caldeira-Leggett model with time-local dissipation, the dissipation kernels are given by $\gamma_{1R}(t,t') + \gamma_{2R}(t,t') = 2\gamma \ [\d\delta(t-t')/\d t'] \theta(t-t')$ and $\gamma_{1I}(t,t') = (2\gamma/\beta_{\rm env}) \delta(t-t')$, where $\beta_{\rm env}$ is the inverse of the Boltzmann constant times the temperature of the thermal environment. We can, therefore, specialize to this model by leaving $\gamma$ as is and replacing $\alpha$ with $4\gamma/\beta_{\rm env}$ in our results below.

We solve the equations of motion\ (\ref{eq:Gpptl}) and (\ref{eq:Gpmtl}) in the first subsection below. In the next two subsections, we obtain the purity of the oscillator and check whether the fluctuation-dissipation relation is satisfied in the late-time limit, and show that $\gamma$ and $\alpha$ can in fact not be independent of one another.

%-------------------------------
\subsection{Green's functions}
\label{subsec:Gtl}
%-------------------------------

Let us first solve Eq.\ (\ref{eq:Gpmtl}) for the retarded Green's function of the theory, $G^{\xi,+-}(t,t')$. This is easily solved in Fourier space \footnote{Since boundary conditions on the retarded Green's function, $G^{\xi,+-}(t,t')$, are set at $t = t'$, it must be a function of $t-t'$. We use the Fourier convention that $G^{\xi,+-}(t,t') = \int_{-\infty}^{\infty} \frac{\d\omega'}{2\pi} e^{-i\omega'(t-t')} \tilde{G}^{\xi,+-}(\omega')$.}, so that $G^{\xi,+-}(t,t')$ is given by the following inverse Fourier transform,
\bea
	G^{\xi,+-}(t,t') & = & \int_{-\infty}^{\infty} \frac{\d\omega'}{2\pi} \frac{i e^{-i\omega'(t-t')}}{\omega'^2 + 2i\gamma\omega' - \omega^2} \, .
\label{eq:GretFT}
\eea
The integrand has two simple poles at $\omega' = -i\gamma \pm \sqrt{\omega^2 - \gamma^2}$, both on the negative imaginary axis for $\omega > \gamma$ or $\omega < \gamma$ and assuming that $\gamma > 0$. Closing the contour from below gives
\bea
    G^{\xi,+-}(t,t') & = & -\frac{i e^{-\gamma(t-t')}}{\Gamma} \sin \[ \Gamma(t-t') \] \theta(t-t') \, , \quad \ \ 
\label{eq:Gret}
\eea
where $\Gamma = \sqrt{\omega^2 - \gamma^2}$, and we can check that this expression has the appropriate $\omega = \gamma$ and $\gamma = 0$ limits as well.

We next solve Eq.\ (\ref{eq:Gpptl}) for the symmetrized two-point correlation, $G^{\xi,++}(t,t')$. Eq.\ (\ref{eq:Gpptl}) is a nonhomogeneous differential equation and, as mentioned earlier, its solution consists of a homogeneous piece $G^{\xi,++[h]}(t,t')$ and a particular solution or noise piece $G^{\xi,++[n]}(t,t')$,
\bea
    G^{\xi,++}(t,t') & = & G^{\xi,++[h]}(t,t') + G^{\xi,++[n]}(t,t') \, , \quad
\label{eq:Gsym}
\eea
that we consider in turn. The homogeneous piece is the solution to the equation
\bea
	\( \frac{\d^2}{\d t^2} + \omega^2 + 2\gamma \frac{\d}{\d t} \) G^{\xi,++[h]}(t,t') & = & 0 \, ,
\label{eq:Gpptlh}
\eea
and, as also mentioned earlier, satisfies the initial conditions imposed on the full solution $G^{\xi,++}(t,t')$. Now since $G^{\xi,++}(t,t')$ is symmetric under the interchange of $t$ and $t'$, it satisfies the same equation of motion in both time coordinates. Also noting that $G^{\xi,++}(t,t')$ is a real function, the solution to Eq.\ (\ref{eq:Gpptlh}) must be of the form
\bea
    G^{\xi,++[h]}(t,t') & = & a h(t) h(t') + a^* h^*(t) h^*(t') \nn \\
    & & \quad + \ b \[ h(t) h^*(t') + h^*(t) h(t') \] , \quad
\label{eq:Gpptlhansatz}
\eea
where $a$ is a complex constant, $b$ is a real constant, and $h(t)$ and $h^*(t)$ are solutions to the equation $\ddot{h}(t) + 2\gamma\dot{h}(t) + \omega^2 h(t) = 0$. 

The constants $a$ and $b$ in Eq.\ (\ref{eq:Gpptlhansatz}) can be fixed by making use of the initial conditions in Eqs.\ (\ref{eq:phiphicorr2}), (\ref{eq:phimomcorr2}), and (\ref{eq:mommomcorr2}), which directly translate into initial conditions on $G^{\xi,++}(t,t')$ and, therefore, on $G^{\xi,++[h]}(t,t')$. We thus have three equations in three constants, which yield $a_R$, $a_I$, and $b$ in terms of $c_{xx}$, $c_{xp}$, and $c_{pp}$ and additionally $h(t_0)$, $\dot{h}(t_0)$, $h^*(t_0)$, and $\dot{h}^*(t_0)$ \footnote{It is worth noting that in the absence of dissipation, one can directly solve for $G^<(t,t')$ by modifying the ansatz in Eq.\ (\ref{eq:Gpptlhansatz}) to have different real constants in front of the $h(t) h^*(t')$ and $h^*(t) h(t')$ terms and adding to the three initial conditions in Eqs.\ (\ref{eq:phiphicorr2}), (\ref{eq:phimomcorr2}), and (\ref{eq:mommomcorr2}) the Wronskian condition in Eq.\ (\ref{eq:wronskiant0}).}. We finally need a solution for $h(t)$ and $h^*(t)$. The solution for $h(t)$ can be written as
\bea
	h(t) & = & -i G_1(t) h(t_0) + i G_2(t) \dot{h}(t_0) \, ,
\label{eq:htsol}
\eea
where $G_1(t)$ and $G_2(t)$ are solutions to the same equation as that of $h(t)$ with the initial conditions $G_1(t_0) = i$, $\dot{G}_1(t_0) = 0$ and $G_2(t_0) = 0$, $\dot{G}_2(t_0) = -i$, and the solution for $h^*(t)$ is similarly given by $h^*(t) = -i G_1(t) h^*(t_0) + i G_2(t) \dot{h}^*(t_0)$. Now plugging in the resulting expressions for the constants $a_R$, $a_I$, and $b$ and the solutions for $h(t)$ and $h^*(t)$ back into Eq.\ (\ref{eq:Gpptlhansatz}) gives us the final expression for $G^{\xi,++[h]}(t,t')$,
\bea
    & & G^{\xi,++[h]}(t,t') \, = \, - G_1(t) G_1(t') c_{xx} + \big[ G_2(t) G_1(t') \nn \\
    & & \qquad + \ G_1(t) G_2(t') \big] c_{xp} - G_2(t) G_2(t') c_{pp} \, .
\label{eq:Gpptlhsolp}
\eea
Lastly, we write down expressions for the functions $G_1(t)$ and $G_2(t)$,
\bea
    G_1(t) & = & ie^{-\gamma(t-t_0)} \( \cos \[ \Gamma(t-t_0) \] + \frac{\gamma}{\Gamma} \sin \[ \Gamma(t-t_0) \] \) , \nn \\
\label{eq:G1sol} \\
    G_2(t) & = & -\frac{ie^{-\gamma(t-t_0)}}{\Gamma} \sin \[ \Gamma(t-t_0) \] .
\label{eq:G2sol}
\eea
To summarize, the homogeneous part of the symmetrized two-point correlation is obtained by plugging Eqs.\ (\ref{eq:G1sol}) and (\ref{eq:G2sol}) into Eq.\ (\ref{eq:Gpptlhsolp}), which gives
\begin{widetext}
\bea
	G^{\xi,++[h]}(t,t') & = & \frac{1}{2 \Gamma^2} e^{-\gamma(t+t'-2t_0)} \big\{ \cos \[ \Gamma(t-t') \] \( \omega^2 c_{xx} + 2\gamma c_{xp} + c_{pp} \) - \cos \[ \Gamma (t+t'-2t_0) \] \big( \gamma^2 c_{xx} - \Gamma^2 c_{xx} \nn \\
	& & \quad + \ 2\gamma c_{xp} + c_{pp} \big) + 2\Gamma \sin \[ \Gamma(t+t'-2t_0) \] \( \gamma c_{xx} + c_{xp} \) \big\} \, .
\label{eq:Gpptlhsol}
\eea
\end{widetext}
We see that $G^{\xi,++[h]}(t,t')$ is proportional to $e^{-\gamma (t+t'-2t_0)}$ and thus vanishes in the limits that $\gamma(t-t_0) \gg 1$ and $\gamma(t'-t_0) \gg 1$ for any choice of initial conditions $c_{xx}$, $c_{xp}$, and $c_{pp}$.

The noise contribution in Eq.\ (\ref{eq:Gsym}) is simpler to calculate and, as mentioned earlier, obtained by convolving the source  term on the right hand side of Eq.\ (\ref{eq:Gpptl}) with the retarded Green's function,
\bea
    G^{\xi,++[n]}(t,t') & = & -\alpha \int_{t''} \[ G^{\xi,+-}(t,t'') G^{\xi,+-}(t',t'') \] . \nn \\
\label{eq:Gpptln}
\eea
We will choose $t \ge t'$ without loss of generality, in which case we find on using Eq.\ (\ref{eq:Gret}) for $G^{\xi,+-}(t,t')$,
\begin{widetext}
\bea
	G^{\xi,++[n]}(t,t') & = & -\frac{\alpha}{4 \gamma \omega^2 \Gamma^2} e^{-\gamma(t+t'-2t_0)} \big( \omega^2 \cos \[ \Gamma(t-t') \] - \gamma^2 \cos \[ \Gamma(t+t'-2t_0) \] + \gamma\Gamma \sin \[ \Gamma(t+t'-2t_0) \] \big) \nn \\
	& & \quad + \ \frac{\alpha}{4\gamma \omega^2 \Gamma} e^{-\gamma(t-t')} \big( \Gamma \cos \[ \Gamma(t-t') \] + \gamma \sin \[ \Gamma(t-t') \] \big) \, .
\label{eq:Gpptlnsol}
\eea
\end{widetext}
Note that this vanishes at the initial time (i.e., at $t' = t_0$), as mentioned earlier. We see that $G^{\xi,++[n]}(t,t')$ also has a piece proportional to $e^{-\gamma(t+t'-2t_0)}$ which vanishes in the limits that $\gamma(t-t_0) \gg 1$ and $\gamma(t'-t_0) \gg 1$, but additionally has a piece proportional to $e^{-\gamma(t-t')}$ that need not vanish unless $\gamma (t-t') \gg 1$ as well. This will be important for the fluctuation-dissipation relation that we discuss later in this section.

With the Green's functions in hand, we can also use Eq.\ (\ref{eq:def1pt}) to write an expression for the one-point function $\la \hat{X}(t) \ra$. Substituting the expressions given after Eq.\ (\ref{eq:defxJs}) for the shifted sources into Eq.\ (\ref{eq:def1pt}), we see that $\la \hat{X}(t) \ra$ can be expressed as
\bea
	\la \hat{X}(t) \ra & = & i \int_{t'} G^{\xi,+-}(t,t') J(t') - i G_1(t) x_0 + i G_2(t) p_0 \, , \nn \\
\eea
which, in the absence of a source and using Eqs.\ (\ref{eq:G1sol}) and (\ref{eq:G2sol}) becomes
\bea
	\la \hat{X}(t) \ra & = & e^{-\gamma(t-t_0)} \( \cos \[ \Gamma(t-t_0) \] + \frac{\gamma}{\Gamma} \sin \[ \Gamma(t-t_0) \] \) \nn \\
    & & \quad \times \, x_0 + \frac{e^{-\gamma(t-t_0)}}{\Gamma} \sin \[ \Gamma(t-t_0) \] p_0 \, . \ \ 
\label{eq:Xtlsol}
\eea
Note that this vanishes in the limit that $\gamma(t-t_0) \gg 1$.

%-------------------------------
\subsection{Purity}
%-------------------------------

As discussed in Sec.\ \ref{subsec:is}, the purity of an oscillator in a Gaussian state can be written in terms of its two-point correlators. At the initial time $t_0$, we found that ${\rm Pu}(t_0) = 0.5/\sqrt{c_{xx} c_{pp} - c_{xp}^2}$. We can similarly write the purity at any time $t$ in terms of two-point correlators at that time using
\bea
    {\rm Pu}(t) & = & \frac{0.5}{\sqrt{\big\la \hat{X}^2(t) \big\ra_c \big\la \hat{P}^2(t) \big\ra_c - (1/4) \big\la \big\{ \hat{X}(t), \hat{P}(t) \big\} \big\ra_c^2}} \, , \nn \\
\eea
with the correlators in turn obtained from $G^{\xi,++}(t,t')$ of the previous subsection,
\bea
	\big\la \hat{X}^2(t) \big\ra_c & = & G^{\xi,++}(t,t) \, ,
\label{eq:phiphicorrt} \\
	\frac{1}{2} \big\la \big\{ \hat{X}(t), \hat{P}(t) \big\} \big\ra_c & = & \partial_{t'} G^{\xi,++}(t,t') \big|_{t'=t} \, ,
\label{eq:phimomcorrt} \\
	\big\la \hat{P}^2(t) \big\ra_c & = & \partial_t \partial_{t'} G^{\xi,++}(t,t') \big|_{t'=t} \, .
\label{eq:mommomcorrt}
\eea
We do not write explicit expressions for the two-point correlators, but they can be obtained using the results from the previous subsection. It is instructive, however, to consider the final expression for the purity, which we find to be
\begin{widetext}
\bea
	{\rm Pu}(t) & = & \bigg[ \frac{\alpha^2}{4 \gamma^2 \omega^2} - \frac{\alpha}{2\gamma^2 \omega^2 \Gamma^2} e^{-2\gamma(t-t_0)} \big\{ \omega^2 \( -2\omega^2 \gamma c_{xx} - 4 \gamma^2 c_{xp} - 2\gamma c_{pp} + \alpha \) \nn \\
	& & \qquad + \ \gamma^2 \( 2\omega^2 \gamma c_{xx} + 4 \omega^2 c_{xp} + 2\gamma c_{pp} - \alpha \) \cos \[ 2\Gamma(t-t_0) \] - 2\gamma^2 \Gamma \( \omega^2 c_{xx} - c_{pp} \) \sin \[ 2\Gamma(t-t_0) \] \big\} \nn \\
	& & \quad + \ \frac{1}{4\gamma^2 \omega^2} e^{-4\gamma(t-t_0)} \( -4\alpha \omega^2 \gamma c_{xx} + 16\omega^2 \gamma^2 c_{xx} c_{pp} - 16\omega^2 \gamma^2 c_{xp}^2 - 4\alpha \gamma c_{pp} + \alpha^2 \) \bigg]^{-1/2} \, . \quad
\label{eq:purityexact}
\eea
\end{widetext}
In the late-time limit, in particular, the above expression reduces to
\bea
	\lim_{\gamma (t-t_0) \gg 1} {\rm Pu}(t) & = & \frac{2\gamma\omega}{\alpha} \, .
\label{eq:puritylt}
\eea
The late-time purity is thus independent of the initial conditions $c_{xx}$, $c_{xp}$, and $c_{pp}$, similar to the late-time correlations, and is in fact constant. Since purity must be between $0$ and $1$ at all times, we find a constraint on the three parameters: $0 \le 2\gamma\omega/\alpha \le 1$, showing that the dissipation parameters $\gamma$ and $\alpha$ are not independent of one another and additionally that $\alpha$ is positive. We also see that $\alpha$ cannot be set to zero as the late-time purity would otherwise diverge. In other words, the $\alpha \rightarrow 0$ limit of Eq.\ (\ref{eq:purityexact}) reads
\bea
	\lim_{\alpha \rightarrow 0} {\rm Pu}(t) & = & e^{2\gamma (t-t_0)} {\rm Pu}(t_0) \, ,
\label{eq:pualpha0}
\eea
which grows exponentially in time and is, therefore, not physical. The fact that $\alpha$ needs to be  non-zero if $\gamma$ is non-zero has to do with properties of the environment that is causing dissipation of the oscillator in the first place.

We can go even further and {\it reconstruct} the late-time density operator. Since our initial state is Gaussian and the dynamics are linear, we expect the density matrix $\la x^+ | \hat{\rho}(t) | x^- \ra$ at any time $t$ to be of the same form as the initial density matrix in Sec.\ \ref{subsec:is}, except that all correlations -- $x_0$, $p_0$, $c_{xx}$, $c_{xp}$, and $c_{pp}$ -- must be replaced by those at time $t$. Upon doing so using the expressions in the previous subsection (in the absence of a source), we find that the late-time density matrix is given by
\bea
    \la x^+ | \hat{\rho}(\infty) | x^- \ra & = & \omega \sqrt{\frac{2\gamma}{\pi \alpha}} \exp \bigg[ -\frac{\omega}{4} \bigg\{ \frac{\alpha}{2\gamma \omega} (x^+ - x^-)^2 \nn \\
    & & \quad + \, \frac{2\gamma \omega}{\alpha} \left(x^+ + x^-\right)^2 \bigg\} \bigg] \, .
\label{AsymptoticKKbar_SHO}
\eea
Let us compare this to the density operator for an isolated harmonic oscillator in a thermal state, $\hat{\rho}_{\beta}(t) = e^{-\beta\hat{H}}$, where $\beta$ is the inverse of the Boltzmann constant times the temperature of the oscillator. In position basis, $\hat{\rho}_{\beta}(t)$ is given by
\bea
    & & \la x^+ | \hat{\rho}_{\beta}(t) | x^- \ra
    \, = \, \sqrt{\frac{\omega \tanh(\beta \omega/2)}{\pi}} \exp \bigg[ -\frac{\omega}{2 \sinh(\beta \omega)} \nn \\
    & & \qquad
    \times \, \Big\{ \[ (x^+)^2 + (x^-)^2 \] \cosh (\beta \omega) - 2 x^+ x^- \Big\} \bigg] \, .
\label{SHO_ThermalDensityMatrix}
\eea
Eqs.\ (\ref{AsymptoticKKbar_SHO}) and (\ref{SHO_ThermalDensityMatrix}) together suggest that the DHO thermalizes at the temperature $\beta = (2/\omega) \tanh^{-1} (2\gamma\omega/\alpha)$. We can arrive at the same conclusion by directly comparing the late-time purity of the DHO in Eq.\ (\ref{eq:puritylt}) with the purity of an oscillator in thermal equilibrium, given by ${\rm Pu}(t) = \tanh(\beta\omega/2)$, as well. Note, however, that although the late-time density operator is thermal, this does not necessarily imply that the fluctuation-dissipation relation is satisfied, and we discuss this further in the next subsection.

\begin{figure*}[!t]
\begin{center}
	\includegraphics[width=3.25in,angle=0]{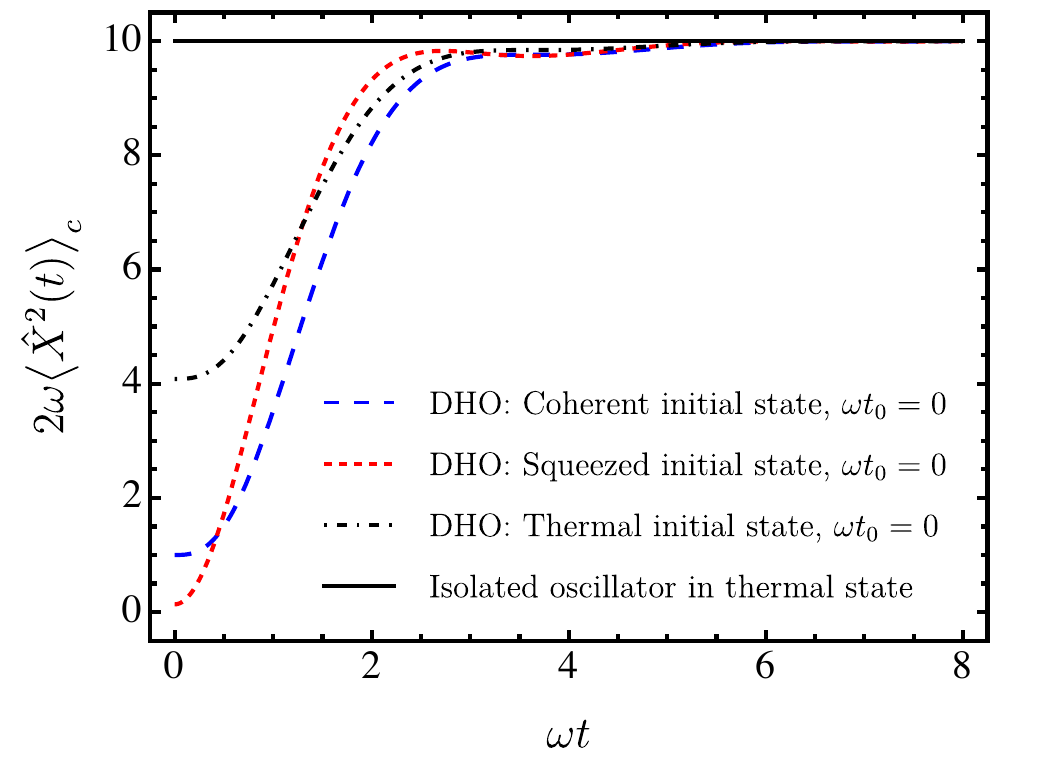}
	\includegraphics[width=3.25in,angle=0]{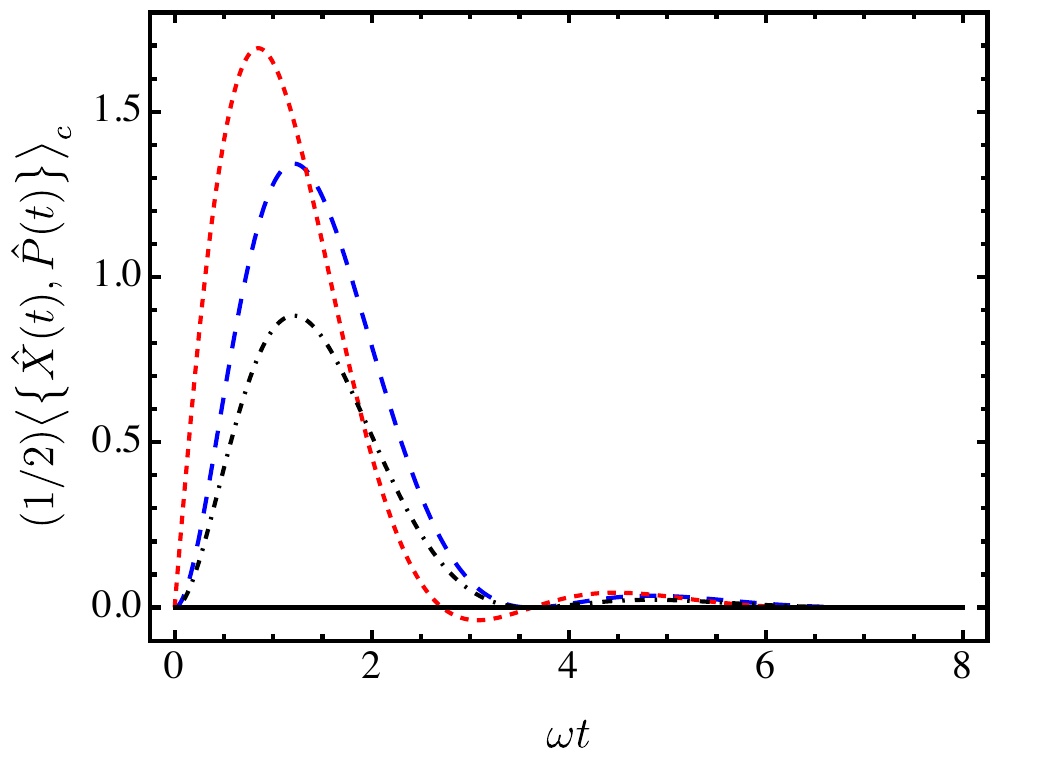}
	\includegraphics[width=3.25in,angle=0]{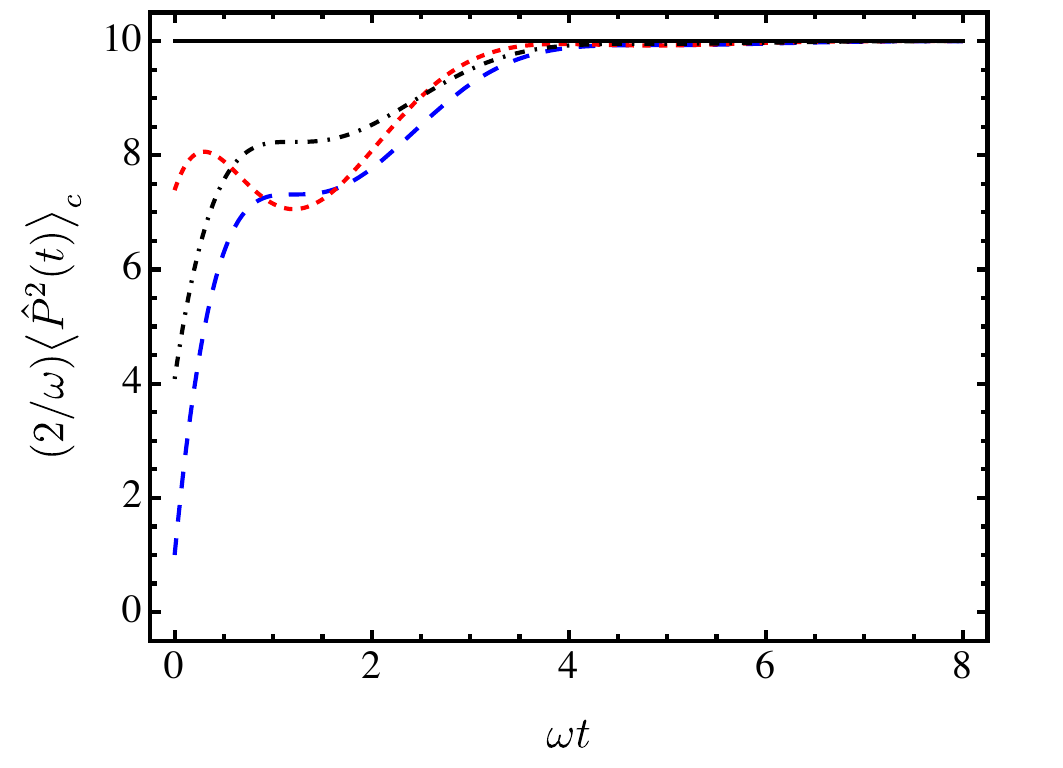}
	\includegraphics[width=3.25in,angle=0]{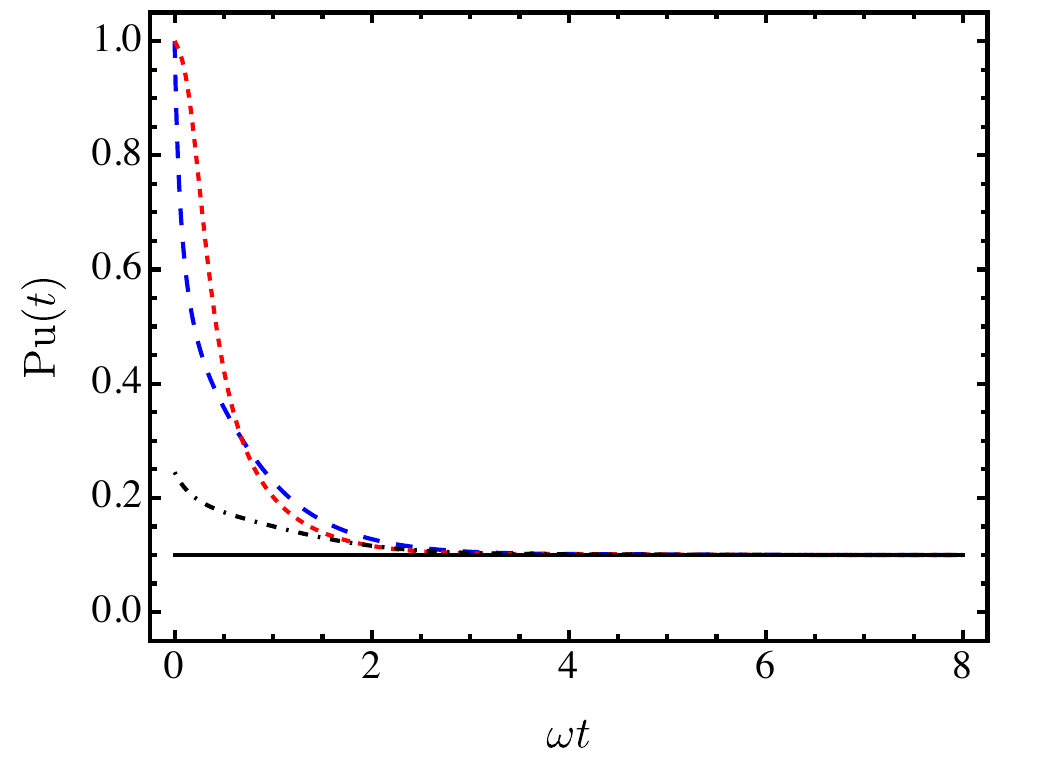}
	\caption{Equal-time two-point correlations $\big\la \hat{X}^2(t) \big\ra_c$ (top, left), $(1/2) \big\la \big\{ \hat{X}(t), \hat{P}(t) \big\} \big\ra_c$ (top, right), and $\big\la \hat{P}^2(t) \big\ra_c$ (bottom, left) and purity (bottom, right) as a function of time for three different choices of initial state -- (i) a coherent state with $c_{xx} = 1/(2\omega)$, $c_{xp} = 0$, and $c_{pp} = \omega/2$, (ii) a squeezed state with $c_{xx} = e^{-2r}/(2\omega)$, $c_{xp} = 0$, $c_{pp} = e^{2r} \omega/2$, and $r = 1$, and (iii) a thermal state with $c_{xx} = \coth(\beta\omega/2)/(2\omega)$, $c_{xp} = 0$, $c_{pp} = \coth(\beta\omega/2) \omega/2$, and $\beta = 1/(2\omega)$ -- in the presence of time-local dissipation with $\gamma = \omega/2$ and $\alpha = 10\omega^2$ in each case and setting $\omega t_0 = 0$ for simplicity. We also show the result for an isolated harmonic oscillator at temperature $\beta = (2/\omega) \tanh^{-1} (2\gamma\omega/\alpha) \approx 0.2/\omega$ for comparison.}
\label{fig:dissex}
\end{center}
\end{figure*}

Before closing this subsection, we also plot in Fig.\ \ref{fig:dissex} the two-point correlators found from Eqs.\ (\ref{eq:phiphicorrt}), (\ref{eq:phimomcorrt}), and (\ref{eq:mommomcorrt}), on substituting for $G^{\xi,++}(t,t')$ from the previous subsection, and the purity in Eq.\ (\ref{eq:purityexact}), for three different choices of initial state and one set of dissipation parameters for illustration. In units of the oscillator frequency $\omega$, we first consider a coherent initial state with $c_{xx} = 1/(2\omega)$, $c_{xp} = 0$, and $c_{pp} = \omega/2$, therefore ${\rm Pu}(t_0) = 1$. We next consider a squeezed initial state with $c_{xx} = e^{-2r}/(2\omega)$, $c_{xp} = 0$, $c_{pp} = e^{2r} \omega/2$, and $r = 1$, therefore ${\rm Pu}(t_0) = 1$ again. And we lastly consider a thermal initial state with $c_{xx} = \coth(\beta\omega/2)/(2\omega)$, $c_{xp} = 0$, $c_{pp} = \coth(\beta\omega/2) \omega/2$, and $\beta = 1/(2\omega)$, therefore ${\rm Pu}(t_0) \approx 0.24$. We choose the dissipation parameters to be $\gamma = \omega/2$ and $\alpha = 10\omega^2$ in all three cases and set $\omega t_0 = 0$ for simplicity. We also show the result for an isolated harmonic oscillator at temperature $\beta = (2/\omega) \tanh^{-1} (2\gamma\omega/\alpha) \approx 0.2/\omega$ for comparison, using Eq.\ (\ref{eq:Glessthermal}) below to find the correlators and the expression from the previous paragraph for the purity. It is evident from the figure that the late-time behavior of the system is independent of the choice of initial state and equal-time results match those of an oscillator in thermal equilibrium.

%========================= FIGURE 1 ==========================
%============================================================

%-------------------------------
\subsection{Fluctuation-dissipation relation}
%-------------------------------

The fluctuation-dissipation relation goes beyond demanding that the density operator be thermal as it is a statement about {\it unequal-time} correlators of the system. Let us first review the fluctuation-dissipation relation for an isolated harmonic oscillator in a thermal state, $\hat{\rho}_{\beta}(t) = e^{-\beta\hat{H}}$. In our notation (see Sec.\ \ref{subsec:npoint}), the two-point correlation $G_{\beta}^<(t,t')$ is given by
\bea
	G_{\beta}^<(t,t') & = & \frac{e^{\beta\omega} e^{i\omega(t-t')} + e^{-i\omega(t-t')}}{2\omega (e^{\beta\omega} - 1)} ,
\label{eq:Glessthermal}
\eea
and $G_{\beta}^>(t,t') = G_{\beta}^{<*}(t,t')$. Since Eq.\ (\ref{eq:Glessthermal}) only depends on the difference of times, we will write $G_{\beta}^<(t,t')$ and $G_{\beta}^>(t,t')$ as $G_{\beta}^<(T)$ and $G_{\beta}^>(T)$ instead, where $T = t-t'$. Denoting the Fourier transforms of $G_{\beta}^<(T)$ and $G_{\beta}^>(T)$ with $\tilde{G}_{\beta}^<(\omega')$ and $\tilde{G}_{\beta}^>(\omega')$, one finds that $\tilde{G}_{\beta}^<(\omega') = e^{-\beta\omega'} \tilde{G}_{\beta}^>(\omega')$, called the Kubo-Martin-Schwinger (KMS) relation.

Let us now consider the Fourier transform of the anti-symmetric two-point correlation $G_{\beta}^>(T) - G_{\beta}^<(T)$, which is equal to twice the real part of the Fourier transform of $\big[ G_{\beta}^>(T) - G_{\beta}^<(T) \big] \theta(T)$. We define
\bea
    {\cal A}_{\beta}(\omega') & = & {\rm Re} \int_{0}^{\infty} \d T \, e^{i\omega' T} \big[ G_{\beta}^>(T) - G_{\beta}^<(T) \big] \, ,
\eea
so that ${\cal A}_{\beta}(\omega') = \big[ \tilde{G}_{\beta}^>(\omega') - \tilde{G}_{\beta}^<(\omega') \big]/2$. Similarly, we denote the Fourier transform of the symmetric two-point correlation with ${\cal S}_{\beta}(\omega')$, so that
\bea
    {\cal S}_{\beta}(\omega') & = & \frac{1}{2} \int_{-\infty}^{\infty} \d T \, e^{i\omega' T} \big[ G_{\beta}^>(T) + G_{\beta}^<(T) \big] \,
\eea
and it is given by $\big[ \tilde{G}_{\beta}^>(\omega') + \tilde{G}_{\beta}^<(\omega') \big]/2$. Using the KMS relation, we can show that ${\cal A}_{\beta}(\omega')$ and ${\cal S}_{\beta}(\omega')$ satisfy
\bea
	{\cal S}_{\beta}(\omega') & = & \coth \( \frac{\beta\omega'}{2} \) {\cal A}_{\beta}(\omega') \, ,
\label{eq:fdr}
\eea
known as the fluctuation-dissipation relation.

%========================= FIGURE 2 ==========================
\begin{figure*}[!t]
\begin{center}
	\includegraphics[width=3.25in,angle=0]{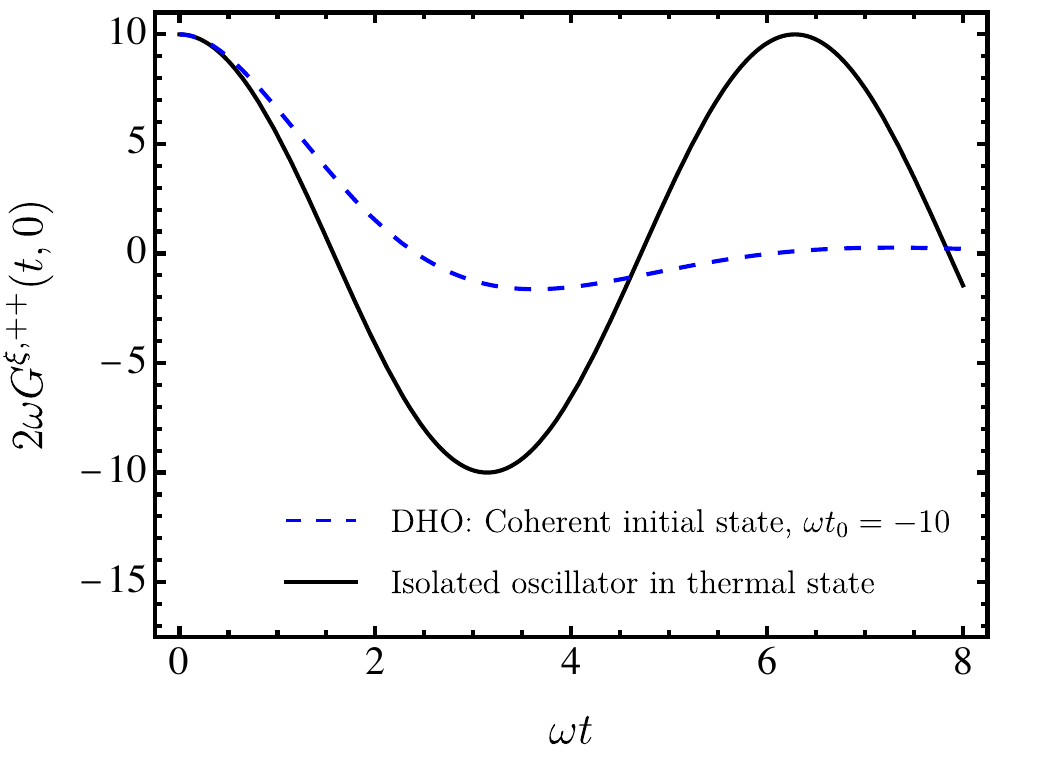}
	\includegraphics[width=3.25in,angle=0]{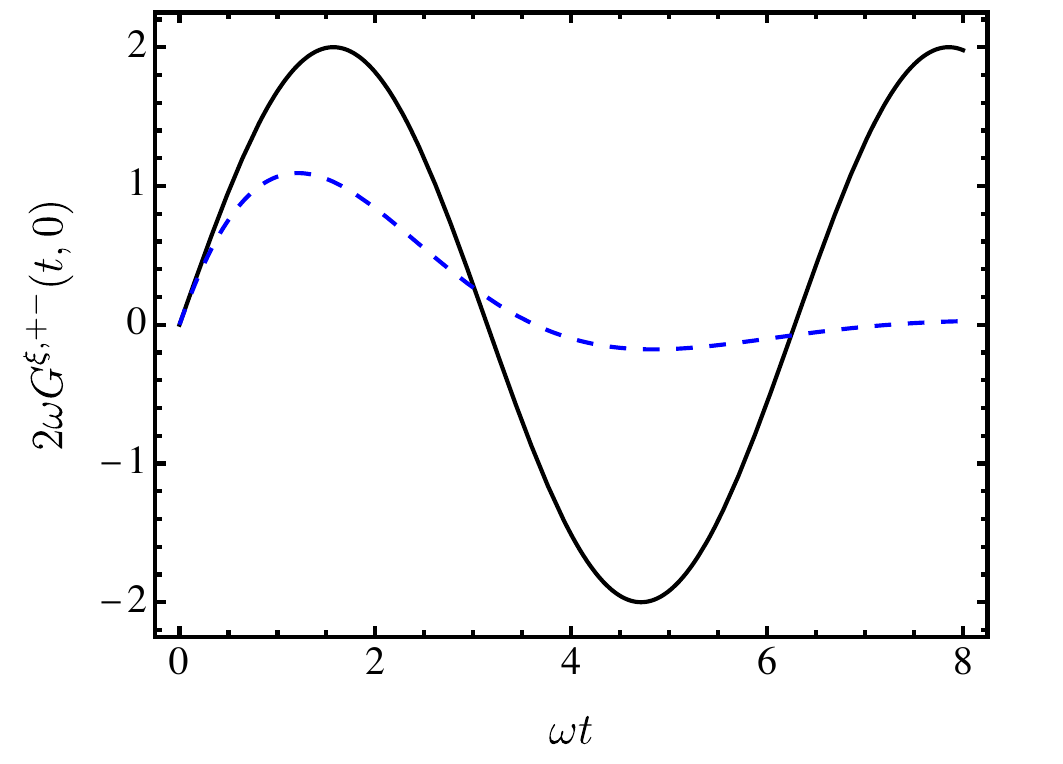}
	\caption{Unequal-time two-point correlations, symmetric (left) and anti-symmetric (right), for a quantum DHO compared to an isolated harmonic oscillator in a thermal state. The DHO is initialized in a coherent state with $c_{xx} = 1/(2\omega)$, $c_{xp} = 0$, and $c_{pp} = \omega/2$ at $\omega t_0 = -10$, and undergoes time-local dissipation with $\gamma = \omega/2$ and $\alpha = 10\omega^2$, while the isolated harmonic oscillator is at the temperature $\beta = (2/\omega) \tanh^{-1} (2\gamma\omega/\alpha) \approx 0.2/\omega$.}
\label{fig:dissexut}
\end{center}
\end{figure*}
%============================================================

We now want to check whether the quantum DHO that we studied in the previous subsections satisfies the fluctuation-dissipation relation in the late-time limit. Consider first ${\cal A}(\omega')$ for the DHO, given by the real part of the Fourier transform of $G^{\xi,+-}(t,t')$ with respect to $T = t-t'$. Reading off the Fourier transform of $G^{\xi,+-}(t,t')$ from Eq.\ (\ref{eq:GretFT}), we find that
\bea
	{\cal A}(\omega') & = & \frac{2\gamma\omega'}{\left| \omega'^2 + 2i\gamma \omega' - \omega^2 \right|^2} \, .
\label{eq:rhofdr}
\eea
Next, consider ${\cal S}(\omega')$, given by the Fourier transform of $G^{\xi,++}(t,t')$. We noted in Sec.\ \ref{subsec:Gtl} that the homogeneous part of $G^{\xi,++}(t,t')$ vanishes in the late-time limit, and therefore we only need to consider its noise part. The Fourier transform of $G^{\xi,++[n]}(t,t')$ in the late-time limit can in turn be obtained from Eq.\ (\ref{eq:Gpptln}) by first replacing $G^{\xi,+-}(t,t'')$ and $G^{\xi,+-}(t',t'')$ with their integral representation in Eq.\ (\ref{eq:GretFT}), then setting the limits of the $t''$ integral to be $t_0 = -\infty$ to $t_f = \infty$, and then using $\int_{-\infty}^{\infty} \d t'' e^{i(\omega'+\omega'') t''} = 2\pi \delta(\omega'+\omega'')$ to collapse one of the two frequency integrals. The resulting expression for ${\cal S}(\omega')$ is given by
\bea
	{\cal S}(\omega') & = & \frac{\alpha}{\left| \omega'^2 + 2i\gamma \omega' - \omega^2 \right|^2} \, ,
\label{eq:kappafdr}
\eea
and ${\cal A}(\omega')$ and ${\cal S}(\omega')$ for the DHO, therefore, satisfy
\bea
	{\cal S}(\omega') & = & \( \frac{\alpha}{2\gamma\omega'} \) {\cal A}(\omega') \, .
\label{eq:dofdr}
\eea
Comparing Eqs.\ (\ref{eq:fdr}) and (\ref{eq:dofdr}), we see that they match for multiple values of $\omega'$ only in the {\it high-temperature regime} where $\beta\omega' \ll 1$ and $\coth \( \beta\omega'/2 \) \approx 2/(\beta\omega')$, and the temperature to which the oscillator thermalizes is given by $\beta = 4\gamma/\alpha$. This agrees with the temperature found towards the end of the previous subsection, $\beta = (2/\omega) \tanh^{-1} (2\gamma\omega/\alpha)$, in the high-temperature limit. It also matches the temperature of the environment for the Caldeira-Leggett model in the regime mentioned at the beginning of this section, where we noted that $\alpha = 4\gamma/\beta_{\rm env}$ in this model.

Eq.\ (\ref{eq:dofdr}) can also be interpreted as a {\it generalized fluctuation-dissipation relation} for the quantum DHO, which when comparied with Eq.\ (\ref{eq:fdr}) gives an effective frequency-dependent temperature, $\beta_{\rm eff}(\omega') = (2/\omega') \tanh^{-1} (2\gamma\omega'/\alpha)$ \cite{Clerk:2008tlb,Britto:2015afa,Scarlatella:2019}. We see that the effective temperature agrees with the temperature $\beta$ obtained earlier at the isolated frequency $\omega' = \omega$; the physical significance of this, however, is unclear to us. We emphasize that the generalized fluctuation-dissipation relation in Eq.\ (\ref{eq:dofdr}) can be used to distinguish a quantum DHO from an isolated harmonic oscillator in a thermal state. Alternatively, the frequency spectrum or {\it unequal-time} measurements of the system can also be used to distinguish between the two cases even in the high-temperature regime. We show the unequal-time correlations explicitly in Fig.\ \ref{fig:dissexut} for clarity, for the same choice of parameters as in Fig.\ \ref{fig:dissex} with a coherent initial state.

It is also worth noting that the late-time limit of $G^{\xi,++}(t,t')$, which is simply the second term in Eq.\ (\ref{eq:Gpptlnsol}), also agrees with the symmetric two-point correlation obtained in \cite{Grabert:1984,Riseborough:1985,Grabert:1988,Hanggi:2005,Weiss:2012} by first assuming that the fluctuation-dissipation relation holds for the quantum DHO, then finding the inverse Fourier transform of $\coth \( \beta\omega'/2 \) \rho(\omega')$, and finally taking the high-temperature and additionally weak-coupling ($\omega > \gamma$) limit of the result.

%%%%%%%%%%%%%%%%%%%%%%%%%%%%%%%%%%%%%%%%%%%%%%%%%%

%-------------------------------
\section{Time-nonlocal dissipation}
\label{sec:tnldissipation}
%-------------------------------

In this section, we briefly consider time-nonlocal dissipation with a specific choice of dissipation kernels, such that the fluctuation-dissipation relation is satisfied in the late-time limit at {\it any} temperature. We choose our dissipation kernels to match those in the Caldeira-Leggett model, specializing again to an Ohmic spectral density with infinite cutoff but without restricting to the high-temperature regime \cite{Breuer:2002pc,Caldeira:1982iu}. In this case, we have $\gamma_{1R}(t,t') + \gamma_{2R}(t,t') = 2\gamma \ [\d\delta(t-t')/\d t'] \theta(t-t')$ as before and
\bea
    \gamma_{1I}(t,t') & = & \gamma \int_{-\infty}^{\infty} \frac{\d\omega'}{2\pi} e^{-i\omega'(t-t')} \omega' \coth \( \frac{\beta_{\rm env}\omega'}{2} \) , \quad \ \ 
\label{eq:gamma1Itnl}
\eea
which simply evaluates to $-(\pi\gamma/\beta_{\rm env}^2) \, {\rm cosech}^2 [\pi (t-t')/\beta_{\rm env}]$. Using the techniques of the previous section, we can now solve for the corresponding anti-symmetric and symmetric two-point correlations of the time-nonlocal DHO.

$G^{\xi,+-}(t,t')$ is, in fact, the same as before, and is given by Eq.\ (\ref{eq:Gret}). Its Fourier transform with respect to $t-t'$ is thus simply given by Eq.\ (\ref{eq:rhofdr}), which we write again for convenience,
\bea
	{\cal A}(\omega') & = & \frac{2\gamma\omega'}{\left| \omega'^2 + 2i\gamma \omega' - \omega^2 \right|^2} \, .
\label{eq:rhofdrtnl}
\eea
The homogeneous piece of $G^{\xi,++}(t,t')$ is also unchanged, given by Eq.\ (\ref{eq:Gpptlhsol}), and again vanishes in the late-time limit. The noise piece of $G^{\xi,++}(t,t')$, on the other hand, is given by
\bea
    G^{\xi,++[n]}(t,t') & = & -2 \int_{t'',t'''} \big[ G^{\xi,+-}(t,t'') \gamma_{1I}(t'',t''') \nn \\
    & & \qquad \times \ G^{\xi,+-}(t',t''') \big] \, ,
\label{eq:Gpptnln}
\eea
with $\gamma_{1I}(t,t')$ given by Eq.\ (\ref{eq:gamma1Itnl}). Since we are only interested in checking the fluctuation-dissipation relation in this section, we only need to consider the Fourier transform of $G^{\xi,++[n]}(t,t')$ in the late-time limit. This can be obtained from Eq.\ (\ref{eq:Gpptnln}) by first replacing $G^{\xi,+-}(t,t'')$, $G^{\xi,+-}(t',t''')$, and $\gamma_{1I}(t,t')$ with their integral representations in Eqs.\ (\ref{eq:GretFT}) and (\ref{eq:gamma1Itnl}), then setting the limits of the $t''$ and $t'''$ integrals to be $t_0 = -\infty$ to $t_f = \infty$, and then using the $\delta$-functions obtained by performing the time integrals to collapse two of the three frequency integrals. The resulting expression for the Fourier transform of $G^{\xi,++}(t,t')$ in the late-time limit is given by
\bea
	{\cal S}(\omega') & = & \coth \( \frac{\beta_{\rm env}\omega'}{2} \) \frac{2\gamma\omega'}{\left| \omega'^2 + 2i\gamma \omega' - \omega^2 \right|^2} \, ,
\label{eq:kappafdrtnl}
\eea
and ${\cal A}(\omega')$ and ${\cal S}(\omega')$ for the time-nonlocal DHO, therefore, satisfy
\bea
	{\cal S}(\omega') & = & \coth \( \frac{\beta_{\rm env}\omega'}{2} \) {\cal A}(\omega') \, .
\label{eq:dofdrtnl}
\eea
Since this matches Eq.\ (\ref{eq:fdr}), we conclude that the fluctuation-dissipation relation is satisfied for a time-nonlocal DHO with the specific dissipation kernels chosen here, and it thermalizes to the temperature of the thermal environment. As in the time-local case, the frequency spectrum or unequal-time measurements of the system can, however, be used to distinguish a quantum DHO from an isolated harmonic oscillator in a thermal state.

Lastly, we comment on how this relates to the time-local case of the previous section. In the high-temperature limit, we can expand the integrand in Eq.\ (\ref{eq:gamma1Itnl}) using $\coth(x) \approx x^{-1} + x/3 + O(x^3)$ as $x \rightarrow 0$. Since the factors of $\omega'$ cancel in the zeroth-order piece, the kernel reduces to the time-local one with $\gamma_{1I}(t,t') = (2\gamma/\beta_{\rm env}) \delta(t-t')$. The next-to-leading order term is equally interesting since it results in a factor of $\omega'^2$ under the integrand, which gives a $-(\gamma\beta_{\rm env}/6) \d^2 \delta(t-t')/\d t'^2$ contribution to the kernel. Note that such a term by itself will need to be handled carefully since we have assumed in our derivation that the dissipation kernels $\gamma_1(t,t')$ and $\gamma_2(t,t')$ are not proportional to $\d^2 \delta(t-t')/\d t'^2$. Our derivation should be valid here, however, since this term arises from expanding the full dissipation kernel in Eq.\ (\ref{eq:gamma1Itnl}). We can see that a $-(\gamma\beta_{\rm env}/6) \d^2 \delta(t-t')/\d t'^2$ contribution would also generate time-local dissipation, but with higher-order time derivatives appearing in the noise term. In the spirit of effective field theory, we can, therefore, replace the time-nonlocal dissipation kernel with a time-local one with successively higher-order time derivatives, in the high-temperature limit. Going back to the frequency domain, we see that the series expansion tracks all the way to Eq.\ (\ref{eq:dofdrtnl}) and the fluctuation-dissipation relation matches that in Eq.\ (\ref{eq:dofdr}) with $\alpha = 4\gamma/\beta_{\rm env}$ to leading order, with higher-order corrections.

%%%%%%%%%%%%%%%%%%%%%%%%%%%%%%%%%%%%%%%%%%%%%%%%%%

%-------------------------------
\section{Beyond Gaussian initial states}
\label{sec:beyondgaussianis}
%-------------------------------

In the previous sections, we folded in the initial conditions arising from the Gaussian density matrix of Eq.\ \eqref{eq:initialS} with the dynamics of the quantum DHO from the outset, and solved for the relevant $2 \times 2$ Green's function $\boldsymbol{G}(t,t')$. We then solved for the $n$-point correlation functions and purity of the system from $\boldsymbol{G}(t,t')$ itself. In this section, we instead compute the time evolution operator of the density matrix for the quantum DHO, without first convolving it against the initial density operator. This approach not only allows us to cleanly separate the influence of dynamics from that of the initial conditions, but also results in expressions for equal-time $n$-point functions and purity that are valid for {\it any} choice of initial state.

As a start, we recall that the path integral -- without dissipation --
\begin{align}
\label{PathIntegral_FullSystem}
    K[t_f,x_f^+;t_0,x_0^+] 
    & \equiv \int_{x_0^+}^{x_f^+} \mathcal{D} x^+ \exp\left[i \int_{t} L[x^+,\dot{x}^+] \right] ,
\end{align}
for an appropriate Lagrangian $L$, evolves any initial state $\ket{\Psi[t_0]}$ (in the \schrodinger \ picture) forward in time, through the relation
\begin{align}
\label{TimeEvolution}
    \big\la x_f^+ \big| \Psi[t_f] \big\ra
    & = \int_{\mathbb{R}} \dd x_0^+ K[t_f,x_f^+;t_0,x_0^+] \braket{x_0^+}{\Psi[t_0]} .
\end{align}
This in turn means that the density matrix of a closed system is related to its initial state via
\begin{align}
\label{ReducedDensityOperator_TimeEvolution}
    \big\la x_f^+ \big| \hat{\rho}[t_f] \big| x_f^- \big\ra
    & \, = \, \int_{\mathbb{R}} \dd x_0^+ \int_{\mathbb{R}} \dd x_0^- 
    \big\la x_0^+ \big| \hat{\rho}[t_0] \big| x_0^- \big\ra \nn\\
    &\qquad \times K\overline{K}[t_f,x_f^+,x_f^-;t_0,x_0^+,x_0^-] \, ,
\end{align}
where the time evolution of $\hat{\rho}[t_f]$ would be the result of Eq.\ \eqref{TimeEvolution} acting upon the ket and the bra separately, namely,
\begin{align}
    K\overline{K}&[t_f,x_f^+,x_f^-;t_0,x_0^+,x_0^-] \nonumber\\
    & = \, K[t_f,x_f^+;t_0,x_0^+] \overline{K}[t_f,x_f^-;t_0,x_0^-] \\
    & = \, \int_{x_0^+}^{x_f^+} \mathcal{D}x^+ \int_{x_0^-}^{x_f^-} \mathcal{D}x^- \nn \\
    & \qquad \times \, \exp\left[i \int_{t} \left( L[x^+,\dot{x}^+] - \overline{L[x^-,\dot{x}^-]} \right) \right] ,
\end{align}
with the over-bar denoting complex conjugation. Note that the time evolution operator here factorizes into two separate ones. As already mentioned in Sec.\ \ref{sec:Zinin}, however, if $\hat{\rho}$ describes a subsystem embedded in a larger environment, then this is in general no longer the case. Then instead,
\begin{align}
\label{KKbar}
    & K\overline{K}[t_f,x_f^+,x_f^-; t_0,x_0^+,x_0^-] \nonumber\\
    & \quad = \, \int_{x_0^+}^{x_f^+} \mathcal{D}x^+ \int_{x_0^-}^{x_f^-} \mathcal{D}x^-
    \exp\left[ i S_T[x^+,x^-] \right] ,
\end{align}
where the total action $S_T$ consists of three separate terms,
\begin{align}
    S_T[x^+,x^-] 
    & \, = \, S[x^+] - \overline{S[x^-]}
    + S_I[x^+,x^-] \, , \\
    S[x^\pm]
    & \, = \, \int_{t} L[x^\pm,\dot{x}^\pm] \, , \\
    S_I[x^+,x^-]
    & \, = \, \int_{t} L_I[x^+,\dot{x}^+;x^-,\dot{x}^-] \, .
\end{align}
That is, there is now a piece of the action, $S_I$, that couples the variables $x^+$ and $x^-$ in the double path integrals as before, rendering the time evolution operator $K\overline{K}$ non-factorizable. Its presence is again due to the `tracing out' of irrelevant or inaccessible states from the full system's density operator.

In the subsections below, we first impose a general constraint on $K\overline{K}$ and next solve for it in the case of time-local dissipation. We then use the resulting solution to time-evolve the $n$-point correlations and purity of the system for any initial state. We emphasize that solving for $K\overline{K}$ rather than $\hat{\rho}[t_f]$ in this section allows us to obtain results for any initial state, whether Gaussian or non-Gaussian.

%-------------------------------
\subsection{Probability conservation} 
%-------------------------------

As long as our subsystem can be embedded in a larger {\it closed} system, the total probability of finding it in some state has to remain unity at all times. And since $K\overline{K}$ arose from tracing over the irrelevant states of the larger closed system, tracing over the remaining states of the subsystem then corresponds to performing the trace over the entirety of the former. We must, therefore, have
\begin{align}
    1
    & \, = \int_{\mathbb{R}} \dd x \braOket{x}{\hat{\rho}[t_f]}{x} \\
    & \, = \int_{\mathbb{R}} \dd x_0^+ \int_{\mathbb{R}} \dd x_0^- \[ \int_{\mathbb{R}} \dd x_f^+ K\overline{K}[t_f,x_f^+=x_f^-;t_0,x_0^+,x_0^-] \] \nonumber\\
    & \qquad\qquad\qquad\qquad \times
    \big\la x_0^+ \big| \hat{\rho}[t_0] \big| x_0^- \big\ra \, .
\end{align}
Now since the initial density operator is arbitrary, the time evolution operator must obey
\begin{align}
\label{KKbar_ProbabilityConservation}
    \int_{\mathbb{R}} \dd x_f^+ K\overline{K}[t_f,x_f^+=x_f^-;t_0,x_0^+,x_0^-]
    & \, = \, \delta[x_0^+ - x_0^-] .
\end{align}
This condition for $K\overline{K}$ that preserves $\Tr{\hat{\rho}[t_f]} = 1$ can be contrasted with the condition $Z_{\rm diss.}[J,J] = 1$ discussed after Eq.\ \eqref{eq:ZininIF}.

%-------------------------------
\subsection{DHO: Setup}
%-------------------------------

Let us compute this time evolutionary operator $K\overline{K}$ for the quantum DHO, defined via the Lagrangians
\begin{align}
\label{DSHO_Lag0}
    L[x^\pm,\dot{x}^\pm] 
    & \, = \, \frac{1}{2} (\dot{x}^\pm)^2 - \frac{1}{2} (\omega^2 - i \alpha) (x^\pm)^2 
\end{align}
and
\begin{align}
\label{DSHO_LagI}
    & L_I[x^+,\dot{x}^+;x^-,\dot{x}^-] \nonumber \\
    & \quad = \, - \gamma \left( x^+ - x^- \right) \left( \dot{x}^+ + \dot{x}^- \right)
    + ig x^+ x^- \, .
\end{align}
This is, of course, the time-local case considered in Sec.\ \ref{sec:tldissipation}, where $\omega$, $\gamma$, and $\alpha$ here are the same as those in that section. That this is the quantum version of the classical DHO system will be explicitly verified when we obtain both the solution and equation of motion of the position operator's expectation value in Eqs.\ \eqref{DampedSHO_X_OnePoint} and \eqref{DampedSHO_X_OnePoint_EoM} below. Physically speaking, $\omega^2-i\alpha$ is the square of the oscillation frequency of the simple harmonic oscillator, where $\omega^2$ and $-\alpha$ are, respectively, its real and imaginary parts. The strength of friction is controlled by the magnitude of $\gamma$. We have included the $g$ term because it is allowed at the quadratic level; in fact, it turns out to be mandatory.

Since $\omega$ appears only as a squared quantity, we will take it to be non-negative without loss of generality. We will also see that in order to prevent a runaway solution to the quantum statistical expectation value of the position, the friction $\gamma$ also needs to non-negative. Moreover, to ensure probability conservation in Eq.\ \eqref{KKbar_ProbabilityConservation}, we will find that $g=-\alpha$. Finally, for non-zero $\gamma$ and $\omega$, $\alpha$ needs to be positive to produce a well-defined density operator in the asymptotic future. To summarize, we will find that
\begin{align}
    \omega > 0 \, ,
    \quad
    (g=-\alpha) < 0 \, ,
    \quad
    \text{and}
    \quad
    \gamma > 0 \, .
\end{align}
As before, we will remain agnostic to how the model in Eqs.\ \eqref{DSHO_Lag0} and \eqref{DSHO_LagI} arises, but simply assume that it is a consistent effective theory with time-independent parameters. For technical convenience, we now perform the change-of-variables \footnote{Note that these are proportional -- but not equal -- to the redefinition in Eq.\ \eqref{eq:xibasis} due to factors of $1/2$ there.}
\begin{align}
    \xi^\pm & \, \equiv \, x^+ \pm x^- \, ,
\end{align}
under which the total action becomes
\begin{align}
    &S_T[\xi^+,\xi^-] 
\label{DampedSHO_TotalAction}
    \, = \, \int_{t} \Bigg\{
    \frac{1}{2} \dot{\xi}^+  \dot{\xi}^- - \frac{\omega^2}{2} \xi^+  \xi^- \nonumber\\
    & \quad + \, i \frac{g + \alpha}{4} (\xi^+)^2
    - i \frac{g - \alpha}{4} (\xi^-)^2 
    - \gamma \xi^-  \dot{\xi}^+ 
\Bigg\} \, ,
\end{align}
on using Eqs.\ (\ref{DSHO_Lag0}) and (\ref{DSHO_LagI}) for $L$ and $L_I$.

%-------------------------------
\subsection{Evaluation of the DHO $K\overline{K}$}
%-------------------------------

The double path integrals occurring in the time-evolution operator $K\overline{K}$ may be evaluated in a similar manner to their single path integral counterpart $K$. First, we seek the classical solutions $\xi^\pm[t] = \xi_c^\pm[t]$ that extremize the total action. Then, we perform a change in path integration variables from $\xi^\pm$ to $\xi_q^\pm$ by expanding around the classical path, namely, $\xi^\pm = \xi_c^\pm + \xi_q^\pm$. We will find that the double path integrals over $\xi_q^\pm$ may be determined by probability conservation.

By varying Eq.\ \eqref{DampedSHO_TotalAction} with respect to $\xi^\pm$, we find that the classical solutions $\xi_c^\pm$ must obey
\begin{align}
\label{DampedSHO_EoM}
    \mathcal{D}^{\phantom{(t)} \text{I}}_{(t) \phantom{\text{I}} \text{J}} \, \xi_c^\text{J}[t]
    & \, = \, 0 \, , 
\end{align}
where the indices I and J run over $\pm$ and the $2 \times 2$ differential operator reads
\begin{align}
    \mathcal{D}_{(t)} 
    & \, \equiv \, \left[ \begin{array}{cc}
    \frac{\dd^2}{\dd t^2} + 2\gamma \frac{\dd}{\dd t} + \omega^2 & -i(\alpha - g) \\
    -i(\alpha+g) & \frac{\dd^2}{\dd t^2} - 2\gamma \frac{\dd}{\dd t} + \omega^2
    \end{array}\right] .
\end{align}
The $x^+$ portion of the $K\overline{K}$ path integrals in Eq.\ \eqref{KKbar} propagates the quantum system from $(t_0,x_0^+)$ to $(t_f,x_f^+)$, whereas the $x^-$ portion propagates the same system from $(t_0,x_0^-)$ to $(t_f,x_f^-)$. This motivates us to impose the following boundary conditions on the classical solutions, keeping in mind that $\xi^\pm \equiv x^+ \pm x^-$,
\begin{align}
\label{DampedSHO_EoM_BCs}
    \xi_c^\pm[t_f] \, = \, x_f^+ \pm x_f^-
    \quad \text{and} \quad
    \xi_c^\pm[t_0] \, = \, x_0^+ \pm x_0^- .
\end{align}
The solutions $\xi_c^\pm$ may be expressed in terms of the retarded Green's function, which we express here as a function of a single time variable $\tau$,
\begin{align}
    G^{\phantom{(R)} \text{I}}_{(\text{R}) \phantom{I} \text{J}}[\tau]
    & \, \equiv \, \theta[\tau] \mathcal{G}^{\text{I}}_{\phantom{I}\text{J}}[\tau] \, ,
\end{align}
of the matrix differential operator $\mathcal{D}^{\phantom{(t)}\text{I}}_{(\tau)\phantom{\text{I}}\text{J}}$ in Eq.\ \eqref{DampedSHO_EoM},
\begin{align}
\label{DampedSHO_EoM_G}
    \mathcal{D}^{\phantom{(\tau)} \text{I}}_{(\tau) \phantom{\text{I}} \text{J}} G^{\phantom{(R)} \text{J}}_{(\text{R}) \phantom{I}\text{M}}[\tau]
    & \, = \, \delta_{\phantom{I}\text{M}}^\text{I} \delta[\tau] \, .
\end{align}
When $\alpha=-g$, Eq.\ \eqref{DampedSHO_EoM_G} is intimately related to Eqs.\ \eqref{eq:Gpptl} and \eqref{eq:Gpmtl}. Unlike the $G^{\xi,++}$ and $G^{\xi,+-}$ there, however, we will solve for $G^{\phantom{(R)} \text{J}}_{(\text{R}) \phantom{I} \text{M}}$ with retarded boundary conditions and use it to construct the classical trajectories $\xi_c^\pm$.

The $2 \times 2$ retarded Green's function admits the integral representation
\begin{align}
    &G^{\phantom{(R)} \text{J}}_{(\text{R}) \phantom{I} \text{M}}[\tau] \nn\\
    & = \, \int_{C} \frac{\dd \omega'}{2\pi} \frac{e^{i\omega' \tau}}{(2\gamma)^2 \omega'^2-g^2+\alpha ^2+\left(\omega'^2-\omega ^2\right)^2} \nonumber\\
    & \quad \times \left[
	\begin{array}{cc}
		-\omega'^2-i 2\gamma \omega'+\omega ^2	& i (\alpha-g) \\
		i (\alpha+g)			    & -\omega'^2+i 2\gamma \omega'+\omega ^2 \\
	\end{array}
	\right]^\text{J}_{\phantom{J}\text{M}} . 
\end{align}
The integrand here is the matrix inverse of the $\mathcal{D}^{\phantom{(\tau)} \text{I}}_{(\tau) \phantom{\text{I}} \text{J}}$ in Eq.\ \eqref{DampedSHO_EoM}, but expressed in frequency $\omega'$-space. The contour $C$ skirts all four poles of the matrix integrand on the complex $\omega'$ plane from {\it below} so that when, and only when, $\tau > 0$ the result is non-zero. Upon summing over the residues, we find
\begin{align}
    2 \chi r_0 ^3 \sigma \mathcal{G}^+_{\phantom{+}+}[\tau]
    & \, = \, \sinh[r_0  \sigma \tau] \Big( \chi (r_0^2 -\omega^2 ) \cos[\chi r_0  \tau] \nonumber\\
    &\qquad - 2\gamma r_0  \sin[\chi r_0  \tau] \Big) \nonumber\\
    &+ \sigma \left(r_0 ^2+\omega
	^2\right) \sin[\chi r_0  \tau] \cosh[r_0  \sigma \tau] \, , \\
    2 \chi r_0 ^3 \sigma \mathcal{G}^+_{\phantom{+}-}[\tau]
    & \, = \, -i (\alpha-g) \Big( \chi \cos[\chi r_0  \tau] \sinh[r_0  \sigma \tau] \nonumber\\
    &\qquad
    - \sigma \sin[\chi r_0  \tau] \cosh[r_0  \sigma \tau] \Big) \, , \\
    2 \chi r_0 ^3 \sigma \mathcal{G}^-_{\phantom{-}+}[\tau]
    & \, = \, -i (\alpha+g) \Big( \chi \cos[\chi r_0  \tau] \sinh[r_0  \sigma \tau] \nonumber\\
    &\qquad - \sigma \sin[\chi r_0  \tau] \cosh[r_0  \sigma \tau] \Big) \, , \\
    2 \chi r_0 ^3 \sigma \mathcal{G}^-_{\phantom{-}-}[\tau]
    & \, = \, \sinh[r_0 \sigma \tau] \Big( 2\gamma r_0  \sin[\chi r_0  \tau] \nonumber\\
    &\qquad
    + \chi (r_0^2 -\omega^2) \cos[\chi r_0  \tau] \Big) \nonumber\\
    &+ \sigma \left(r_0 ^2+\omega ^2\right) \sin[\chi r_0  \tau] \cosh[r_0  \sigma \tau] \, ,
\end{align}
with the relations
\begin{align}
    r_0 & = \omega \sqrt[4]{1+(\alpha^2-g^2)/\omega^4} \, , \\
    \binom{r_0 \chi}{r_0 \sigma}
	& = \omega \sqrt{\frac{1}{2} \left( \sqrt{1+(\alpha^2-g^2)/\omega^4} \pm \left(1-\frac{(2\gamma)^2}{2 \omega^2}\right) \right)} \, .
\end{align}
A direct calculation further reveals that
\begin{align}
\label{DampedSHO_G_P1}
    \mathcal{D}^{\phantom{(t)} \text{I}}_{(\tau) \phantom{\text{I}} \text{J}} \mathcal{G}^{\text{J}}_{\phantom{J} \text{M}}[\tau]
    & = 0 \, , \\
\label{DampedSHO_G_P2}
    \mathcal{G}^{\text{J}}_{\phantom{J} \text{M}}[0] 
    & = 0 \, , \\
\label{DampedSHO_G_P3}
    \partial_\tau \mathcal{G}^{\text{J}}_{\phantom{J}\text{M}}[\tau = 0] 
    & = \delta^{\text{J}}_{\phantom{J} \text{M}} \, .
\end{align}
In terms of $\mathcal{G}$, the homogeneous solution portion of 
%the retarded $G^{\phantom{(R)}\text{J}}_{(\text{R})\phantom{I}\text{M}}$,
the classical trajectories is
\begin{align}
\label{DampedSHO_Classical_zpm}
    & \xi_c^\text{J}[t_0 \leq t \leq t_f] \nonumber\\
    & \ \ = \, \mathcal{G}^\text{J}_{\phantom{J}\text{K}}[t-t_0] (\mathcal{G}^{-1})^\text{K}_{\phantom{\text{K}}\text{M}}[t_f-t_0] \left[ \begin{array}{c}
    x_f^+ + x_f^- \\
    x_f^+ - x_f^-
    \end{array} \right]^\text{M} \nonumber\\
    & \quad \ \ \ + \Big( \mathcal{G}^\text{J}_{\phantom{\text{J}}\text{K}}[t-t_0] (\mathcal{G}^{-1})^\text{K}_{\phantom{\text{K}}\text{L}}[t_f-t_0] \partial_{t_0} \mathcal{G}^\text{L}_{\phantom{\text{L}}\text{M}}[t_f-t_0] \nonumber\\
    & \qquad\qquad- \partial_{t_0} \mathcal{G}^\text{J}_{\phantom{\text{J}}\text{M}}[t-t_0] \Big) \left[ \begin{array}{c}
    x_0^+ + x_0^- \\
    x_0^+ - x_0^-
    \end{array} \right]^\text{M} .
\end{align}
That Eq.\ \eqref{DampedSHO_Classical_zpm} solves its equation of motion in \eqref{DampedSHO_EoM} is because of Eq.\ \eqref{DampedSHO_G_P1}, and that it obeys the appropriate boundary conditions in Eq.\ \eqref{DampedSHO_EoM_BCs} is because of Eqs.\ \eqref{DampedSHO_G_P2} and \eqref{DampedSHO_G_P3}. We will also need the first derivatives of these trajectories at the boundary times. Invoking Eq.\ \eqref{DampedSHO_G_P3} hands us
{\allowdisplaybreaks
\begin{align}
    & \dot{\xi}_c^\text{J}[t_0]
    \, = \, (\mathcal{G}^{-1})^\text{J}_{\phantom{\text{J}}\text{M}}[t_f-t_0] \left[ \begin{array}{c}
	x_f^+ + x_f^- \nn \\
	x_f^+ - x_f^-
    \end{array} \right]^\text{M} \\
    & \quad + \Big( (\mathcal{G}^{-1})^\text{J}_{\phantom{\text{J}}\text{L}}[t_f-t_0] \partial_{t_0} \mathcal{G}^\text{L}_{\phantom{\text{L}}\text{M}}[t_f-t_0] \nonumber\\
    & \qquad \quad
    - \partial_t \partial_{t_0} \mathcal{G}^\text{J}_{\phantom{\text{J}}\text{M}}[t=t_0] \Big) \left[ \begin{array}{c}
	x_0^+ + x_0^- \\
	x_0^+ - x_0^-
    \end{array} \right]^\text{M} ,
\label{DampedSHO_Classical_zpmdot_tp}
\end{align}}%
while $\dot{\xi}_c^\text{J}[t_f]$ is simply the $t$-derivative of Eq.\ \eqref{DampedSHO_Classical_zpm} with $t$ replaced with $t_f$.

Note that even though the total action involves the time integral over $t \in [t_0,t_f]$, it may be converted into a difference of boundary terms when it is evaluated on the classical solutions $\xi_c^\pm$. For, upon integration-by-parts, the action in Eq.\ \eqref{DampedSHO_TotalAction} is transformed into
\begin{align}
    & S_T[\xi^\pm = \xi_c^\pm] \nonumber\\
    & \quad = \, \frac{1}{4} \left[ \xi_c^+  \dot{\xi}_c^- + \dot{\xi}_c^+ \xi_c^- - 2\gamma \xi_c^- \xi_c^+ \right]_{t=t_0}^{t=t_f} \nonumber\\
    & \qquad \quad - \, \frac{1}{4} \int_t \left( \xi_c^+ \mathcal{D}^{\phantom{(t)}-}_{(t)\phantom{-}\text{J}} \xi_c^\text{J}
    + \xi_c^- \mathcal{D}^{\phantom{(t)}+}_{(t)\phantom{+}\text{J}} \xi_c^\text{J} \right) \\
\label{DampedSHO_TotalAction_Classical}
    & \quad = \, \frac{1}{4} \Big( (x_f^+ + x_f^-)  \dot{\xi}_c^-[t_f] 
    + \dot{\xi}_c^+[t_f]  (x_f^+ - x_f^-) \nonumber\\
    & \qquad \qquad \quad
    - 2\gamma \, \big\{ (x_f^+)^2-(x_f^-)^2 \big\} \Big) \nonumber\\
    & \qquad \quad
    - \frac{1}{4} \Big( (x_0^+ + x_0^-)  \dot{\xi}_c^-[t_0] + \dot{\xi}_c^+[t_0]  (x_0^+ - x_0^-) \nonumber\\
    & \qquad \qquad \quad
    - 2\gamma \left\{ (x_0^+)^2 - (x_0^-)^2 \right\} \Big) \, ,
\end{align}
where the integral terms vanish because of Eq.\ \eqref{DampedSHO_EoM} and we used the boundary conditions in Eq.\ \eqref{DampedSHO_EoM_BCs}. 

We now explicitly shift the path integration variables using
\begin{align}
    \xi^\pm \, \equiv \, \xi_c^\pm + \xi_q^\pm \, .
\end{align}
Since the boundary conditions for the path integration variables, $\xi^\pm[t_f] = x_f^+ \pm x_f^-$ and $\xi^\pm[t_0] = x_0^+ \pm x_0^-$, are already accounted for by the $\xi_c^\pm$ in Eq.\ \eqref{DampedSHO_EoM_BCs}, we must impose
\begin{align}
\label{DampedSHO_Quantum_BCs}
    \xi_q^\pm[t_f] \, = \, 0 \, = \, \xi_q^\pm[t_0] \, ,
\end{align}
which in turn implies that
{\allowdisplaybreaks
\begin{align}
\label{DampedSHO_KKbar_Limits}
    \int_{x_0^+}^{x_f^+} \mathcal{D} x^+
    \int_{x_0^-}^{x_f^-} \mathcal{D} x^-
    & \to \int_{x_0^+ + x_0^-}^{x_f^+ + x_f^-} \mathcal{D} \xi^+ \int_{x_0^+ - x_0^-}^{x_f^+ - x_f^-} \mathcal{D} \xi^- \nonumber\\
    & \to \int_{0}^{0} \mathcal{D} \xi_q^+ \int_{0}^{0} \mathcal{D} \xi_q^- \, .
\end{align}}%
Eq.\ \eqref{DampedSHO_TotalAction} now reads
\begin{align}
\label{DampedSHO_TotalAction_v2}
    S_T[\xi^\pm = \xi_c^\pm+\xi_q^\pm]
    & \, = \, S_T[\xi_c^\pm] + S_T[\xi_q^\pm] \, ,
\end{align}
where $S_T[\xi_c^\pm]$ is the total action evaluated solely on $\xi_c^\pm$ and given by Eq.\ \eqref{DampedSHO_TotalAction_Classical} while $S_T[\xi_q^\pm]$ is evaluated solely on $\xi_q^\pm$ but with the limits implied by Eq.\ \eqref{DampedSHO_Quantum_BCs}. There are no cross terms between $\xi_c^\pm$ and $\xi_q^\pm$ as the action is quadratic and the cross terms are, therefore, themselves necessarily linear in $\xi_q$. Since the action itself is extremized by the solution to Eq.\ \eqref{DampedSHO_EoM}, these linear-in-$\xi_q$ terms must vanish upon integrating-by-parts all the derivatives acting on $\xi_q^\pm$ and employing Eq.\ \eqref{DampedSHO_Quantum_BCs} to set to zero the associated boundary terms. 

With the new form of the total action in Eq.\ \eqref{DampedSHO_TotalAction_v2} and keeping in mind the integration limits in Eq.\ \eqref{DampedSHO_KKbar_Limits}, we can now write the time evolution operator in Eq.\ \eqref{KKbar} as
\begin{align}
\label{DampedSHO_KKbar}
    & K\overline{K}[t_f,x_f^+,x_f^-; t_0,x_0^+,x_0^-] \nonumber\\
    & \quad = \, K\overline{K}[t_f,0,0; t_0,0,0] \exp\left[ i S_T[\xi_c^\pm] \right] .
\end{align}
Let us now consider again the trace of $\hat{\rho}[t_f]$, which involves setting $x_f^+ = x_f^- \equiv x$ in Eq.\ \eqref{DampedSHO_KKbar}, followed by integrating over all real $x$. This procedure will, however, yield an $S_T[\xi_c^\pm]$ that is quadratic in $x$, because of the $(x_f^+ + x_f^-) \dot{\xi}_c^-[t_f]$ term in Eq.\ \eqref{DampedSHO_TotalAction_Classical}. More explicitly, a direct calculation tells us that these quadratic-in-$x$ terms in $S_T[\xi_c^\pm]$ are 
\begin{widetext}
\begin{align}
\label{DampedSHO_TotalAction_X2}
    x^2 (g+\alpha)
    & r_0 ^2 \Big(2\gamma \left(1-\sigma^2 \cos[2 r_0 \chi (t_f-t_0)]\right)-\chi (2\gamma \chi \cosh[2 r_0  \sigma (t_f-t_0)] \nonumber\\
    &\qquad\qquad
    +2 r_0  \sigma (\chi \sinh[2 r_0  \sigma (t_f-t_0)] - \sigma \sin[2 r_0 \chi 
    (t_f-t_0)]))\Big) \nonumber\\
    & \times\Big\{ 2 \sinh^2[r_0  \sigma (t_f-t_0)] \left(\chi^2 \cos ^2[r_0 \chi (t_f-t_0)] \left(-\alpha ^2+g^2-\left(r_0 ^2-\omega ^2\right)^2\right)+(2\gamma)^2 r_0 ^2 \sin ^2[r_0 \chi (t_f-t_0)]\right) \nn \\
    & \qquad +2 \sigma^2 \sin ^2[r_0 \chi (t_f-t_0)] \left(-\alpha ^2+g^2-\left(r_0 ^2+\omega ^2\right)^2\right) \cosh ^2[r_0  \sigma (t_f-t_0)] \nonumber\\
    & \qquad\qquad +\chi \sigma \sin[2 \chi r_0  (t_f-t_0)] \left(\alpha
	^2-g^2-r_0 ^4+\omega ^4\right) \sinh[2 r_0  \sigma (t_f-t_0)] \Big\}^{-1} .
\end{align}
\end{widetext}
If we were to substitute the above into the integral on the left hand side of the probability conservation statement in Eq.\ \eqref{KKbar_ProbabilityConservation}, we would end up with a Gaussian integral and would {\it not} obtain the required $\delta$-function on the right hand side. We must, therefore, eliminate these $x^2$ terms completely, which can be done by setting
\begin{align}
\label{DampedSHO_gEqualsalpha}
    g \, = \, -\alpha \, ,
\end{align}
as indicated by the overall factor of $(g+\alpha)$ in Eq.\ \eqref{DampedSHO_TotalAction_X2}. This, in fact, recovers the time-local version of the $\gamma_{1I}(t,t') = \gamma_{2I}(t,t')$ condition that we deduced from Eq.\ \eqref{GammsAreEqual}. Notice also that we had to perform an explicit calculation of the density operator's time evolution operator $K\overline{K}$ in this section to obtain the condition in Eq.\ \eqref{DampedSHO_gEqualsalpha}, whereas it arose directly from the Green's function equation in Sec.\ \ref{sec:greensfunctions}.

With the condition in Eq.\ \eqref{DampedSHO_gEqualsalpha} and the definitions
\begin{align}
\label{DampedSHO_Gamma_Def}
    \Gamma \, \equiv \, \sqrt{\omega^2-\gamma^2} 
    \quad \text{and} \quad
    T \, \equiv \, t_f-t_0
\end{align}
-- the $T$ here not to be confused with the $T \equiv t-t'$ of Sec.\ \ref{sec:tldissipation} -- we may now record the following. The homogeneous solution portion of the $2 \times 2$ Green's function and its inverse, written as a function of a single time variable $\tau$, are
\begin{widetext}
{\allowdisplaybreaks
\begin{align}
\label{RetardedG_Homogeneous}
    \mathcal{G}^{\text{J}}_{\phantom{J} \text{K}}[\tau]
    & \, = \, \left[
    \begin{array}{cc}
        \mathcal{G}_\gamma[\tau] & \frac{i \alpha}{\omega^2} \left( \Gamma^{-1}  \cosh\left[\gamma \tau\right] \sin\left[ \Gamma  \tau \right] - \frac{1}{\gamma} \cos \left[\Gamma  \tau \right] \sinh \left[\gamma \tau\right] \right) \\
        0 & \mathcal{G}_{-\gamma}[\tau]
    \end{array}
    \right] , \\
\label{RetardedG_Homogeneous_Inverse}
    (\mathcal{G}^{-1})^\text{J}_{\phantom{J} \text{K}}[\tau] 
    & \, = \, \left[
    \begin{array}{cc}
        \mathcal{G}_\gamma[\tau]^{-1}
        & \frac{-i \alpha \Gamma^2}{\omega^2 \sin^2\left[\Gamma  \tau\right]} \left( \Gamma^{-1} \cosh\left[\gamma \tau\right] \sin\left[ \Gamma  \tau \right] - \frac{1}{\gamma} \cos \left[\Gamma  \tau\right] \sinh \left[\gamma \tau\right] \right) \\
        0 & \mathcal{G}_{-\gamma}[\tau]^{-1}
    \end{array}
    \right] ,
\end{align}}
\end{widetext}
where we have identified the Green's function of the classical DHO as
\begin{align}
\label{DSHO_GreensFunction_IofII}
    G_\gamma[\tau] 
    & \, \equiv \, \theta[\tau] \mathcal{G}_\gamma[\tau] \, , \\
\label{DSHO_GreensFunction_IIofII}
    \mathcal{G}_\gamma[\tau] 
    & \, \equiv \, e^{-\gamma \tau} \Gamma^{-1} \sin[\Gamma \tau] \, ,
\end{align}
and its anti-damped counterpart $G_{-\gamma}[\tau]$ and $\mathcal{G}_{-\gamma}[\tau]$ given by the same expressions, except with $\gamma$ replaced by $-\gamma$. The DHO Green's function obeys
{\allowdisplaybreaks
\begin{align}
    \left(\frac{\dd^2}{\dd \tau^2} + 2\gamma \frac{\dd}{\dd \tau} + \omega^2 \right) G_\gamma[\tau] 
    & \, = \, \delta[\tau] \, , \\
    \left(\frac{\dd^2}{\dd \tau^2} + 2\gamma \frac{\dd}{\dd \tau} + \omega^2 \right)\mathcal{G}_\gamma[\tau] 
    & \, = \, 0 \, ,
\end{align}}%
and the anti-damped one again obeys the same equations, except with $\gamma$ replaced by $-\gamma$. Let us also observe that $\mathcal{G}^+_{\phantom{+}+}$ in Eq.\ \eqref{RetardedG_Homogeneous} is, up to a factor of $-i$, the same as $G^{\xi,+-}$ in Eq.\ \eqref{eq:Gret}.

The action evaluated on the classical trajectory now reads
\begin{widetext}
{\allowdisplaybreaks\begin{align}
\label{DampedSHO_TotalAction_Classical_alphaOnly}
    i S_T[\xi_c^\pm]
    & \, = \, \frac{\alpha  \csc ^2[\Gamma T]}{32\gamma \omega^2} \left( (x_f^+-x_f^-)^2-(x_0^+-x_0^-)^2\right)  \left((2\gamma)^2 \cos [2 \Gamma  T]-4 \omega ^2\right) \nonumber\\
    & \qquad - \frac{\alpha  \Gamma ^2 e^{-2\gamma T}}{8\gamma \omega ^2 \sin^2[\Gamma  T]} \left(e^{4\gamma T} (x_0^+-x_0^-)^2-(x_f^+-x_f^-)^2\right) 
    +\frac{\alpha  \Gamma ^2  \cos[\Gamma  T]}{2\gamma \omega ^2  \sin^2[\Gamma  T]} \sinh[\gamma T] (x_f^+-x_f^-)  (x_0^+-x_0^-) \nonumber\\
    & \qquad +\frac{\alpha \Gamma}{4 \omega ^2 \sin[\Gamma T]} \left[ \cos[\Gamma  T]
    \left((x_f^+-x_f^-)^2+(x_0^+-x_0^-)^2\right)-2 (x_f^+-x_f^-)  (x_0^+-x_0^-) \cosh \left[ \gamma T \right] \right] \nonumber\\
    & \qquad + \frac{i}{2} \frac{\Gamma}{\sin[\Gamma  T]} \bigg[ \sinh \left[ \gamma T \right] 2( x_f^+  x_0^- - x_f^-  x_0^+)+\cosh \left[\gamma T\right] 2(
    x_f^-  x_0^- - x_f^+  x_0^+) \nonumber\\
    & \qquad\qquad\qquad\qquad
    + \cos[\Gamma  T] \big\{ (x_f^+)^2-(x_f^-)^2 + (x_0^+)^2 - (x_0^-)^2 \big\} \bigg] \nonumber\\
    & \qquad - \frac{i \gamma}{2} \left[ \big\{ (x_f^+)^2 - (x_f^-)^2 \big\} - \big\{ (x_0^+)^2 - (x_0^-)^2 \big\} \right] \, ,
\end{align}}
\end{widetext}
and setting $x_f^+ = x_f^- \equiv x$ in Eq.\ \eqref{DampedSHO_TotalAction_Classical_alphaOnly} gives
{\allowdisplaybreaks
\begin{align}
\label{DampedSHO_TotalAction_Classical_alphaOnly_x=xp}
    & i S_T[\xi_c^\pm; x_f^+ = x_f^- \equiv x] \nonumber\\
    & \quad = \, - \, i \frac{x  \left(x_0^+ - x_0^-\right)}{\mathcal{G}_\gamma[T]} \nonumber\\
    & \quad \qquad + \frac{i}{2} \big\{ (x_0^+)^2 - (x_0^-)^2 \big\} \left( \gamma + \Gamma \cot\left[ \Gamma  T \right] \right) \nn \\
    & \quad \qquad -\frac{\alpha  (x_0^+ - x_0^-)^2}{8\gamma \omega^2\sin^2\left[ \Gamma T \right]} \Big( \Gamma^2  e^{2\gamma  T} - \omega^2 
	+ \gamma^2 \cos\left[ 2 \Gamma  T \right] \nonumber\\
    & \quad \qquad \qquad
    - \gamma \Gamma \sin\left[ 2 \Gamma T \right] \Big) \, .
\end{align}}%
The probability conservation of Eq.\ \eqref{KKbar_ProbabilityConservation} applied to the time evolution operator in Eq.\ \eqref{DampedSHO_KKbar} further yields
\begin{align}
    K\overline{K}[t_f,0,0; t_0,0,0] &\int_{\mathbb{R}} \dd x \exp\left[ i S_T[\xi_c^\pm;x=x' \equiv x] \right] \nonumber\\
    & = \, \delta[x_0^+ - x_0^-] \, .
\end{align}
Employing Eq.\ \eqref{DampedSHO_TotalAction_Classical_alphaOnly_x=xp} thus hands us the relation
\begin{align}
    K\overline{K}[t_f,0,0; t_0,0,0] 
    & \, = \, |2\pi\mathcal{G}_\gamma[T]|^{-1} \, .
\end{align}
As alluded to earlier, probability conservation allows us to evaluate the remaining double path integrals involving $\xi_q^\pm$, namely, $K\overline{K}[t_f,0,0; t_0,0,0]$. We may, at this point, conclude that the time evolution operator for the DHO system in Eq.\ \eqref{DampedSHO_KKbar} with $g=-\alpha$ is given by
\begin{align}
\label{DampedSHO_KKbar_alphaOnly}
    & K\overline{K}[t_f,x_f^+,x_f^-; t_0,x_0^+,x_0^-] \nonumber\\
    & \qquad = \, |2\pi\mathcal{G}_\gamma[T]|^{-1} \exp\left[ i S_T[\xi_c^\pm] \right] ,
\end{align}
with the classical action specified in Eq.\ \eqref{DampedSHO_TotalAction_Classical_alphaOnly}. Furthermore, armed with this $K\overline{K}$ for the DHO, we may immediately integrate it against a {\it Gaussian} initial density matrix $\braOket{x^+}{\hat{\rho}(t_0)}{x^-} = N {\rm exp} \left\{ iS_0[x^+, x^-] \right\}$, with $S_0$ given in Eq.\ \eqref{eq:initialS} -- i.e., compute Eq.\ \eqref{ReducedDensityOperator_TimeEvolution} -- to discover that the final density matrix takes the same form as the initial one, except that all relevant `initial' parameters are replaced with the time-dependent final ones. For instance, the initial position and momentum are replaced as
\begin{align}
    x_0 & \to \big\la \hat{X}[t_f] \big\ra \, , \\
    p_0 & \to \frac{\dd}{\dd t_f} \big\la \hat{X}[t_f] \big\ra \, = \, \big\la \hat{P}[t_f] \big\ra \, ,
\end{align}
and the parameters $A$ and $B$, related to the initial two-point correlations $c_{xx}$, $c_{xp}$, and $c_{pp}$ by Eqs.\ (\ref{eq:phiphicorr}), (\ref{eq:phimomcorr}), and (\ref{eq:mommomcorr}), are similarly replaced by their counterparts at $t_f$.

%-------------------------------
\subsection{DHO `equal-time' $n$-point functions}
\label{subsec:DSHOnpoint}
%-------------------------------

We now turn to calculating the quantum statistical average of a product of $n \geq 1$ position operators at time $t_f$. An application of Eqs.\ \eqref{ReducedDensityOperator_TimeEvolution} and \eqref{DampedSHO_KKbar_alphaOnly} gives us
\begin{align}
\label{NPointFunction_Def}
    & \big\la \hat{X}^{n}[t_f] \big\ra 
    \, \equiv \, \int_{\mathbb{R}} \dd x \big\la x \big| \hat{X}^n \hat{\rho}[t_f] \big| x \big\ra \nonumber\\
    & \ \ = \, \int_{\mathbb{R}} \frac{\dd x}{|2\pi \mathcal{G}_\gamma[T]|} x^n \int_{\mathbb{R}} \dd x_0^+ \int_{\mathbb{R}} \dd x_0^- \nonumber\\
    & \qquad \times \exp \Big[ i S_T[\xi_c^\pm;x_f^+=x_f^- \equiv x] \Big] \big\la x_0^+ \big| \hat{\rho}[t_0] \big| x_0^- \big\ra \, . 
\end{align}
To proceed, we consider the generating functional
\begin{align}
\label{NPointFunction_Z}
    & Z[J] 
    \, \equiv \, \int_{\mathbb{R}} \frac{\dd x}{|2\pi \mathcal{G}_\gamma[T]|} e^{iJ  x} \int_{\mathbb{R}} \dd x_0^+ \int_{\mathbb{R}} \dd x_0^- \nonumber\\
    & \quad \times \exp\left[ i S_T[\xi_c^\pm;x_f^+ = x_f^- \equiv x] \right] \braOket{x_0^+}{\hat{\rho}[t_0]}{x_0^-} .
\end{align}
The time-independent $J$ here is similar in spirit to, but not to be confused with, the time-dependent $J^\pm$ introduced in the previous sections. From Eq.\ \eqref{NPointFunction_Z}, the $n$-point function in Eq.\ \eqref{NPointFunction_Def} may be extracted via
\begin{align}
\label{NPointFunction_FromZ}
    \big\la \hat{X}^n[t_f] \big\ra
    & \, = \, \left. \frac{1}{i^n} \frac{\partial^n Z[J]}{\partial J^n} \right\vert_{J=0} ,
\end{align}
or, equivalently, by reading off the coefficient in the Taylor series of $Z[J]$ containing exactly $n$ powers of $J$. Now, employing Eq.\ \eqref{DampedSHO_TotalAction_Classical_alphaOnly_x=xp} in Eq.\ \eqref{NPointFunction_Z}, we would discover that $x_0^+ - x_0^- = \mathcal{G}_\gamma[T] J$ is enforced by the $\delta$-functions obtained by integrating over $x$. The generating functional is then found to be
\begin{align}
\label{DampedSHO_Z}
    Z[J] 
    & \, = \, \exp\Bigg[ \frac{\alpha  (i J)^2}{8 \gamma \omega^2} \Bigg( 1 + \frac{e^{-2\gamma T}}{\Gamma^2} \Big\{ \gamma^2 \cos\left[ 2 \Gamma  T \right] \nn\\
    & \qquad \quad
    - \gamma  \Gamma \sin\left[ 2 \Gamma T \right] - \omega^2 \Big\} \Bigg) \Bigg] \nn\\
    & \, \times \int_{\mathbb{R}} \dd x \left.\left\langle x - \frac{i}{2} \mathcal{G}_\gamma[T]  (i J) \right\vert \hat{\rho}[t_0] \left\vert x + \frac{i}{2} \mathcal{G}_\gamma[T]  (i J) \right\rangle\right. \nn\\
    &\qquad \quad \times \exp\big[ x \mathcal{G}_\gamma[T] (i J) \left( \gamma + \Gamma \cot\left[ \Gamma  T \right] \right) \big] \, . 
\end{align}
We note, parenthetically, that the $x$ integral in Eq.\ \eqref{DampedSHO_Z} bears a resemblance to the Wigner function in quantum mechanics. Moreover, we have written $iJ$ as a group in the above expressions since $n$-point correlations are generated by differentiating $Z[J]$ with respect to it.

Applying Eq.\ \eqref{NPointFunction_FromZ} to Eq.\ \eqref{DampedSHO_Z}, we first find the one-point function to not only depend on the initial position and momentum one-point functions,
\begin{align}
\label{DampedSHO_X_OnePoint}
    \big\la \hat{X}[t_f] \big\ra
    & = e^{-\gamma T} \left( \cos\left[ \Gamma  T \right] + \frac{\gamma}{\Gamma} \sin[\Gamma  T] \right) \big\la \hat{X}[t_0] \big\ra \nn \\
    & \qquad + \, \frac{e^{-\gamma T}}{\Gamma} \sin[\Gamma  T] \big\la \hat{P}[t_0] \big\ra \, ,
\end{align}
which is, of course, Eq.\ \eqref{eq:Xtlsol} with $t$ replaced with $t_f$, but also to obey the {\it classical} DHO equation,
\begin{align}
\label{DampedSHO_X_OnePoint_EoM}
    \left( \frac{\dd^2}{\dd t^2} + 2\gamma \frac{\dd}{\dd t} + \omega^2 \right) \big\la \hat{X}[t=t_f] \big\ra \, = \, 0 \, ,
\end{align}
since the pair $e^{-\gamma T} \sin[\Gamma  T]$ and $e^{-\gamma T} \cos[\Gamma  T]$ do. Eq.\ \eqref{DampedSHO_X_OnePoint} is thus the quantum analog of the initial value formulation of the classical DHO differential equation. Notice that it does not depend on the additional parameter $\alpha$, the imaginary portion of the DHO's frequency-squared, occurring in our dynamics defined in Eqs.\ \eqref{DSHO_Lag0} and \eqref{DSHO_LagI} (with $g=-\alpha$). This suggests that $\alpha$ is tied to the underlying quantum and/or statistical aspect(s) of the DHO system at hand.

As in Sec.\ \ref{sec:tldissipation}, we choose $\gamma \geq 0$ in order for the $e^{-\gamma T}$ factors in Eq.\ \eqref{DampedSHO_X_OnePoint} to not describe runaway trajectories. Also, $\big\la \hat{X}[t_0] \big\ra$ is readily recovered by setting $t_f=t_0$ in Eq.\ \eqref{DampedSHO_X_OnePoint}, while taking one time derivative before setting $t_f=t_0$ gives us $\big\la \hat{P}[t_0] \big\ra$,
\begin{align}
    \big\la \hat{P}[t_0] \big\ra
    & \, = \, \frac{\dd}{\dd t} \big\la \hat{X}[t] \big\ra \Big|_{t=t_0} \, .
\end{align}
In fact, a direct calculation shows that the general momentum one-point function,
{\allowdisplaybreaks
\begin{align}
    & \big\la \hat{P}[t_f] \big\ra
    \, \equiv \, \int_{\mathbb{R}} \dd x \left( -i\partial_{x} \langle x \vert \right) \hat{\rho}[t_f] \vert x \rangle \\
    & \quad = \, \int_{\mathbb{R}^{2}} \dd x_0^+ \dd x_0^- \braOket{x_0^+}{\hat{\rho}[t_0]}{x_0^-} \nn \\
    & \qquad \times \int_{\mathbb{R}^{2}} \frac{\dd x_f^+ \dd x_f^-}{|2\pi \mathcal{G}_\gamma[T]|} \left( -i \partial_{x_f^+} e^{i S_T[\xi_c^\pm]} \right) \delta[x_f^+ - x_f^-] \, ,
\end{align}}%
is exactly the time derivative of the position one-point function in Eq.\ \eqref{DampedSHO_X_OnePoint}, i.e.,
\begin{align}
    \big\la \hat{P}[t_f] \big\ra \, = \, \frac{\dd}{\dd t_f} \big\la \hat{X}[t_f] \big\ra ,
\end{align}
again suggesting that $\hat{P}(t) = \dot{\hat{X}}(t)$ for the problem we consider.

Turning next to the two-point function, application of Eq.\ \eqref{NPointFunction_FromZ} returns
\begin{widetext}
\begin{align}
\label{DampedSHO_X_TwoPoint}
    & \big\la \hat{X}^2[t_f] \big\ra
    \, = \, \frac{\alpha}{4\gamma \omega^2}
    \left( 1
    + \frac{e^{-2\gamma T}}{\Gamma^2} \left\{  \gamma^2 \cos\left[ 2 \Gamma  T \right] - \gamma \Gamma \sin\left[ 2 \Gamma T \right] - \omega^2 \right\} \right) \nn \\
    & \qquad + \, \mathcal{G}_\gamma[T]^2 \[ (\gamma + \Gamma \cot[\Gamma T])^2 \big\la \hat{X}^2[t_0] \big\ra + \frac{1}{2} (\gamma + \Gamma \cot[\Gamma T]) \big\la \big\{ \hat{X}[t_0], \hat{P}[t_0] \big\} \big\ra + \big\la \hat{P}^2[t_0] \big\ra \] .
\end{align}
\end{widetext}
This recovers Eq.\ \eqref{eq:phiphicorrt} with $G^{\xi,++}(t,t)$ there given by the sum of Eqs.\ (\ref{eq:Gpptlhsol}) and (\ref{eq:Gpptlnsol}) with $t'=t$, except that the initial two-point correlations here are {\it arbitrary}. We also note that the $\alpha$ term is independent of the initial conditions -- a feature already present in Eq.\ (\ref{eq:Gpptlnsol}) -- whereas the remaining terms depend linearly on the initial two-point functions involving both position and momentum operators. In fact, in the asymptotic future, $T \to \infty$, the $\exp(-2\gamma T)$ in $\mathcal{G}_\gamma[T]^2$ -- recall Eq.\ \eqref{DSHO_GreensFunction_IIofII} -- implies that the entire dependence on the initial two-point functions is damped out and all that remains is the $\alpha$-term,
\begin{align}
\label{DampedSHO_X_TwoPoint_InfiniteFuture}
    & \big\la \hat{X}^2[\infty] \big\ra
    \, = \, \frac{1}{2 \omega} \left(\frac{\alpha}{2\gamma  \omega}\right) .
\end{align} 
It is worth reiterating that although our dynamics as encoded within Eq.\ \eqref{DampedSHO_TotalAction} are purely quadratic -- i.e., with linear equations of motion -- we have not made any assumptions regarding the initial density operator $\hat{\rho}[t_0]$ in Eq.\ \eqref{NPointFunction_FromZ} and, therefore, regarding the initial $n$-point functions in Eqs.\ \eqref{DampedSHO_X_OnePoint} and \eqref{DampedSHO_X_TwoPoint}.

%-------------------------------
\subsection{Purity of the DHO system and the $\alpha \to 0$ limit}
%-------------------------------

We now turn to computing the purity of the DHO system without assumptions on the initial state. Using the time evolution in Eq.\ \eqref{ReducedDensityOperator_TimeEvolution} and the form of $K\overline{K}$ for the DHO in Eq.\ \eqref{DampedSHO_KKbar_alphaOnly}, the purity at time $t_f$ is given by
\begin{align}
    \text{Pu}[t_f] 
    & \, \equiv \, \Tr{\hat{\rho}^2[t_f]} \nonumber\\
\label{Purity_Def}
    & \, = \, \int_{\mathbb{R}}\dd x_0^+ \int_{\mathbb{R}} \dd x_0^- \int_{\mathbb{R}} \dd y_0^+ \int_{\mathbb{R}} \dd y_0^- \nonumber\\
    & \qquad \quad
    \times K\overline{K}K\overline{K}[t_f;x_0^+,x_0^-,y_0^+,y_0^-] \nonumber \\
    & \qquad \quad
    \times \big\la x_0^+ \big| \hat{\rho}[t_0] \big| x_0^- \big\ra \big\la y_0^+ \big| \hat{\rho}[t_0] \big| y_0^- \big\ra \, , 
\end{align}
where the time evolution operator associated with purity is
\begin{align}
    & K\overline{K}K\overline{K}[t_f;x_0^+,x_0^-,y_0^+,y_0^-] \nonumber\\
    & \quad \equiv \, \int_{\mathbb{R}^2} \frac{\dd x_f^+ \dd x_f^-}{|2\pi \mathcal{G}_\gamma[T]|^2} \exp\Big[ i S_T[\xi_c^\pm;x_f^+,x_f^-,x_0^+,x_0^-] \nonumber\\
    & \qquad\qquad\qquad
    + \, i S_T[\xi_c^\pm;x_f^-,x_f^+,y_0^+,y_0^-] \Big] \, .
\end{align}
Let us now take the $\alpha \to 0$ limit of $\text{Pu}[t_f]$'s time evolution operator $K\overline{K}K\overline{K}$. Upon switching to the $\xi^\pm \equiv x^+ \pm x^-$ variables, the exponent is found to be linear in both $\xi^\pm$,
{\allowdisplaybreaks
\begin{align}
    & K\overline{K}K\overline{K}[t_f;x_0^+,x_0^-,y_0^+,y_0^-] \nonumber\\
    & \quad = \, \exp\bigg[ \frac{i}{2} \left( \gamma + \Gamma  \cot[\Gamma  T] \right) \nonumber\\ 
    & \qquad\qquad\qquad
    \times \left\{ (x_0^+)^2 - (x_0^-)^2 + (y_0^+)^2 - (y_0^-)^2 \right\} \bigg] \nonumber\\
    & \qquad \, \times
    \frac{1}{2} \int_{\mathbb{R}^2} \frac{\dd \xi^+ \dd \xi^-}{|2\pi \mathcal{G}_\gamma[T]|^2}
    \exp \bigg[ - \frac{i}{2} \frac{\Gamma}{\sin[\Gamma  T]} \nn \\
    & \qquad\qquad
    \times \Big\{ e^{-\gamma T} \xi^-  (x_0^+ + x_0^- - y_0^+ - y_0^-) \nonumber\\
    & \qquad\qquad\qquad
    + e^{\gamma T} \xi^+  (x_0^+ - x_0^- + y_0^+ - y_0^-) \Big\} \bigg] \, .
\label{DSHO_Purity_KKKK_alphaZero_Step1}
\end{align}}%
The integral representation of the Dirac $\delta$-function together with the identity $\delta[a z] = \delta[z]/|a|$, which holds for integration variable $z$ and constant $a$, tells us that $K\overline{K}K\overline{K}$ is proportional to
\begin{align}
    & \delta[x_0^+ + x_0^- - y_0^+ - y_0^-] \delta[x_0^+ - x_0^- + y_0^+ - y_0^-] \nonumber\\
    & \qquad
    = \, \frac{1}{2} \delta[x_0^+ - y_0^-] \delta[x_0^- - y_0^+] \, ,
\end{align}
where we have used that $x_0^+ + x_0^- = y_0^+ + y_0^-$ and $x_0^+ - x_0^- = -(y_0^+ - y_0^-)$ together imply $x_0^+ = y_0^-$ and $x_0^- = y_0^+$. Eq.\ \eqref{DSHO_Purity_KKKK_alphaZero_Step1}, therefore, simplifies to
\begin{align}
\label{DSHO_Purity_KKKK_alphaZero}
    K\overline{K}K\overline{K}[\alpha = 0]
    & \, = \, e^{2\gamma  T} \delta[x_0^+ - y_0^-] \delta[x_0^- - y_0^+] \, .
\end{align}
Plugging Eq.\ \eqref{DSHO_Purity_KKKK_alphaZero} back into Eq.\ \eqref{Purity_Def} informs us that, in the $\alpha=0$ case, the purity at time $t_f$ is the initial purity multiplied by an exponential growth factor,
\begin{align}
\label{DSHO_Purity_alphaZero}
    \text{Pu}[t_f]_{\alpha=0} \, = \, e^{2\gamma T} \text{Pu}[t_0] \, .
\end{align}
We conclude that the {\it quantum} DHO system cannot have a purely real frequency-squared, i.e., $\alpha \neq 0$ as long as $\gamma > 0$. This is because purity is constrained to lie within $[0,1]$, while the $e^{2\gamma T}$ factor in Eq.\ \eqref{DSHO_Purity_alphaZero} indicates that the $\alpha=0$ DHO system has an infinite purity in the asymptotic future. This was already noted in Eq.\ (\ref{eq:pualpha0}), but once again we see that it holds for arbitrary initial conditions.

Before closing this discussion on purity, we record here the $\alpha \neq 0$ integral representation of purity at time $t_f$,
\begin{widetext}
{\allowdisplaybreaks
\begin{align}
    \text{Pu}[t_f]
    & \, = \, e^{\gamma T} \omega \sqrt{-\frac{\gamma \Gamma^2}{\pi \alpha  \Upsilon[T]}} \int_{\mathbb{R}^{3}} \dd u^- \dd \xi^- \dd \xi'^- \delta[\xi^- + \xi'^-]
    \exp \bigg[ \frac{1}{\Upsilon[T]} \bigg\{ - 2 \alpha \frac{(\xi'^-)^2}{\gamma}  \left(\omega^2 - \gamma^2\cos[2 \Gamma  T] - \Gamma^2 \cosh[2\gamma T]\right) \nonumber\\
    & \qquad\qquad\qquad\qquad\qquad
    + \frac{(u^-)^2}{\alpha} e^{-2\gamma T} (4 \Gamma^2 \gamma \omega ^2) + (u^- \xi'^-) \left(8 i \gamma \omega ^2 \sin^2[\Gamma  T]\right) \bigg\} \bigg] \nonumber\\
    & \quad \ \times \int_{\mathbb{R}} \dd u^+ \left\langle \frac{1}{2} (u^+ + u^- - \xi'^-) \right\vert \hat{\rho}[t_0] \left\vert \frac{1}{2} (u^+ + u^- + \xi'^-) \right\rangle
    \left\langle \frac{1}{2} (u^+ - u^- + \xi'^-) \right\vert \hat{\rho}[t_0] \left\vert \frac{1}{2} (u^+ - u^- - \xi'^-) \right\rangle ,
\end{align}}
\end{widetext}
where the time-dependent $\Upsilon$ is defined as
\begin{align}
    \Upsilon[T] 
    & \, \equiv \,
    4 \Gamma^2 e^{-2\gamma T} + 4 \gamma^2 \cos[2 \Gamma  T] \nn\\
    & \qquad \qquad \quad 
    + 4 \Gamma \gamma \sin[2 \Gamma  T] - 4 \omega^2 \, .
\end{align}
In terms of the original integration variables occurring in Eq.\ \eqref{Purity_Def}, $\xi^- \equiv x_0^+ - x_0^-$, $\xi'^- \equiv y_0^+ - y_0^-$, $u^- \equiv (x_0^+ + x_0^- - y_0^+ - y_0^-)/2$, and $u^+ \equiv (x_0^+ + x_0^- + y_0^+ + y_0^-)/2$.

%%%%%%%%%%%%%%%%%%%%%%%%%%%%%%%%%%%%%%%%%%%%%%%%%%

%-------------------------------
\section{Discussion}
\label{sec:discussion}
%-------------------------------

The quantum DHO is a prototypical example of an exactly solvable dissipative quantum system. We revisited this simple system in this paper, with a goal of developing an effective field theory-inspired and influence functional-based method to describe its dynamics. We were particularly interested in what constraints one can impose on various terms that can appear in the influence functional. We thus considered a quadratic influence functional without describing the system-environment interaction that led us to it. We first found that the dissipation kernels we introduced are related in such a way that only two real functions are needed to describe the system's dynamics. We next restricted to time-local dissipation and solved for the exact Green's functions of the system in this case. We used the resulting Green's functions to show that late-time correlations and purity are independent of the initial conditions for both, Gaussian and more general initial states, and argued that the two dissipation kernels must in fact be constrained in order to obtain a physically meaningful purity in the late-time limit. We briefly considered time-nonlocal dissipation as well, and showed that the fluctuation-dissipation relation is satisfied for a specific choice of dissipation kernels.

It is worth highlighting that we discussed two different approaches to solve the problem -- in Secs.\ \ref{sec:Zinin}--\ref{sec:tnldissipation}, we focused on the in-in formalism, and in Sec.\ \ref{sec:beyondgaussianis}, we focused on a double in-out formalism. We found that each approach has its own advantages. For example, the former allowed the calculation of unequal-time correlations, which are crucial to understand whether the fluctuation-dissipation relation holds at late times. The latter, on the other hand, allowed us to go beyond Gaussian initial states and obtain results that are non-perturbative in the initial state. Which approach is ``better'' to use thus depends on the problem at hand, but it may be interesting to understand how to extend each approach.

We would also like to highlight a few other extensions of the methods developed in this paper that could be useful. First, we restricted our calculation to dissipation kernels $\gamma_1(t,t')$ and $\gamma_2(t,t')$ that are not proportional to $\d^2 \delta(t-t')/\d t'^2$. We further specialized first to time-local dissipation with $\gamma_{1R}(t,t') + \gamma_{2R}(t,t') = \gamma \d\delta(t-t')/\d t'$ and $\gamma_{1I}(t,t') = (\alpha/2) \delta(t-t')$, with $\gamma$ and $\alpha$ constant, and then to time-nonlocal dissipation with the specific choice of $\gamma_{1I}(t,t')$ given in Eq.\ (\ref{eq:gamma1Itnl}). It would be interesting to lift these restrictions to consider more general influence functionals -- time-dependent, generally nonlocal, and arising from derivative system-environment interactions. Second, it would be interesting to adapt these methods to describe a dissipative QFT in Minkowski spacetime as well as time-dependent spacetimes such as de Sitter. Such an effective field theory-inspired approach would again complement existing work on open QFTs. And third, it would be interesting to study loop corrections in dissipative QFTs in the presence of interactions and/or nonlinear dissipation, replacing the usual Green's functions in loop corrections with their dissipative analogs. This may, for example, help us to better understand the secular growth that is often found in loop corrections to observables in time-dependent quantum systems and QFTs; see \cite{Seery:2010kh,Tanaka:2013caa,Hu:2018nxy,Woodard:2023vcw} for reviews and \cite{Burgess:2009bs,Boyanovsky:2012qs,Pimentel:2012tw,Chen:2016nrs,Trunin:2018egi,Gorbenko:2019rza,Baumgart:2019clc,Trunin:2021lwg,Chaykov:2022zro,Chaykov:2022pwd} for related work and other possible resolutions.

%%%%%%%%%%%%%%%%%%%%%%%%%%%%%%%%%%%%%%%%%%%%%%%%%%

\acknowledgments

It is a pleasure to thank Brenden Bowen, Spasen Chaykov, Sachin Jain, and especially Archana Kamal for useful discussions. N.~A. was supported in part by the Department of Energy under Awards No. DE-SC0019515 and No. DE-SC0020360. Y.-Z.~C. was supported by the Taiwan NSTC Grant No. 111-2112-M-008-003.

%%%%%%%%%%%%%%%%%%%%%%%%%%%%%%%%%%%%%%%%%%%%%%%%%%

%------------------------------------------------
\appendix
\section{Generating functional in Eq.\ (\ref{eq:Zinin})}
\label{app:Zinin}
%------------------------------------------------

\renewcommand{\theequation}{A\arabic{equation}}
\setcounter{equation}{0}

In this appendix, we derive the generating functional in the absence of dissipation that was shown in Eq.\ (\ref{eq:Zinin}). This is a standard calculation, reproduced here for completeness and to highlight a few important points. As mentioned earlier, the generating functional is defined as $Z[J^+,J^-] = {\rm Tr} \[ \hat{\rho}(t_f) \]_{J^+,J^-}$, which can be written as
\bea
    Z[J^+,J^-] & = & {\rm Tr} \[ \hat{U}_{J^+}(t_f,t_0) \hat{\rho}(t_0) \hat{U}_{J^-}^{\dagger}(t_f,t_0) \] , \quad
\label{eq:Zinindef}
\eea
where $\hat{U}_{J^{\pm}}(t_f,t_0) = T \exp \[ -i \int_t \big( \hat{H} - J^{\pm} \hat{X} \big) \]$ is the time evolution operator for the harmonic oscillator problem we consider in this paper, with $T$ denoting time-ordering. The subscripts $J^{\pm}$ indicate that the forward time evolution is performed in the presence of the source $J^+(t)$ and the backward time evolution in the presence of the source $J^-(t)$.

The trace in Eq.\ (\ref{eq:Zinindef}) can be written as the integral $\int_{\mathbb{R}} \d x_f \la x_f | [ \ldots ] |x_f \ra$, where the subscript $f$ is simply a label to denote states at the final time $t_f$ and the dots indicate the operator on the right hand side of Eq.\ (\ref{eq:Zinindef}). Let us now insert a complete set of states on either side of $\hat{\rho}(t_0)$, say $\int_{\mathbb{R}} \d x^+_0 | x^+_0 \ra \la x^+_0 |$ on the left and $\int_{\mathbb{R}} \d x^-_0 | x^-_0 \ra \la x^-_0 |$ on the right, where the subscript $0$ is again a label that denotes states at the initial time $t_0$, and use the standard path integral representation of transition amplitudes,
\bea
	\la x_f | \hat{U}_{J^+}(t_f,t_0) | x^+_0 \ra & = & \int_{x^+_0}^{x_f} {\cal D}x^+ \, {\rm exp} \bigg[ i \bigg\{ S[x^+] \nn \\
    & & \quad + \ \int_t J^+ x^+ \bigg\} \bigg]
\label{eq:UJplus}
\eea
and
\bea
	\la x^-_0 | \hat{U}_{J^-}^{\dagger}(t_f,t_0) | x_f \ra & = & \int_{x^-_0}^{x_f} {\cal D}x^- \, {\rm exp} \bigg[ -i \bigg\{ S[x^-] \nn \\
    & & \quad + \ \int_t J^- x^- \bigg\} \bigg] .
\label{eq:UJminus}
\eea
We are now left with integrals over $x_f$, $x^+_0$, and $x^-_0$, and the matrix element $\la x^+_0 | \hat{\rho}(t_0) | x^-_0 \ra$, that we will denote $\rho[x^+_0, x^-_0, t_0]$. The integral over $x_f$ can be used to remove the upper limits from the path integrals in Eqs.\ (\ref{eq:UJplus}) and (\ref{eq:UJminus}), while inserting a $\delta$-function to impose that $x^+(t_f) = x^-(t_f)$. The integrals over $x^+_0$ and $x^-_0$, on the other hand, can be used to remove the lower limits and write the initial density matrix element as $\rho[x^+, x^-, t_0]$, with the functions $x^{\pm}(t)$ evaluated at $t_0$. The generating functional then becomes
\bea
    & & Z[J^+,J^-] \, = \, \int {\cal D}x^+ {\cal D}x^- \rho[x^+, x^-, t_0] \quad \nn \\
    & & \quad \times \, {\rm exp} \bigg[ i \left\{ S[x^+] + \int_t J^+ x^+ - S[x^-] - \int_t J^- x^- \right\} \bigg] \nn \\
    & & \quad \times \, \delta \[ x^+(t_f) - x^-(t_f) \] ,
\label{eq:Zininapp}
\eea
as given in Eq.\ (\ref{eq:Zinin}).

The $\delta$-function at the turn-around time $t_f$ also imposes the constraint $\dot{x}^+(t_f) = \dot{x}^-(t_f)$ that is needed to cancel the boundary terms at $t_f$, as discussed in Sec.\ \ref{sec:greensfunctions}. This was shown in \cite{Kaya:2012ks} and we reproduce their argument here for completeness. Following \cite{Kaya:2012ks}, let us first discretize the $x^+$ path integral and action as $\int {\cal D}x^+ = \lim_{\delta t \rightarrow 0} \int \d x^+_f \d x^+_{t_f-\delta t} \ldots \d x^+_0$ and
\bea
    S[x^{\pm}] \ = \ \frac{1}{2} \sum_t \delta t \, \Bigg[ \Bigg( \frac{x^{\pm}_t - x^{\pm}_{t-\delta t}}{\delta t} \Bigg)^2 - \omega^2 \( x^{\pm}_t \)^2 \Bigg] \, , \quad
\eea
where the summation runs over all $t$ between $t_0 + \delta t$ and $t_f$ in steps of $\delta t$, and similarly discretize the $x^-$ path integral and action. Using the $\delta$-function in Eq.\ (\ref{eq:Zininapp}), we can also write $\int_{\mathbb{R}^2} \d x^+_f \d x^-_f \delta \big( x^+_f - x^-_f \big)$ as a single integral that we denote $\int_{\mathbb{R}} \d x_f$, replacing $x^{\pm}_f$ everywhere with $x_f$. Let us now specifically consider the $x_f$ integral. The terms proportional to $\omega^2$ (and multiplied by $x_f^2$) cancel between $S[x^+]$ and $S[x^-]$. We can also set $J^{\pm}(t_f) = 0$ since $t_f$ is chosen to be later than any times at which we are interested in calculating correlation functions. Then the $x_f$ integral in the discretized version of the generating functional in Eq.\ (\ref{eq:Zininapp}) becomes
\bea
	\lim_{\delta t \rightarrow 0} \int_{\mathbb{R}} \d x_f \, {\rm exp} \[ -i x_f \( \frac{x^+_{t_f-\delta t} - x^-_{t_f-\delta t}}{\delta t} \) \] ,
\label{eq:deltaxdot}
\eea
which is simply $2\pi \delta \( \frac{x^+_{t_f-\delta t} - x^-_{t_f-\delta t}}{\delta t} \)$. This can be further simplified to $\delta \big( \dot{x}^+_f - \dot{x}^-_f \big)$ by Taylor expanding $x^{\pm}_{t_f-\delta t} = x_f - \dot{x}^{\pm}_f \delta t$. The path integral, therefore, imposes the constraint $\dot{x}^+(t_f) = \dot{x}^-(t_f)$.

Lastly, we note that the constraint $\dot{x}^+(t_f) = \dot{x}^-(t_f)$ may not necessarily hold in the presence of dissipation. Specifically, if the dissipation kernels $\gamma_1(t,t')$ and $\gamma_2(t,t')$ that we introduced in Eq.\ (\ref{eq:dissaction}) are proportional to $\d^2 \delta(t-t')/\d t'^2$, then the argument in the previous paragraph needs to be revisited as Eq.\ (\ref{eq:deltaxdot}) could potentially contain other terms. If, on the other hand, the dissipation kernels are proportional to $\d \delta(t-t')/\d t'$, as considered in the main text, then the constraint $\dot{x}^+(t_f) = \dot{x}^-(t_f)$ continues to hold.

%------------------------------------------------
\section{$n$-point correlations from $Z[J^+, J^-]$}
\label{app:npoint}
%------------------------------------------------

\renewcommand{\theequation}{B\arabic{equation}}
\setcounter{equation}{0}

In this appendix, we use the generating functional derived in Appendix\ \ref{app:Zinin} in the absence of dissipation to define the one- and two-point correlations of $\hat{X}(t)$. Consider first the one-point function $\la \hat{X}(t) \ra$, with angular brackets denoting the expectation value in $\hat{\rho}(t_0)$. Explicitly, $\la \hat{X}(t) \ra = {\rm Tr} \big[ \hat{\rho}(t_0) \hat{X}(t) \big] = {\rm Tr} \big[ \hat{\rho}(t_0) \hat{U}^{\dagger}(t,t_0) \hat{X}_S \hat{U}(t,t_0) \big]$, which can further be written as ${\rm Tr} \big[ \hat{U}(t_f,t) \hat{X}_S \hat{U}(t,t_0) \hat{\rho}(t_0) \hat{U}^{\dagger}(t_f,t_0) \big]$ by inserting $\hat{U}^{\dagger}(t_f,t) \hat{U}(t_f,t) = \hat{I}$ before $\hat{X}_S$ and rearranging operators inside the trace. Note that all of these steps can be done in the presence of sources $J^{\pm}(t)$ as well, as long as we set them to be equal at the end of the calculation. We can write the resulting expression in path integral form using a generalization of Eq.\ (\ref{eq:UJplus}),
\bea
	& & \la x_{t_f} | \hat{U}_{J^+}(t_f,t) \hat{X}_S \hat{U}_{J^+}(t,t_0) | x^+_{t_0} \ra \, = \, \int_{x^+_{t_0}}^{x_{t_f}} {\cal D}x^+ \, x^+(t) \nn \\
	& & \qquad \qquad \times \, {\rm exp} \bigg[ i \left\{ S[x^+] + \int_t J^+ x^+ \right\} \bigg] \, , \quad
\label{eq:UJpluswithX}
\eea
and Eq.\ (\ref{eq:UJminus}), and we see that $\la \hat{X}(t) \ra$ can, therefore, be written as
\bea
	\la \hat{X}(t) \ra & = & -i \frac{\delta}{\delta J^+(t)} \, Z[J^+,J^-] \Big|_{J^{\pm}=J} \, ,
\label{eq:def1ptappendix}
\eea
in the presence of a source $J(t)$.

Let us next consider the time-ordered two-point correlation $\big\la T \hat{X}(t) \hat{X}(t') \big\ra$, which can be written using similar manipulations as we did for the one-point function as
\bea
	\big\la T \hat{X}(t) \hat{X}(t') \big\ra & = & (-i)^2 \frac{\delta}{\delta J^+(t)} \frac{\delta}{\delta J^+(t')} \nn \\
	& & \quad \times \, Z[J^+,J^-] \Big|_{J^{\pm}=J} \, ,
\label{eq:def2ptwithT}
\eea
again in the presence of a source $J(t)$. Taking the Hermitian conjugate of the above expression further gives us the anti-time-ordered correlation $\big\la \bar{T} \hat{X}(t) \hat{X}(t') \big\ra$ in terms of functional derivatives with respect to $J^-(t)$ and $J^-(t')$ instead. Lastly, let us consider the two-point correlation $\big\la \hat{X}(t') \hat{X}(t) \big\ra$ without time-ordering. This can be written as ${\rm Tr} \big[ \big\{ \hat{U}(t_f,t) \hat{X}_S \hat{U}(t,t_0) \big\} \hat{\rho}(t_0) \big\{ \hat{U}^{\dagger}(t',t_0) \hat{X}_S \hat{U}^{\dagger}(t_f,t') \big\} \big]$, where we have grouped together terms such that the first $\hat{X}_S$ is inserted at time $t$ and the second at $t'$, without specifying their ordering. This can be expressed in terms of the generating functional as
\bea
	\big\la \hat{X}(t') \hat{X}(t) \big\ra & = & (-i) i \frac{\delta}{\delta J^+(t)} \frac{\delta}{\delta J^-(t')} \nn \\
	& & \quad \times \, Z[J^+,J^-] \Big|_{J^{\pm}=J} \, .
\label{eq:def2pt}
\eea
Taking the Hermitian conjugate of this expression similarly gives us the correlation $\big\la \hat{X}(t) \hat{X}(t') \big\ra$ in terms of functional derivatives with respect to $J^-(t)$ and $J^+(t')$ instead. The in-in generating functional, therefore, allows the calculation of non-time-ordered two-point correlations in addition to the usual time-ordered ones. 

Lastly, we note that replacing $Z[J^+,J^-]$ in the above expressions with $Z_{\rm diss.}[J^+,J^-]$, which we defined in Eq.\ (\ref{eq:ZininIF}), yields instead correlation functions in the presence of dissipation, as explained in Sec.\ \ref{subsec:npoint}.

%%%%%%%%%%%%%%%%%%%%%%%%%%%%%%%%%%%%%%%%%%%%%%%%%%

\bibliography{references}

\begin{thebibliography}{55}%
\makeatletter
\providecommand \@ifxundefined [1]{%
 \@ifx{#1\undefined}
}%
\providecommand \@ifnum [1]{%
 \ifnum #1\expandafter \@firstoftwo
 \else \expandafter \@secondoftwo
 \fi
}%
\providecommand \@ifx [1]{%
 \ifx #1\expandafter \@firstoftwo
 \else \expandafter \@secondoftwo
 \fi
}%
\providecommand \natexlab [1]{#1}%
\providecommand \enquote  [1]{``#1''}%
\providecommand \bibnamefont  [1]{#1}%
\providecommand \bibfnamefont [1]{#1}%
\providecommand \citenamefont [1]{#1}%
\providecommand \href@noop [0]{\@secondoftwo}%
\providecommand \href [0]{\begingroup \@sanitize@url \@href}%
\providecommand \@href[1]{\@@startlink{#1}\@@href}%
\providecommand \@@href[1]{\endgroup#1\@@endlink}%
\providecommand \@sanitize@url [0]{\catcode `\\12\catcode `\$12\catcode
  `\&12\catcode `\#12\catcode `\^12\catcode `\_12\catcode `\%12\relax}%
\providecommand \@@startlink[1]{}%
\providecommand \@@endlink[0]{}%
\providecommand \url  [0]{\begingroup\@sanitize@url \@url }%
\providecommand \@url [1]{\endgroup\@href {#1}{\urlprefix }}%
\providecommand \urlprefix  [0]{URL }%
\providecommand \Eprint [0]{\href }%
\providecommand \doibase [0]{http://dx.doi.org/}%
\providecommand \selectlanguage [0]{\@gobble}%
\providecommand \bibinfo  [0]{\@secondoftwo}%
\providecommand \bibfield  [0]{\@secondoftwo}%
\providecommand \translation [1]{[#1]}%
\providecommand \BibitemOpen [0]{}%
\providecommand \bibitemStop [0]{}%
\providecommand \bibitemNoStop [0]{.\EOS\space}%
\providecommand \EOS [0]{\spacefactor3000\relax}%
\providecommand \BibitemShut  [1]{\csname bibitem#1\endcsname}%
\let\auto@bib@innerbib\@empty
%</preamble>
\bibitem [{\citenamefont {Schwinger}(1961)}]{Schwinger:1960qe}%
  \BibitemOpen
  \bibfield  {author} {\bibinfo {author} {\bibfnamefont {J.~S.}\ \bibnamefont
  {Schwinger}},\ }\bibfield  {title} {\enquote {\bibinfo {title} {{Brownian
  motion of a quantum oscillator}},}\ }\href {\doibase 10.1063/1.1703727}
  {\bibfield  {journal} {\bibinfo  {journal} {J. Math. Phys.}\ }\textbf
  {\bibinfo {volume} {2}},\ \bibinfo {pages} {407--432} (\bibinfo {year}
  {1961})}\BibitemShut {NoStop}%
%%CITATION = JMAPA,2,407;%%
\bibitem [{\citenamefont {Bakshi}\ and\ \citenamefont
  {Mahanthappa}(1963{\natexlab{a}})}]{Bakshi:1962dv}%
  \BibitemOpen
  \bibfield  {author} {\bibinfo {author} {\bibfnamefont {P.~M.}\ \bibnamefont
  {Bakshi}}\ and\ \bibinfo {author} {\bibfnamefont {K.~T.}\ \bibnamefont
  {Mahanthappa}},\ }\bibfield  {title} {\enquote {\bibinfo {title}
  {{Expectation value formalism in quantum field theory. 1.}}}\ }\href
  {\doibase 10.1063/1.1703883} {\bibfield  {journal} {\bibinfo  {journal} {J.
  Math. Phys.}\ }\textbf {\bibinfo {volume} {4}},\ \bibinfo {pages} {1--11}
  (\bibinfo {year} {1963}{\natexlab{a}})}\BibitemShut {NoStop}%
%%CITATION = JMAPA,4,1;%%
\bibitem [{\citenamefont {Bakshi}\ and\ \citenamefont
  {Mahanthappa}(1963{\natexlab{b}})}]{Bakshi:1963bn}%
  \BibitemOpen
  \bibfield  {author} {\bibinfo {author} {\bibfnamefont {P.~M.}\ \bibnamefont
  {Bakshi}}\ and\ \bibinfo {author} {\bibfnamefont {K.~T.}\ \bibnamefont
  {Mahanthappa}},\ }\bibfield  {title} {\enquote {\bibinfo {title}
  {{Expectation value formalism in quantum field theory. 2.}}}\ }\href
  {\doibase 10.1063/1.1703879} {\bibfield  {journal} {\bibinfo  {journal} {J.
  Math. Phys.}\ }\textbf {\bibinfo {volume} {4}},\ \bibinfo {pages} {12--16}
  (\bibinfo {year} {1963}{\natexlab{b}})}\BibitemShut {NoStop}%
%%CITATION = JMAPA,4,12;%%
\bibitem [{\citenamefont {Keldysh}(1964)}]{Keldysh:1964ud}%
  \BibitemOpen
  \bibfield  {author} {\bibinfo {author} {\bibfnamefont {L.~V.}\ \bibnamefont
  {Keldysh}},\ }\bibfield  {title} {\enquote {\bibinfo {title} {{Diagram
  technique for nonequilibrium processes}},}\ }\href@noop {} {\bibfield
  {journal} {\bibinfo  {journal} {Zh. Eksp. Teor. Fiz.}\ }\textbf {\bibinfo
  {volume} {47}},\ \bibinfo {pages} {1515--1527} (\bibinfo {year}
  {1964})}\BibitemShut {NoStop}%
%%CITATION = ZETFA,47,1515;%%
\bibitem [{\citenamefont {Jordan}(1986)}]{Jordan:1986ug}%
  \BibitemOpen
  \bibfield  {author} {\bibinfo {author} {\bibfnamefont {R.~D.}\ \bibnamefont
  {Jordan}},\ }\bibfield  {title} {\enquote {\bibinfo {title} {{Effective field
  equations for expectation values}},}\ }\href {\doibase
  10.1103/PhysRevD.33.444} {\bibfield  {journal} {\bibinfo  {journal} {Phys.
  Rev.}\ }\textbf {\bibinfo {volume} {D33}},\ \bibinfo {pages} {444--454}
  (\bibinfo {year} {1986})}\BibitemShut {NoStop}%
%%CITATION = PHRVA,D33,444;%%
\bibitem [{\citenamefont {Feynman}\ and\ \citenamefont
  {Vernon}(1963)}]{Feynman:1963}%
  \BibitemOpen
  \bibfield  {author} {\bibinfo {author} {\bibfnamefont {R.~P.}\ \bibnamefont
  {Feynman}}\ and\ \bibinfo {author} {\bibfnamefont {F.~L.}\ \bibnamefont
  {Vernon}},\ }\bibfield  {title} {\enquote {\bibinfo {title} {{The theory of a
  general quantum system interacting with a linear dissipative system}},}\
  }\href {\doibase 10.1016/0003-4916(63)90068-X} {\bibfield  {journal}
  {\bibinfo  {journal} {Annals Phys.}\ }\textbf {\bibinfo {volume} {24}},\
  \bibinfo {pages} {118} (\bibinfo {year} {1963})}\BibitemShut {NoStop}%
\bibitem [{\citenamefont {Feynman}\ and\ \citenamefont
  {Hibbs}(2012)}]{Feynman:2012}%
  \BibitemOpen
  \bibfield  {author} {\bibinfo {author} {\bibfnamefont {R.~P.}\ \bibnamefont
  {Feynman}}\ and\ \bibinfo {author} {\bibfnamefont {A.~R.}\ \bibnamefont
  {Hibbs}},\ }\bibfield  {title} {\enquote {\bibinfo {title} {{Quantum
  mechanics and path integrals: Emended edition}},}\ }\href@noop {} {\bibfield
  {journal} {\bibinfo  {journal} {Dover Publications}\ } (\bibinfo {year}
  {2012})}\BibitemShut {NoStop}%
\bibitem [{\citenamefont {Breuer}\ and\ \citenamefont
  {Petruccione}(2002)}]{Breuer:2002pc}%
  \BibitemOpen
  \bibfield  {author} {\bibinfo {author} {\bibfnamefont {H.~P.}\ \bibnamefont
  {Breuer}}\ and\ \bibinfo {author} {\bibfnamefont {F.}~\bibnamefont
  {Petruccione}},\ }\bibfield  {title} {\enquote {\bibinfo {title} {{The theory
  of open quantum systems}},}\ }\href@noop {} {\bibfield  {journal} {\bibinfo
  {journal} {Oxford University Press}\ } (\bibinfo {year} {2002})}\BibitemShut
  {NoStop}%
\bibitem [{\citenamefont {Calzetta}\ and\ \citenamefont
  {Hu}(2008)}]{Calzetta:2008}%
  \BibitemOpen
  \bibfield  {author} {\bibinfo {author} {\bibfnamefont {E.~A.}\ \bibnamefont
  {Calzetta}}\ and\ \bibinfo {author} {\bibfnamefont {B.-L.~B.}\ \bibnamefont
  {Hu}},\ }\bibfield  {title} {\enquote {\bibinfo {title} {Nonequilibrium
  quantum field theory},}\ }\href@noop {} {\bibfield  {journal} {\bibinfo
  {journal} {Cambridge University Press}\ } (\bibinfo {year}
  {2008})}\BibitemShut {NoStop}%
\bibitem [{\citenamefont {Boyanovsky}(2015)}]{Boyanovsky:2015xoa}%
  \BibitemOpen
  \bibfield  {author} {\bibinfo {author} {\bibfnamefont {D.}~\bibnamefont
  {Boyanovsky}},\ }\bibfield  {title} {\enquote {\bibinfo {title} {{Effective
  field theory out of equilibrium: Brownian quantum fields}},}\ }\href
  {\doibase 10.1088/1367-2630/17/6/063017} {\bibfield  {journal} {\bibinfo
  {journal} {New J. Phys.}\ }\textbf {\bibinfo {volume} {17}},\ \bibinfo
  {pages} {063017} (\bibinfo {year} {2015})},\ \Eprint
  {http://arxiv.org/abs/1503.00156} {arXiv:1503.00156 [hep-ph]} \BibitemShut
  {NoStop}%
\bibitem [{\citenamefont {Weiss}(2012)}]{Weiss:2012}%
  \BibitemOpen
  \bibfield  {author} {\bibinfo {author} {\bibfnamefont {U.}~\bibnamefont
  {Weiss}},\ }\bibfield  {title} {\enquote {\bibinfo {title} {Quantum
  dissipative systems},}\ }\href@noop {} {\bibfield  {journal} {\bibinfo
  {journal} {World Scientific}\ } (\bibinfo {year} {2012})}\BibitemShut
  {NoStop}%
\bibitem [{\citenamefont {Hu}\ \emph {et~al.}(1992)\citenamefont {Hu},
  \citenamefont {Paz},\ and\ \citenamefont {Zhang}}]{Hu:1991di}%
  \BibitemOpen
  \bibfield  {author} {\bibinfo {author} {\bibfnamefont {B.~L.}\ \bibnamefont
  {Hu}}, \bibinfo {author} {\bibfnamefont {J.~P.}\ \bibnamefont {Paz}}, \ and\
  \bibinfo {author} {\bibfnamefont {Y.}~\bibnamefont {Zhang}},\ }\bibfield
  {title} {\enquote {\bibinfo {title} {{Quantum Brownian motion in a general
  environment: 1. Exact master equation with nonlocal dissipation and colored
  noise}},}\ }\href {\doibase 10.1103/PhysRevD.45.2843} {\bibfield  {journal}
  {\bibinfo  {journal} {Phys. Rev. D}\ }\textbf {\bibinfo {volume} {45}},\
  \bibinfo {pages} {2843--2861} (\bibinfo {year} {1992})}\BibitemShut {NoStop}%
\bibitem [{\citenamefont {Hu}\ \emph {et~al.}(1993)\citenamefont {Hu},
  \citenamefont {Paz},\ and\ \citenamefont {Zhang}}]{Hu:1993vs}%
  \BibitemOpen
  \bibfield  {author} {\bibinfo {author} {\bibfnamefont {B.~L.}\ \bibnamefont
  {Hu}}, \bibinfo {author} {\bibfnamefont {J.~P.}\ \bibnamefont {Paz}}, \ and\
  \bibinfo {author} {\bibfnamefont {Y.}~\bibnamefont {Zhang}},\ }\bibfield
  {title} {\enquote {\bibinfo {title} {{Quantum Brownian motion in a general
  environment. 2: Nonlinear coupling and perturbative approach}},}\ }\href
  {\doibase 10.1103/PhysRevD.47.1576} {\bibfield  {journal} {\bibinfo
  {journal} {Phys. Rev. D}\ }\textbf {\bibinfo {volume} {47}},\ \bibinfo
  {pages} {1576--1594} (\bibinfo {year} {1993})}\BibitemShut {NoStop}%
\bibitem [{\citenamefont {Magazz{\`u}}\ and\ \citenamefont
  {Grifoni}(2022)}]{Magazzu:2022}%
  \BibitemOpen
  \bibfield  {author} {\bibinfo {author} {\bibfnamefont {L.}~\bibnamefont
  {Magazz{\`u}}}\ and\ \bibinfo {author} {\bibfnamefont {M.}~\bibnamefont
  {Grifoni}},\ }\bibfield  {title} {\enquote {\bibinfo {title} {Feynman-vernon
  influence functional approach to quantum transport in interacting
  nanojunctions: An analytical hierarchical study},}\ }\href {\doibase
  10.1103/PhysRevB.105.125417} {\bibfield  {journal} {\bibinfo  {journal}
  {Phys. Rev. B}\ }\textbf {\bibinfo {volume} {105}},\ \bibinfo {pages}
  {125417} (\bibinfo {year} {2022})},\ \Eprint
  {http://arxiv.org/abs/2104.14497} {arXiv:2104.14497 [cond-mat]} \BibitemShut
  {NoStop}%
\bibitem [{\citenamefont {Koksma}\ \emph {et~al.}(2010)\citenamefont {Koksma},
  \citenamefont {Prokopec},\ and\ \citenamefont {Schmidt}}]{Koksma:2009wa}%
  \BibitemOpen
  \bibfield  {author} {\bibinfo {author} {\bibfnamefont {J.~F.}\ \bibnamefont
  {Koksma}}, \bibinfo {author} {\bibfnamefont {T.}~\bibnamefont {Prokopec}}, \
  and\ \bibinfo {author} {\bibfnamefont {M.~G.}\ \bibnamefont {Schmidt}},\
  }\bibfield  {title} {\enquote {\bibinfo {title} {{Decoherence in an
  interacting quantum field theory: The vacuum case}},}\ }\href {\doibase
  10.1103/PhysRevD.81.065030} {\bibfield  {journal} {\bibinfo  {journal} {Phys.
  Rev. D}\ }\textbf {\bibinfo {volume} {81}},\ \bibinfo {pages} {065030}
  (\bibinfo {year} {2010})},\ \Eprint {http://arxiv.org/abs/0910.5733}
  {arXiv:0910.5733 [hep-th]} \BibitemShut {NoStop}%
\bibitem [{\citenamefont {Koksma}\ \emph {et~al.}(2011)\citenamefont {Koksma},
  \citenamefont {Prokopec},\ and\ \citenamefont {Schmidt}}]{Koksma:2011dy}%
  \BibitemOpen
  \bibfield  {author} {\bibinfo {author} {\bibfnamefont {J.~F.}\ \bibnamefont
  {Koksma}}, \bibinfo {author} {\bibfnamefont {T.}~\bibnamefont {Prokopec}}, \
  and\ \bibinfo {author} {\bibfnamefont {M.~G.}\ \bibnamefont {Schmidt}},\
  }\bibfield  {title} {\enquote {\bibinfo {title} {{Decoherence in an
  interacting quantum field theory: Thermal case}},}\ }\href {\doibase
  10.1103/PhysRevD.83.085011} {\bibfield  {journal} {\bibinfo  {journal} {Phys.
  Rev. D}\ }\textbf {\bibinfo {volume} {83}},\ \bibinfo {pages} {085011}
  (\bibinfo {year} {2011})},\ \Eprint {http://arxiv.org/abs/1102.4713}
  {arXiv:1102.4713 [hep-th]} \BibitemShut {NoStop}%
\bibitem [{\citenamefont {Lombardo}\ and\ \citenamefont
  {Lopez~Nacir}(2005)}]{Lombardo:2005iz}%
  \BibitemOpen
  \bibfield  {author} {\bibinfo {author} {\bibfnamefont {F.~C.}\ \bibnamefont
  {Lombardo}}\ and\ \bibinfo {author} {\bibfnamefont {D.}~\bibnamefont
  {Lopez~Nacir}},\ }\bibfield  {title} {\enquote {\bibinfo {title}
  {{Decoherence during inflation: The generation of classical
  inhomogeneities}},}\ }\href {\doibase 10.1103/PhysRevD.72.063506} {\bibfield
  {journal} {\bibinfo  {journal} {Phys. Rev. D}\ }\textbf {\bibinfo {volume}
  {72}},\ \bibinfo {pages} {063506} (\bibinfo {year} {2005})},\ \Eprint
  {http://arxiv.org/abs/gr-qc/0506051} {arXiv:gr-qc/0506051} \BibitemShut
  {NoStop}%
\bibitem [{\citenamefont {Boyanovsky}(2016)}]{Boyanovsky:2016exa}%
  \BibitemOpen
  \bibfield  {author} {\bibinfo {author} {\bibfnamefont {D.}~\bibnamefont
  {Boyanovsky}},\ }\bibfield  {title} {\enquote {\bibinfo {title} {{Fermionic
  influence on inflationary fluctuations}},}\ }\href {\doibase
  10.1103/PhysRevD.93.083507} {\bibfield  {journal} {\bibinfo  {journal} {Phys.
  Rev. D}\ }\textbf {\bibinfo {volume} {93}},\ \bibinfo {pages} {083507}
  (\bibinfo {year} {2016})},\ \Eprint {http://arxiv.org/abs/1602.05609}
  {arXiv:1602.05609 [gr-qc]} \BibitemShut {NoStop}%
\bibitem [{\citenamefont {Boyanovsky}(2018)}]{Boyanovsky:2018soy}%
  \BibitemOpen
  \bibfield  {author} {\bibinfo {author} {\bibfnamefont {D.}~\bibnamefont
  {Boyanovsky}},\ }\bibfield  {title} {\enquote {\bibinfo {title} {{Imprint of
  entanglement entropy in the power spectrum of inflationary fluctuations}},}\
  }\href {\doibase 10.1103/PhysRevD.98.023515} {\bibfield  {journal} {\bibinfo
  {journal} {Phys. Rev. D}\ }\textbf {\bibinfo {volume} {98}},\ \bibinfo
  {pages} {023515} (\bibinfo {year} {2018})},\ \Eprint
  {http://arxiv.org/abs/1804.07967} {arXiv:1804.07967 [astro-ph.CO]}
  \BibitemShut {NoStop}%
\bibitem [{\citenamefont {Lombardo}\ and\ \citenamefont
  {Mazzitelli}(1996)}]{Lombardo:1995fg}%
  \BibitemOpen
  \bibfield  {author} {\bibinfo {author} {\bibfnamefont {F.}~\bibnamefont
  {Lombardo}}\ and\ \bibinfo {author} {\bibfnamefont {F.~D.}\ \bibnamefont
  {Mazzitelli}},\ }\bibfield  {title} {\enquote {\bibinfo {title} {{Coarse
  graining and decoherence in quantum field theory}},}\ }\href {\doibase
  10.1103/PhysRevD.53.2001} {\bibfield  {journal} {\bibinfo  {journal} {Phys.
  Rev. D}\ }\textbf {\bibinfo {volume} {53}},\ \bibinfo {pages} {2001--2011}
  (\bibinfo {year} {1996})},\ \Eprint {http://arxiv.org/abs/hep-th/9508052}
  {arXiv:hep-th/9508052} \BibitemShut {NoStop}%
\bibitem [{\citenamefont {Agon}\ \emph {et~al.}(2018)\citenamefont {Agon},
  \citenamefont {Balasubramanian}, \citenamefont {Kasko},\ and\ \citenamefont
  {Lawrence}}]{Agon:2014uxa}%
  \BibitemOpen
  \bibfield  {author} {\bibinfo {author} {\bibfnamefont {C.}~\bibnamefont
  {Agon}}, \bibinfo {author} {\bibfnamefont {V.}~\bibnamefont
  {Balasubramanian}}, \bibinfo {author} {\bibfnamefont {S.}~\bibnamefont
  {Kasko}}, \ and\ \bibinfo {author} {\bibfnamefont {A.}~\bibnamefont
  {Lawrence}},\ }\bibfield  {title} {\enquote {\bibinfo {title} {{Coarse
  grained quantum dynamics}},}\ }\href {\doibase 10.1103/PhysRevD.98.025019}
  {\bibfield  {journal} {\bibinfo  {journal} {Phys. Rev. D}\ }\textbf {\bibinfo
  {volume} {98}},\ \bibinfo {pages} {025019} (\bibinfo {year} {2018})},\
  \Eprint {http://arxiv.org/abs/1412.3148} {arXiv:1412.3148 [hep-th]}
  \BibitemShut {NoStop}%
\bibitem [{\citenamefont {Jana}\ \emph {et~al.}(2020)\citenamefont {Jana},
  \citenamefont {Loganayagam},\ and\ \citenamefont {Rangamani}}]{Jana:2020vyx}%
  \BibitemOpen
  \bibfield  {author} {\bibinfo {author} {\bibfnamefont {C.}~\bibnamefont
  {Jana}}, \bibinfo {author} {\bibfnamefont {R.}~\bibnamefont {Loganayagam}}, \
  and\ \bibinfo {author} {\bibfnamefont {M.}~\bibnamefont {Rangamani}},\
  }\bibfield  {title} {\enquote {\bibinfo {title} {{Open quantum systems and
  Schwinger-Keldysh holograms}},}\ }\href {\doibase 10.1007/JHEP07(2020)242}
  {\bibfield  {journal} {\bibinfo  {journal} {JHEP}\ }\textbf {\bibinfo
  {volume} {07}},\ \bibinfo {pages} {242} (\bibinfo {year} {2020})},\ \Eprint
  {http://arxiv.org/abs/2004.02888} {arXiv:2004.02888 [hep-th]} \BibitemShut
  {NoStop}%
\bibitem [{\citenamefont {BenTov}(2021)}]{BenTov:2021jsf}%
  \BibitemOpen
  \bibfield  {author} {\bibinfo {author} {\bibfnamefont {Y.}~\bibnamefont
  {BenTov}},\ }\bibfield  {title} {\enquote {\bibinfo {title}
  {{Schwinger-Keldysh path integral for the quantum harmonic oscillator}},}\
  }\href@noop {} {\  (\bibinfo {year} {2021})},\ \Eprint
  {http://arxiv.org/abs/2102.05029} {arXiv:2102.05029 [hep-th]} \BibitemShut
  {NoStop}%
\bibitem [{\citenamefont {Um}\ \emph {et~al.}(2002)\citenamefont {Um},
  \citenamefont {Yeon},\ and\ \citenamefont {George}}]{Um:2002ab}%
  \BibitemOpen
  \bibfield  {author} {\bibinfo {author} {\bibfnamefont {C.-I.}\ \bibnamefont
  {Um}}, \bibinfo {author} {\bibfnamefont {K.-H.}\ \bibnamefont {Yeon}}, \ and\
  \bibinfo {author} {\bibfnamefont {T.~F.}\ \bibnamefont {George}},\ }\bibfield
   {title} {\enquote {\bibinfo {title} {{The quantum damped harmonic
  oscillator}},}\ }\href {\doibase 10.1016/S0370-1573(01)00077-1} {\bibfield
  {journal} {\bibinfo  {journal} {Phys. Rept.}\ }\textbf {\bibinfo {volume}
  {362}},\ \bibinfo {pages} {63--192} (\bibinfo {year} {2002})}\BibitemShut
  {NoStop}%
\bibitem [{\citenamefont {M\'arkus}\ and\ \citenamefont
  {Gamb\'ar}(2022)}]{Markus:2022zbu}%
  \BibitemOpen
  \bibfield  {author} {\bibinfo {author} {\bibfnamefont {F.}~\bibnamefont
  {M\'arkus}}\ and\ \bibinfo {author} {\bibfnamefont {K.}~\bibnamefont
  {Gamb\'ar}},\ }\bibfield  {title} {\enquote {\bibinfo {title} {{A
  Potential-Based Quantization Procedure of the Damped Oscillator}},}\ }\href
  {\doibase 10.3390/quantum4040028} {\bibfield  {journal} {\bibinfo  {journal}
  {Quantum Rep.}\ }\textbf {\bibinfo {volume} {4}},\ \bibinfo {pages}
  {390--400} (\bibinfo {year} {2022})},\ \Eprint
  {http://arxiv.org/abs/2204.02893} {arXiv:2204.02893 [quant-ph]} \BibitemShut
  {NoStop}%
\bibitem [{Note1()}]{Note1}%
  \BibitemOpen
  \bibinfo {note} {We restrict to a time-independent $\omega $ in this paper.
  The problem with a more general $\omega (t)$, if not exactly solvable, may be
  solvable with the JWKB approximation if $\omega (t)$ is a slowly-varying
  function.}\BibitemShut {Stop}%
\bibitem [{\citenamefont {Berges}(2005)}]{Berges:2004yj}%
  \BibitemOpen
  \bibfield  {author} {\bibinfo {author} {\bibfnamefont {J.}~\bibnamefont
  {Berges}},\ }\bibfield  {title} {\enquote {\bibinfo {title} {{Introduction to
  nonequilibrium quantum field theory}},}\ }\href {\doibase 10.1063/1.1843591}
  {\bibfield  {journal} {\bibinfo  {journal} {AIP Conf. Proc.}\ }\textbf
  {\bibinfo {volume} {739}},\ \bibinfo {pages} {3--62} (\bibinfo {year}
  {2005})},\ \Eprint {http://arxiv.org/abs/hep-ph/0409233}
  {arXiv:hep-ph/0409233 [hep-ph]} \BibitemShut {NoStop}%
%%CITATION = HEP-PH/0409233;%%
\bibitem [{\citenamefont {Agarwal}\ \emph {et~al.}(2013)\citenamefont
  {Agarwal}, \citenamefont {Holman}, \citenamefont {Tolley},\ and\
  \citenamefont {Lin}}]{Agarwal:2012mq}%
  \BibitemOpen
  \bibfield  {author} {\bibinfo {author} {\bibfnamefont {N.}~\bibnamefont
  {Agarwal}}, \bibinfo {author} {\bibfnamefont {R.}~\bibnamefont {Holman}},
  \bibinfo {author} {\bibfnamefont {A.~J.}\ \bibnamefont {Tolley}}, \ and\
  \bibinfo {author} {\bibfnamefont {J.}~\bibnamefont {Lin}},\ }\bibfield
  {title} {\enquote {\bibinfo {title} {{Effective field theory and
  non-Gaussianity from general inflationary states}},}\ }\href {\doibase
  10.1007/JHEP05(2013)085} {\bibfield  {journal} {\bibinfo  {journal} {JHEP}\
  }\textbf {\bibinfo {volume} {05}},\ \bibinfo {pages} {085} (\bibinfo {year}
  {2013})},\ \Eprint {http://arxiv.org/abs/1212.1172} {arXiv:1212.1172
  [hep-th]} \BibitemShut {NoStop}%
%%CITATION = ARXIV:1212.1172;%%
\bibitem [{\citenamefont {Weinberg}(2005)}]{Weinberg:2005vy}%
  \BibitemOpen
  \bibfield  {author} {\bibinfo {author} {\bibfnamefont {S.}~\bibnamefont
  {Weinberg}},\ }\bibfield  {title} {\enquote {\bibinfo {title} {{Quantum
  contributions to cosmological correlations}},}\ }\href {\doibase
  10.1103/PhysRevD.72.043514} {\bibfield  {journal} {\bibinfo  {journal} {Phys.
  Rev.}\ }\textbf {\bibinfo {volume} {D72}},\ \bibinfo {pages} {043514}
  (\bibinfo {year} {2005})},\ \Eprint {http://arxiv.org/abs/hep-th/0506236}
  {arXiv:hep-th/0506236 [hep-th]} \BibitemShut {NoStop}%
%%CITATION = HEP-TH/0506236;%%
\bibitem [{\citenamefont {Caldeira}\ and\ \citenamefont
  {Leggett}(1983)}]{Caldeira:1982iu}%
  \BibitemOpen
  \bibfield  {author} {\bibinfo {author} {\bibfnamefont {A.~O.}\ \bibnamefont
  {Caldeira}}\ and\ \bibinfo {author} {\bibfnamefont {A.~J.}\ \bibnamefont
  {Leggett}},\ }\bibfield  {title} {\enquote {\bibinfo {title} {{Path integral
  approach to quantum Brownian motion}},}\ }\href {\doibase
  10.1016/0378-4371(83)90013-4} {\bibfield  {journal} {\bibinfo  {journal}
  {Physica A}\ }\textbf {\bibinfo {volume} {121}},\ \bibinfo {pages} {587--616}
  (\bibinfo {year} {1983})}\BibitemShut {NoStop}%
\bibitem [{Note2()}]{Note2}%
  \BibitemOpen
  \bibinfo {note} {Since boundary conditions on the retarded Green's function,
  $G^{\xi ,+-}(t,t')$, are set at $t = t'$, it must be a function of $t-t'$. We
  use the Fourier convention that $G^{\xi ,+-}(t,t') = \DOTSI \intop \ilimits@
  _{-\infty }^{\infty } \protect \frac {\protect \mathrm {d}\omega '}{2\pi }
  e^{-i\omega '(t-t')} \protect \tilde {G}^{\xi ,+-}(\omega ')$.}\BibitemShut
  {Stop}%
\bibitem [{Note3()}]{Note3}%
  \BibitemOpen
  \bibinfo {note} {It is worth noting that in the absence of dissipation, one
  can directly solve for $G^<(t,t')$ by modifying the ansatz in Eq.\ (\ref
  {eq:Gpptlhansatz}) to have different real constants in front of the $h(t)
  h^*(t')$ and $h^*(t) h(t')$ terms and adding to the three initial conditions
  in Eqs.\ (\ref {eq:phiphicorr2}), (\ref {eq:phimomcorr2}), and (\ref
  {eq:mommomcorr2}) the Wronskian condition in Eq.\ (\ref
  {eq:wronskiant0}).}\BibitemShut {Stop}%
\bibitem [{\citenamefont {Clerk}\ \emph {et~al.}(2010)\citenamefont {Clerk},
  \citenamefont {Devoret}, \citenamefont {Girvin}, \citenamefont {Marquardt},\
  and\ \citenamefont {Schoelkopf}}]{Clerk:2008tlb}%
  \BibitemOpen
  \bibfield  {author} {\bibinfo {author} {\bibfnamefont {A.~A.}\ \bibnamefont
  {Clerk}}, \bibinfo {author} {\bibfnamefont {M.~H.}\ \bibnamefont {Devoret}},
  \bibinfo {author} {\bibfnamefont {S.~M.}\ \bibnamefont {Girvin}}, \bibinfo
  {author} {\bibfnamefont {F.}~\bibnamefont {Marquardt}}, \ and\ \bibinfo
  {author} {\bibfnamefont {R.~J.}\ \bibnamefont {Schoelkopf}},\ }\bibfield
  {title} {\enquote {\bibinfo {title} {{Introduction to quantum noise,
  measurement, and amplification}},}\ }\href {\doibase
  10.1103/revmodphys.82.1155} {\bibfield  {journal} {\bibinfo  {journal} {Rev.
  Mod. Phys.}\ }\textbf {\bibinfo {volume} {82}},\ \bibinfo {pages}
  {1155--1208} (\bibinfo {year} {2010})},\ \Eprint
  {http://arxiv.org/abs/0810.4729} {arXiv:0810.4729 [cond-mat.mes-hall]}
  \BibitemShut {NoStop}%
\bibitem [{\citenamefont {Britto}\ \emph {et~al.}(2015)\citenamefont {Britto},
  \citenamefont {Das},\ and\ \citenamefont {Frenkel}}]{Britto:2015afa}%
  \BibitemOpen
  \bibfield  {author} {\bibinfo {author} {\bibfnamefont {A.~L.~M.}\
  \bibnamefont {Britto}}, \bibinfo {author} {\bibfnamefont {A.~K.}\
  \bibnamefont {Das}}, \ and\ \bibinfo {author} {\bibfnamefont
  {J.}~\bibnamefont {Frenkel}},\ }\bibfield  {title} {\enquote {\bibinfo
  {title} {{Generalized fluctuation-dissipation theorem in a soluble out of
  equilibrium model}},}\ }\href {\doibase 10.1103/PhysRevD.92.025020}
  {\bibfield  {journal} {\bibinfo  {journal} {Phys. Rev. D}\ }\textbf {\bibinfo
  {volume} {92}},\ \bibinfo {pages} {025020} (\bibinfo {year} {2015})},\
  \Eprint {http://arxiv.org/abs/1507.00896} {arXiv:1507.00896 [hep-th]}
  \BibitemShut {NoStop}%
\bibitem [{\citenamefont {Scarlatella}\ \emph {et~al.}(2019)\citenamefont
  {Scarlatella}, \citenamefont {Clerk},\ and\ \citenamefont
  {Schiro}}]{Scarlatella:2019}%
  \BibitemOpen
  \bibfield  {author} {\bibinfo {author} {\bibfnamefont {O.}~\bibnamefont
  {Scarlatella}}, \bibinfo {author} {\bibfnamefont {A.~A.}\ \bibnamefont
  {Clerk}}, \ and\ \bibinfo {author} {\bibfnamefont {M.}~\bibnamefont
  {Schiro}},\ }\bibfield  {title} {\enquote {\bibinfo {title} {Spectral
  functions and negative density of states of a driven-dissipative nonlinear
  quantum resonator},}\ }\href {\doibase 10.1088/1367-2630/ab0ce9} {\bibfield
  {journal} {\bibinfo  {journal} {New J. Phys.}\ }\textbf {\bibinfo {volume}
  {21}},\ \bibinfo {pages} {043040} (\bibinfo {year} {2019})},\ \Eprint
  {http://arxiv.org/abs/1811.03518} {arXiv:1811.03518 [quant-th]} \BibitemShut
  {NoStop}%
\bibitem [{\citenamefont {Grabert}\ \emph {et~al.}(1984)\citenamefont
  {Grabert}, \citenamefont {Weiss},\ and\ \citenamefont
  {Talkner}}]{Grabert:1984}%
  \BibitemOpen
  \bibfield  {author} {\bibinfo {author} {\bibfnamefont {H.}~\bibnamefont
  {Grabert}}, \bibinfo {author} {\bibfnamefont {U.}~\bibnamefont {Weiss}}, \
  and\ \bibinfo {author} {\bibfnamefont {P.}~\bibnamefont {Talkner}},\
  }\bibfield  {title} {\enquote {\bibinfo {title} {Quantum theory of the damped
  harmonic oscillator},}\ }\href {\doibase 10.1007/BF01307505} {\bibfield
  {journal} {\bibinfo  {journal} {Z. Phys. B}\ }\textbf {\bibinfo {volume}
  {55}},\ \bibinfo {pages} {87--94} (\bibinfo {year} {1984})}\BibitemShut
  {NoStop}%
\bibitem [{\citenamefont {Riseborough}\ \emph {et~al.}(1985)\citenamefont
  {Riseborough}, \citenamefont {H\"{a}nggi},\ and\ \citenamefont
  {Weiss}}]{Riseborough:1985}%
  \BibitemOpen
  \bibfield  {author} {\bibinfo {author} {\bibfnamefont {P.~S.}\ \bibnamefont
  {Riseborough}}, \bibinfo {author} {\bibfnamefont {P.}~\bibnamefont
  {H\"{a}nggi}}, \ and\ \bibinfo {author} {\bibfnamefont {U.}~\bibnamefont
  {Weiss}},\ }\bibfield  {title} {\enquote {\bibinfo {title} {Exact results for
  a damped quantum-mechanical harmonic oscillator},}\ }\href {\doibase
  10.1103/PhysRevA.31.471} {\bibfield  {journal} {\bibinfo  {journal} {Phys.
  Rev. A}\ }\textbf {\bibinfo {volume} {31}},\ \bibinfo {pages} {471} (\bibinfo
  {year} {1985})}\BibitemShut {NoStop}%
\bibitem [{\citenamefont {Grabert}\ \emph {et~al.}(1988)\citenamefont
  {Grabert}, \citenamefont {Schramm},\ and\ \citenamefont
  {Ingold}}]{Grabert:1988}%
  \BibitemOpen
  \bibfield  {author} {\bibinfo {author} {\bibfnamefont {H.}~\bibnamefont
  {Grabert}}, \bibinfo {author} {\bibfnamefont {P.}~\bibnamefont {Schramm}}, \
  and\ \bibinfo {author} {\bibfnamefont {G.-L.}\ \bibnamefont {Ingold}},\
  }\bibfield  {title} {\enquote {\bibinfo {title} {Quantum brownian motion: The
  functional integral approach},}\ }\href {\doibase
  10.1016/0370-1573(88)90023-3} {\bibfield  {journal} {\bibinfo  {journal}
  {Phys. Rep.}\ }\textbf {\bibinfo {volume} {168}},\ \bibinfo {pages}
  {115--207} (\bibinfo {year} {1988})}\BibitemShut {NoStop}%
\bibitem [{\citenamefont {H\"{a}nggi}\ and\ \citenamefont
  {Ingold}(2005)}]{Hanggi:2005}%
  \BibitemOpen
  \bibfield  {author} {\bibinfo {author} {\bibfnamefont {P.}~\bibnamefont
  {H\"{a}nggi}}\ and\ \bibinfo {author} {\bibfnamefont {G.-L.}\ \bibnamefont
  {Ingold}},\ }\bibfield  {title} {\enquote {\bibinfo {title} {Fundamental
  aspects of quantum brownian motion},}\ }\href {\doibase 10.1063/1.1853631}
  {\bibfield  {journal} {\bibinfo  {journal} {Chaos}\ }\textbf {\bibinfo
  {volume} {15}},\ \bibinfo {pages} {026105} (\bibinfo {year} {2005})},\
  \Eprint {http://arxiv.org/abs/quant-ph/0412052} {quant-ph/0412052 [quant-ph]}
  \BibitemShut {NoStop}%
\bibitem [{Note4()}]{Note4}%
  \BibitemOpen
  \bibinfo {note} {Note that these are proportional -- but not equal -- to the
  redefinition in Eq.\ \protect \textup {\hbox {\mathsurround \z@ \protect
  \normalfont (\ignorespaces \ref {eq:xibasis}\unskip \@@italiccorr )}} due to
  factors of $1/2$ there.}\BibitemShut {Stop}%
\bibitem [{\citenamefont {Seery}(2010)}]{Seery:2010kh}%
  \BibitemOpen
  \bibfield  {author} {\bibinfo {author} {\bibfnamefont {D.}~\bibnamefont
  {Seery}},\ }\bibfield  {title} {\enquote {\bibinfo {title} {{Infrared effects
  in inflationary correlation functions}},}\ }\href {\doibase
  10.1088/0264-9381/27/12/124005} {\bibfield  {journal} {\bibinfo  {journal}
  {Class. Quant. Grav.}\ }\textbf {\bibinfo {volume} {27}},\ \bibinfo {pages}
  {124005} (\bibinfo {year} {2010})},\ \Eprint {http://arxiv.org/abs/1005.1649}
  {arXiv:1005.1649 [astro-ph.CO]} \BibitemShut {NoStop}%
%%CITATION = ARXIV:1005.1649;%%
\bibitem [{\citenamefont {Tanaka}\ and\ \citenamefont
  {Urakawa}(2013)}]{Tanaka:2013caa}%
  \BibitemOpen
  \bibfield  {author} {\bibinfo {author} {\bibfnamefont {T.}~\bibnamefont
  {Tanaka}}\ and\ \bibinfo {author} {\bibfnamefont {Y.}~\bibnamefont
  {Urakawa}},\ }\bibfield  {title} {\enquote {\bibinfo {title} {{Loops in
  inflationary correlation functions}},}\ }\href {\doibase
  10.1088/0264-9381/30/23/233001} {\bibfield  {journal} {\bibinfo  {journal}
  {Class. Quant. Grav.}\ }\textbf {\bibinfo {volume} {30}},\ \bibinfo {pages}
  {233001} (\bibinfo {year} {2013})},\ \Eprint {http://arxiv.org/abs/1306.4461}
  {arXiv:1306.4461 [hep-th]} \BibitemShut {NoStop}%
%%CITATION = ARXIV:1306.4461;%%
\bibitem [{\citenamefont {Hu}(2018)}]{Hu:2018nxy}%
  \BibitemOpen
  \bibfield  {author} {\bibinfo {author} {\bibfnamefont {B.-L.}\ \bibnamefont
  {Hu}},\ }\bibfield  {title} {\enquote {\bibinfo {title} {{Infrared behavior
  of quantum fields in inflationary cosmology -- Issues and approaches: An
  overview}},}\ }\href@noop {} {\  (\bibinfo {year} {2018})},\ \Eprint
  {http://arxiv.org/abs/1812.11851} {arXiv:1812.11851 [gr-qc]} \BibitemShut
  {NoStop}%
%%CITATION = ARXIV:1812.11851;%%
\bibitem [{\citenamefont {Woodard}(2023)}]{Woodard:2023vcw}%
  \BibitemOpen
  \bibfield  {author} {\bibinfo {author} {\bibfnamefont {R.~P.}\ \bibnamefont
  {Woodard}},\ }\bibfield  {title} {\enquote {\bibinfo {title} {{Big Steve and
  the state of the Universe}},}\ }\href {\doibase 10.3390/sym15040856} {\
  (\bibinfo {year} {2023}),\ 10.3390/sym15040856},\ \Eprint
  {http://arxiv.org/abs/2303.05111} {arXiv:2303.05111 [hep-th]} \BibitemShut
  {NoStop}%
\bibitem [{\citenamefont {Burgess}\ \emph {et~al.}(2010)\citenamefont
  {Burgess}, \citenamefont {Leblond}, \citenamefont {Holman},\ and\
  \citenamefont {Shandera}}]{Burgess:2009bs}%
  \BibitemOpen
  \bibfield  {author} {\bibinfo {author} {\bibfnamefont {C.~P.}\ \bibnamefont
  {Burgess}}, \bibinfo {author} {\bibfnamefont {L.}~\bibnamefont {Leblond}},
  \bibinfo {author} {\bibfnamefont {R.}~\bibnamefont {Holman}}, \ and\ \bibinfo
  {author} {\bibfnamefont {S.}~\bibnamefont {Shandera}},\ }\bibfield  {title}
  {\enquote {\bibinfo {title} {{Super-Hubble de Sitter fluctuations and the
  dynamical RG}},}\ }\href {\doibase 10.1088/1475-7516/2010/03/033} {\bibfield
  {journal} {\bibinfo  {journal} {JCAP}\ }\textbf {\bibinfo {volume} {1003}},\
  \bibinfo {pages} {033} (\bibinfo {year} {2010})},\ \Eprint
  {http://arxiv.org/abs/0912.1608} {arXiv:0912.1608 [hep-th]} \BibitemShut
  {NoStop}%
%%CITATION = ARXIV:0912.1608;%%
\bibitem [{\citenamefont {Boyanovsky}(2012)}]{Boyanovsky:2012qs}%
  \BibitemOpen
  \bibfield  {author} {\bibinfo {author} {\bibfnamefont {D.}~\bibnamefont
  {Boyanovsky}},\ }\bibfield  {title} {\enquote {\bibinfo {title} {{Condensates
  and quasiparticles in inflationary cosmology: Mass generation and decay
  widths}},}\ }\href {\doibase 10.1103/PhysRevD.85.123525} {\bibfield
  {journal} {\bibinfo  {journal} {Phys. Rev.}\ }\textbf {\bibinfo {volume}
  {D85}},\ \bibinfo {pages} {123525} (\bibinfo {year} {2012})},\ \Eprint
  {http://arxiv.org/abs/1203.3903} {arXiv:1203.3903 [hep-ph]} \BibitemShut
  {NoStop}%
%%CITATION = ARXIV:1203.3903;%%
\bibitem [{\citenamefont {Pimentel}\ \emph {et~al.}(2012)\citenamefont
  {Pimentel}, \citenamefont {Senatore},\ and\ \citenamefont
  {Zaldarriaga}}]{Pimentel:2012tw}%
  \BibitemOpen
  \bibfield  {author} {\bibinfo {author} {\bibfnamefont {G.~L.}\ \bibnamefont
  {Pimentel}}, \bibinfo {author} {\bibfnamefont {L.}~\bibnamefont {Senatore}},
  \ and\ \bibinfo {author} {\bibfnamefont {M.}~\bibnamefont {Zaldarriaga}},\
  }\bibfield  {title} {\enquote {\bibinfo {title} {{On Loops in inflation III:
  Time independence of $\zeta$ in single clock inflation}},}\ }\href {\doibase
  10.1007/JHEP07(2012)166} {\bibfield  {journal} {\bibinfo  {journal} {JHEP}\
  }\textbf {\bibinfo {volume} {07}},\ \bibinfo {pages} {166} (\bibinfo {year}
  {2012})},\ \Eprint {http://arxiv.org/abs/1203.6651} {arXiv:1203.6651
  [hep-th]} \BibitemShut {NoStop}%
%%CITATION = ARXIV:1203.6651;%%
\bibitem [{\citenamefont {Chen}\ \emph {et~al.}(2016)\citenamefont {Chen},
  \citenamefont {Wang},\ and\ \citenamefont {Xianyu}}]{Chen:2016nrs}%
  \BibitemOpen
  \bibfield  {author} {\bibinfo {author} {\bibfnamefont {X.}~\bibnamefont
  {Chen}}, \bibinfo {author} {\bibfnamefont {Y.}~\bibnamefont {Wang}}, \ and\
  \bibinfo {author} {\bibfnamefont {Z.-Z.}\ \bibnamefont {Xianyu}},\ }\bibfield
   {title} {\enquote {\bibinfo {title} {{Loop corrections to standard model
  fields in inflation}},}\ }\href {\doibase 10.1007/JHEP08(2016)051} {\bibfield
   {journal} {\bibinfo  {journal} {JHEP}\ }\textbf {\bibinfo {volume} {08}},\
  \bibinfo {pages} {051} (\bibinfo {year} {2016})},\ \Eprint
  {http://arxiv.org/abs/1604.07841} {arXiv:1604.07841 [hep-th]} \BibitemShut
  {NoStop}%
%%CITATION = ARXIV:1604.07841;%%
\bibitem [{\citenamefont {Trunin}(2018)}]{Trunin:2018egi}%
  \BibitemOpen
  \bibfield  {author} {\bibinfo {author} {\bibfnamefont {D.~A.}\ \bibnamefont
  {Trunin}},\ }\bibfield  {title} {\enquote {\bibinfo {title} {{Comments on the
  adiabatic theorem}},}\ }\href {\doibase 10.1142/S0217751X18501403} {\bibfield
   {journal} {\bibinfo  {journal} {Int. J. Mod. Phys. A}\ }\textbf {\bibinfo
  {volume} {33}},\ \bibinfo {pages} {1850140} (\bibinfo {year} {2018})},\
  \Eprint {http://arxiv.org/abs/1805.04856} {arXiv:1805.04856 [hep-th]}
  \BibitemShut {NoStop}%
\bibitem [{\citenamefont {Gorbenko}\ and\ \citenamefont
  {Senatore}(2019)}]{Gorbenko:2019rza}%
  \BibitemOpen
  \bibfield  {author} {\bibinfo {author} {\bibfnamefont {V.}~\bibnamefont
  {Gorbenko}}\ and\ \bibinfo {author} {\bibfnamefont {L.}~\bibnamefont
  {Senatore}},\ }\bibfield  {title} {\enquote {\bibinfo {title} {{$\lambda
  \phi^4$ in dS}},}\ }\href@noop {} {\  (\bibinfo {year} {2019})},\ \Eprint
  {http://arxiv.org/abs/1911.00022} {arXiv:1911.00022 [hep-th]} \BibitemShut
  {NoStop}%
\bibitem [{\citenamefont {Baumgart}\ and\ \citenamefont
  {Sundrum}(2020)}]{Baumgart:2019clc}%
  \BibitemOpen
  \bibfield  {author} {\bibinfo {author} {\bibfnamefont {M.}~\bibnamefont
  {Baumgart}}\ and\ \bibinfo {author} {\bibfnamefont {R.}~\bibnamefont
  {Sundrum}},\ }\bibfield  {title} {\enquote {\bibinfo {title} {{De Sitter
  diagrammar and the resummation of time}},}\ }\href {\doibase
  10.1007/JHEP07(2020)119} {\bibfield  {journal} {\bibinfo  {journal} {JHEP}\
  }\textbf {\bibinfo {volume} {07}},\ \bibinfo {pages} {119} (\bibinfo {year}
  {2020})},\ \Eprint {http://arxiv.org/abs/1912.09502} {arXiv:1912.09502
  [hep-th]} \BibitemShut {NoStop}%
\bibitem [{\citenamefont {Trunin}(2021)}]{Trunin:2021lwg}%
  \BibitemOpen
  \bibfield  {author} {\bibinfo {author} {\bibfnamefont {D.~A.}\ \bibnamefont
  {Trunin}},\ }\bibfield  {title} {\enquote {\bibinfo {title} {{Particle
  creation in nonstationary large N quantum mechanics}},}\ }\href {\doibase
  10.1103/PhysRevD.104.045001} {\bibfield  {journal} {\bibinfo  {journal}
  {Phys. Rev. D}\ }\textbf {\bibinfo {volume} {104}},\ \bibinfo {pages}
  {045001} (\bibinfo {year} {2021})},\ \Eprint
  {http://arxiv.org/abs/2105.01647} {arXiv:2105.01647 [hep-th]} \BibitemShut
  {NoStop}%
\bibitem [{\citenamefont {Chaykov}\ \emph
  {et~al.}(2023{\natexlab{a}})\citenamefont {Chaykov}, \citenamefont {Agarwal},
  \citenamefont {Bahrami},\ and\ \citenamefont {Holman}}]{Chaykov:2022zro}%
  \BibitemOpen
  \bibfield  {author} {\bibinfo {author} {\bibfnamefont {S.}~\bibnamefont
  {Chaykov}}, \bibinfo {author} {\bibfnamefont {N.}~\bibnamefont {Agarwal}},
  \bibinfo {author} {\bibfnamefont {S.}~\bibnamefont {Bahrami}}, \ and\
  \bibinfo {author} {\bibfnamefont {R.}~\bibnamefont {Holman}},\ }\bibfield
  {title} {\enquote {\bibinfo {title} {{Loop corrections in Minkowski spacetime
  away from equilibrium. Part I. Late-time resummations}},}\ }\href {\doibase
  10.1007/JHEP02(2023)093} {\bibfield  {journal} {\bibinfo  {journal} {JHEP}\
  }\textbf {\bibinfo {volume} {02}},\ \bibinfo {pages} {093} (\bibinfo {year}
  {2023}{\natexlab{a}})},\ \Eprint {http://arxiv.org/abs/2206.11288}
  {arXiv:2206.11288 [hep-th]} \BibitemShut {NoStop}%
\bibitem [{\citenamefont {Chaykov}\ \emph
  {et~al.}(2023{\natexlab{b}})\citenamefont {Chaykov}, \citenamefont {Agarwal},
  \citenamefont {Bahrami},\ and\ \citenamefont {Holman}}]{Chaykov:2022pwd}%
  \BibitemOpen
  \bibfield  {author} {\bibinfo {author} {\bibfnamefont {S.}~\bibnamefont
  {Chaykov}}, \bibinfo {author} {\bibfnamefont {N.}~\bibnamefont {Agarwal}},
  \bibinfo {author} {\bibfnamefont {S.}~\bibnamefont {Bahrami}}, \ and\
  \bibinfo {author} {\bibfnamefont {R.}~\bibnamefont {Holman}},\ }\bibfield
  {title} {\enquote {\bibinfo {title} {{Loop corrections in Minkowski spacetime
  away from equilibrium. Part II. Finite-time results}},}\ }\href {\doibase
  10.1007/JHEP02(2023)094} {\bibfield  {journal} {\bibinfo  {journal} {JHEP}\
  }\textbf {\bibinfo {volume} {02}},\ \bibinfo {pages} {094} (\bibinfo {year}
  {2023}{\natexlab{b}})},\ \Eprint {http://arxiv.org/abs/2206.11289}
  {arXiv:2206.11289 [hep-th]} \BibitemShut {NoStop}%
\bibitem [{\citenamefont {Kaya}(2015)}]{Kaya:2012ks}%
  \BibitemOpen
  \bibfield  {author} {\bibinfo {author} {\bibfnamefont {A.}~\bibnamefont
  {Kaya}},\ }\bibfield  {title} {\enquote {\bibinfo {title} {{The functional
  measure for the in-in path integral}},}\ }\href {\doibase
  10.1088/0264-9381/32/9/095008} {\bibfield  {journal} {\bibinfo  {journal}
  {Class. Quant. Grav.}\ }\textbf {\bibinfo {volume} {32}},\ \bibinfo {pages}
  {095008} (\bibinfo {year} {2015})},\ \Eprint {http://arxiv.org/abs/1212.3066}
  {arXiv:1212.3066 [hep-th]} \BibitemShut {NoStop}%
%%CITATION = ARXIV:1212.3066;%%
\end{thebibliography}%

\end{document}